\documentclass[12pt]{article}
\usepackage{graphicx,subfig}
\usepackage{feynmf}
\usepackage{amssymb}
\usepackage{amsmath}
\usepackage[normalem]{ulem}
\usepackage[dvipsnames]{xcolor}
\usepackage{cellspace}
\usepackage{cite}
\newcommand{\msbar}{\ensuremath{\overline{{\rm MS}}}}
\newcommand{\dilog}{\mbox{Li}_2}

\newcommand{\mK}{\mathcal{K}}

\newcommand{\bD}{\bar{D}}
\newcommand{\bA}{\bar{A}}

\newcommand{\cala}{\mathcal A}
\newcommand{\calb}{\mathcal B}
\newcommand{\calc}{\mathcal C}
\newcommand{\calf}{\mathcal F}
\newcommand{\calh}{\mathcal H}
\newcommand{\calj}{\mathcal J}

\newcommand{\calt}{\mathcal T}

\newcommand{\bd}{\bar{d}}
\newcommand{\yNmax}{y_{N \, \text{max}}}
\newcommand{\yNmin}{y_{N \, \text{min}}}
\newcommand{\ytNmax}{\tilde{y}_{N \, \text{max}}}
\newcommand{\ytNmin}{\tilde{y}_{N \, \text{min}}}
\newcommand{\ptm}{p_{T \, \text{m}}}
\newcommand{\xtm}{x_{T \, \text{m}}}
\newcommand{\xtmax}[1]{x_{T \, #1 \, \text{max}}}
\newcommand{\xtmaxp}[1]{x_{T \, #1 \, \text{max}}^{\prime}}

\newcommand{\xt}[1]{x_{T \, #1}}
\newcommand{\Xt}[1]{X_{T \, #1}}
\newcommand{\dyM}{\Delta \, Y_M}
\newcommand{\dytM}{\Delta \, \tilde{Y}_M}
\newcommand{\dym}{\Delta \, Y_m}

\newcommand{\pt}[1]{p_{T \, #1}}
\newcommand{\vpt}[1]{\vec{p}_{T \, #1}}
\newcommand{\kt}[1]{K_{T \, #1}}
\newcommand{\vkt}[1]{\vec{K}_{T \, #1}}

\newcommand{\zmin}[1]{z_{#1 \, \text{min}}}

\newcommand{\xmax}[1]{x_{#1 \, \text{max}}}

\newcommand{\rth}{R_{\text{th}}}

\newcommand{\Deltaymax}{\Delta \tilde{y}_{\text{max}}}
\newcommand{\zm}[1]{z_{#1 \, \text{m}}}
\newcommand{\zmj}{\zeta_{\text{jet} \, \text{m}}}

\newcommand{\bx}[1]{\bar{x}_{#1}}

\newcommand{\Hn}[1]{H^{(n)}_{#1}}
\newcommand{\Hq}[1]{H^{(4)}_{#1}}
\newcommand{\piij}{\Pi_{ij}}
\newcommand{\pitij}{\Pi_{T \, ij}}
\newcommand{\vpitij}{\vec{\Pi}_{T \, ij}}
\newcommand{\mtij}{m_{T \, ij}}
\newcommand{\Jso}{J^{\text{soft}}_{ij}}
\newcommand{\Hyp}{{}_2\text{F}_1}
\newcommand{\bup}{\bar{u}_{+}}
\newcommand{\bum}{\bar{u}_{-}}
\newcommand{\bupm}{\bar{u}_{\pm}}
\newcommand{\tup}{\tilde{u}_{+}}
\newcommand{\tum}{\tilde{u}_{-}}
\newcommand{\tupm}{\tilde{u}_{\pm}}
\newcommand{\ystij}{y^{\star}_{ij}}
\newcommand{\vpthat}[1]{\vec{\hat{p}}_{T \, #1}}
\newcommand{\hp}{\hat{p}}

\newcommand{\czm}[1]{\zeta_{#1 \, \text{m}}}
\newcommand{\ktj}{K_{T \, \text{jet}}}
\newcommand{\yj}{y_{\text{jet}}}
\newcommand{\vktj}{\vec{K}_{T \, \text{jet}}}
\newcommand{\Xtj}{X_{T \, \text{jet}}}

\newcommand{\psla}{\mbox{$\not{\! p}$}}
\newcommand{\Tmat}[3]{\left(T^{#1}\right)_{#2 #3}}

\newcommand{\reac}[2]{[#1]_{#2}}
\newcommand{\indic}[2]{(#1)_{#2}}
\newcommand{\seq}[2]{\{#1\}_{#2}}
\newcommand{\indict}[4]{(#1|#2:#3)_{#4}}
\newcommand{\seqt}[4]{\left\{#1|#2:#3\right\}_{#4}}
\newcommand{\react}[4]{[#1|#2:#3]_{#4}}

\newcommand{\PSN}{d \text{PS}_N^{(n)}}

\newcommand{\PSH}[1]{d \, \text{PS}_{N-1 \,\text{h}}^{(n)}(#1)}
\newcommand{\PSHQ}[1]{d \, \text{PS}_{N-1 \,\text{h}}^{(4)}(#1)}
\newcommand{\PSHP}[2]{d \, \text{PS}_{#1 \,\text{h}}^{(n)}(#2)}
\newcommand{\PSHQP}[2]{d \, \text{PS}_{#1 \,\text{h}}^{(4)}(#2)}
\newcommand{\Ti}[1]{T^{(#1)}}

\newcommand{\cTiin}[1]{\calt_{\text{in}}^{(#1)}}
\newcommand{\cTiins}[1]{\calt_{\text{in}}^{(#1) \, \text{soft}}}
\newcommand{\cTiinc}[1]{\calt_{\text{in}}^{(#1) \, \text{coll}}}
\newcommand{\cTiinci}[1]{\calt_{\text{in}}^{(#1) \, \text{coll ini}}}
\newcommand{\cTiincf}[1]{\calt_{\text{in}}^{(#1) \, \text{coll fin}}}
\newcommand{\cTiindiv}[1]{\calt_{\text{in}}^{(#1) \, \text{div}}}
\newcommand{\cTioutc}[2]{\calt_{\text{out}}^{(#1,#2) \, \text{coll}}}
\newcommand{\cTiout}[1]{\calt_{\text{out}}^{(#1)}}
\newcommand{\cTioutdiv}[1]{\calt_{\text{out}}^{(#1) \, \text{div}}}

\newcommand{\cTitotdiv}{\calt_{\text{tot}}^{\text{div}}}
\newcommand{\cTitotsoftdiv}{\calt_{\text{tot}}^{\text{ soft}}}

\numberwithin{equation}{section}

\usepackage{ulem,lipsum}
\newcommand\xoutpars[1]{\let\helpcmd\xout\parhelp#1\par\relax\relax}
\newcommand\soutpars[1]{\let\helpcmd\sout\parhelp#1\par\relax\relax}
\long\def\parhelp#1\par#2\relax{%
\helpcmd{#1}\ifx\relax#2\else\par\parhelp#2\relax\fi%
}

\usepackage{pifont}

\usepackage[many]{tcolorbox}
\newtcolorbox{cross}{blank,breakable,parbox=false,
  overlay={\draw[red,line width=1pt] (interior.south west)--(interior.north east);
    \draw[red,line width=1pt] (interior.north west)--(interior.south east);}}

\begin{document}

\setlength{\unitlength}{1mm}
\begin{fmffile}{simplepics}

\begin{titlepage}

\vspace{1.cm}

\long\def\symbolfootnote[#1]#2{\begingroup%
\def\thefootnote{\fnsymbol{footnote}}\footnote[#1]{#2}\endgroup}

\begin{center}

{\large \bf A subtraction scheme for processes involving fragmentation functions at NLO}\\[2cm]

{\large M.~S.~Zidi$^{a}$, J.~Ph.~Guillet$^{b}$, I.~Schienbein$^{c}$ and H.~Zaraket$^{c,d}$} \\[.5cm]
\normalsize
{$^{a}$ LPTh, Department of Physics, University of Jijel,
B.P. 98 Ouled Aissa, 18000 Jijel, Algeria}\\
{$^{b}$ LAPTH, Univ. Savoie Mont Blanc, CNRS, F-74000 Annecy, France}\\
{$^{c}$}{Laboratoire de Physique Subatomique et de Cosmologie, Université Grenoble-Alpes, CNRS/IN2P3,
53 Avenue des Martyrs, 38026 Grenoble, France}\\
{$^{d}$}{Multi-Disciplinary Physics Laboratory, Optics and Fiber Optics Group, Faculty of
Sciences, Lebanese University, Lebanon}\\
\today
\end{center}

\vspace{2cm}

\begin{abstract}
\noindent
We present a novel subtraction method to remove the soft and collinear divergences at next-to-leading order for processes involving an arbitrary number of fragmentation functions, where this method acts directly in the hadronic centre-of-mass frame. We provide the analytical formulae of the subtraction terms in the general case where all the final state partons fragment to hadrons and for the two special cases when one of the partons of the final state does not fragment, i.e. it is a photon or involved in a jet. 
\end{abstract}

\vspace{1cm}

\begin{flushright}
LAPTH-013/24\\
\end{flushright}

\vspace{2cm}

\end{titlepage}

\newpage

\tableofcontents

\newpage

\section{Introduction}
Among the processes constituting the Standard Model background, those involving fragmentation functions (FFs) play a crucial role, with prompt photon production being a well-known example. In such production, two components stand out: the direct component, where the photon is produced directly in the hard sub-process, and the fragmentation component, where the photon is emitted collinearly by a hard parton. While the latter component can be significantly reduced by implementing isolation criteria, it cannot be completely eliminated due to finite resolutions in energy and angle of the detectors. Given the precision of experimental data at the LHC, accounting for this contribution is imperative.
For instance, studies on di-photon production at NLO, as demonstrated in references \cite{Binoth:1999qq} and \cite{Binoth:2000zt}, have revealed the significance of including fragmentation components. These components provide a qualitative understanding of the data by considering additional topologies within the collinear approximation, which were absent when solely considering NLO corrections to the direct component. Consequently, this approach has led to improved data descriptions concerning distributions such as the azimuthal angle between the two photons or the transverse momentum of the photon pair.
To achieve a quantitative understanding, computations of the direct part must progress to next-to-next-to-leading order (NNLO) accuracy \cite{Catani:2018krb,Gehrmann:2020oec}, which encompasses these topologies beyond the collinear approximation. Notably, recent advancements have reached NNLO accuracy for the fragmentation component of inclusive photon production \cite{Gehrmann:2022cih,Chen:2022gpk}, complementing the direct contribution \cite{Campbell:2016lzl,Chen:2019zmr}.
A second example, also well-known, pertains to the production of heavy quarks, particularly charm ($c$) and bottom ($b$) quarks, at high transverse momentum. In this kinematic regime, where the transverse momentum significantly exceeds the heavy quark's mass, perturbative calculations exhibit the emergence of large collinear logarithms at each order.
Such collinear logarithms can be subtracted from the fixed order calculations and resummed 
to all orders by the introduction of 
heavy quark parton densities and
renormalisation group evolved fragmentation functions 
of light quarks, gluons and heavy quarks into heavy quark flavoured hadrons ($B, D, \Lambda_c$).
Such FFs have been determined either in Mellin moment $N$-space 
\cite{Mele:1990yq,Mele:1990cw,Nason:1999zj,Cacciari:2005uk,Anderle:2017cgl,Czakon:2022pyz}
or directly in $x$-space
\cite{Binnewies:1997xq,Kniehl:2005de,Kniehl:2006mw,Kniehl:2007erq,Kneesch:2007ey,Soleymaninia:2017xhc,Salajegheh:2019ach,Salajegheh:2019nea,Kniehl:2020szu,Delpasand:2020vlb}.
For transverse momenta significantly exceeding the heavy quark mass, this procedure becomes indispensable to reinstate the convergence of the perturbative expansion. Nonetheless, even for transverse momenta only moderately larger than the heavy quark mass, resumming the collinear logarithms while retaining finite mass terms $m^2/p_T^2$ in the hard process \cite{Cacciari:1998it,Kniehl:2004fy,Kniehl:2005mk,Helenius:2018uul} yields improved theoretical predictions. These predictions exhibit reduced theoretical uncertainties stemming from scale variations and demonstrate better agreement with experimental data. For instance, see \cite{Kniehl:2012ti,Kniehl:2015fla,Benzke:2019usl} for comprehensive studies on inclusive $D$ and $B$ meson production at the LHC.
\\

\noindent
The NLO QCD corrections to processes involving Fragmentation Functions (FFs) have a long history, dating back to the late 1970s. Initially, computations focused on inclusive cross-sections for single hadron production in $e^+ e^-$ collisions, considering both massless \cite{Altarelli:1979kv} and massive quarks \cite{Mele:1990yq,Mele:1990cw}. Subsequently, similar calculations were extended to hadron collisions \cite{Aversa:1988vb} and deep-inelastic scattering \cite{Kretzer:1998nt}. The NLO computations also encompass di-hadron production in $e^+ e^-$ collisions \cite{Cacciari:2001cw} and hadron collisions \cite{Binoth:2001vm,Owens:2001rr}. Initially, these computations were often tailored to specific observables, whereas more recent efforts strive for flexibility to describe a broader range of observables. Achieving this flexibility requires addressing the soft and collinear divergences arising from real emissions across a general phase space. It's worth noting that at NLO, only one parton can be soft and/or collinear, with the divergences typically being logarithmic at most.
\\

\noindent
There are two main methods, with variations, used to handle singularities in terms of the space-time regulator 
$\varepsilon\equiv (4-n)/2$ (where $n$ is the space-time dimension).
The first method involves slicing the phase space into small regions in which these divergences show up and a region free of divergences. 
Within these small regions, the integration over the soft/collinear parton is carried out analytically, retaining only the most singular terms as the size of the regions approaches zero.
In other words, this method involves neglecting terms that vanish as the size of these small regions approaches zero and retaining only the size-dependent logarithmic terms. This approach is commonly referred to as the phase space slicing method. Within this framework, general algorithms have been developed to address jet and hadron production in $e^+ e^-$ and hadronic collisions \cite{Giele:1991vf,Giele:1993dj,Keller:1998tf,Harris:2001sx}.
The other method, known as the subtraction method, consists of adding and subtracting certain integrands. 
The sum of these integrands retains the same divergences as the original integrand when a parton becomes soft and/or collinear, 
but they are simplified enough to allow for analytic integration over the phase space of the soft/collinear parton. 
Similar to the phase space slicing method, general algorithms have also been developed for the subtraction method, primarily focusing on jets \cite{Frixione:1995ms,Catani:1996vz,Frixione:1997np,Kosower:1997zr}.
Subtraction methods have proven their efficiency compared to phase space slicing methods to deal with the soft and collinear singularities by avoiding important numerical cancellations between large positive terms coming from the real emission and negative ones coming from the soft and collinear terms.
The general methods developed so far require boosts to transition from the laboratory frame to a dedicated frame chosen to simplify the analytical computation of the subtracted terms. However, these boosts can be computationally costly. Therefore, we are exploring the feasibility of performing the analytical integration of the subtraction terms directly in the laboratory frame.
To be more specific, we focus on the case of hadron collisions, where the laboratory frame is the hadronic centre-of-mass frame.
In these collisions, the standard subtraction methods typically parameterise the phase space using energy, polar angle, and azimuthal angle.
However, these variables are not the most natural for describing hadronic collisions. Instead, the natural variables are the transverse momentum, rapidity (or pseudo-rapidity, given that all masses are neglected), and azimuthal angle.
Consequently, incorporating cuts in this parameterisation becomes more complicated.
An initial attempt towards this objective was made in reference \cite{Kunszt:1992tn}, limited to $2 \rightarrow 2$ reactions 
and focused on two-jet production. 
However, the generalisation of the method presented in \cite{Kunszt:1992tn} to reactions involving more particles in the final state 
was deemed complicated (cf. ref. \cite{Frixione:1995ms}).
A similar challenge arose in another effort \cite{Chiappetta:1996wp} focused on two-hadron production.
\\

\noindent
The objective of this article is to introduce a novel subtraction method dedicated to reactions involving fragmentation functions.
Specifically, we address the general scenario where the Leading Order (LO) processes are $2 \rightarrow N-3$ reactions (or $N-1$ body reactions), and where $M$ partons fragment with $M \leq N-3$.
The new method presented in this work incorporates the two features outlined in the previous paragraph: 1) the integration of 
the subtraction terms is carried out in the hadronic centre-of-mass frame, and 2) the phase space is characterised 
by the "natural" variables of hadronic collisions, namely the transverse momentum, rapidity, and azimuthal angle.
While it is certainly possible to adapt existing general subtraction methods such as FKS or Catani-Seymour to accommodate reactions involving 
multiple fragmentation functions, the desired features would not be present.
This is the case of the work described in the reference  \cite{Liu:2023fsq}, which consists in adapting the tool {\tt aMC@NLO} 
dedicated to jet processes to fragmentation ones. 
Nevertheless, their way to proceed is a combination of the subtraction and phase space slicing methods.
\\

\noindent
The outline of the article is the following.
Section \ref{presmeth} provides an overview of the method. We consider the case of a hadronic reaction involving two hadrons 
yielding $N-3$ hadrons ($M = N-3$), and present the formulas for the hadronic cross section at both LO and NLO accuracies
after having taken into account the constraints on energy and longitudinal momentum conservation.
Additionally, we outline the structure of initial and final state collinear divergences, obtained from the LO formula by expressing the evolved parton density functions (PDFs) and fragmentation functions in terms of the bare ones.
The subtraction strategy is explained in a general manner, with detailed calculations postponed to Section \ref{detcalsubterm}. 
This subtraction is performed within a cylinder in transverse momentum around the beam axis and outside, in $N-3$ cones centred 
on the direction of the $N-3$ hard partons.
\\

\noindent
In Section \ref{detcalsubterm}, we provide the detailed construction of the subtraction terms and their analytical integration. We consider the different regions, namely inside the cylinder and inside the various cones (i.e., outside the cylinder). The divergences in terms of the regulator $\varepsilon$ are discussed.
\\

\noindent
In Section \ref{divterm}, we collect the different divergent terms resulting from the analytical integration of the subtracted terms. These terms are used to construct the different parts of the cross sections containing the initial state collinear divergences, the final state collinear divergences, and the soft divergences. We demonstrate that the collinear divergences fit the structure derived in Section \ref{presmeth}, allowing them to be reabsorbed into a redefinition of the PDFs or FFs. Additionally, we show that the soft divergences cancel against those coming from the virtual contribution.
\\

\noindent
In Section \ref{one_frag}, we apply the subtraction method to the case where some hard partons do not fragment, i.e., $M < N-3$. Due to space constraints, we focus specifically on the case where $M=N-4$. We investigate two scenarios: \textit{(i)} when the non-fragmenting parton is a photon, and \textit{(ii)} when the non-fragmenting parton is involved in a jet. We demonstrate that the method presented in the preceding sections works effectively in these cases as well.
\\

\noindent
Finally, we conclude this article with a summary and prospects for future research. While we have removed many detailed calculations from the main text to improve readability, we believe they remain valuable for readers. Appendix \ref{APkernels} provides a summary of the expressions of DGLAP kernels at the lowest order. In Appendix \ref{softint}, we present the detailed computation of the soft integral using azimuthal angle and rapidity. The computation of collinear integrals (both inside and outside the cylinder) is detailed in Appendices \ref{collint} and \ref{intir} respectively. In Appendix \ref{gutsdivterm}, we outline the steps to obtain the results presented in Section \ref{divterm}. Additionally, Appendix \ref{qqbgg} illustrates the discussion in the main text regarding the soft limit in QCD using the specific reaction $q + \bar{q} \rightarrow g + g$. Lastly, Appendix \ref{notation} provides a recap of the different notations used throughout the article to facilitate understanding of the formulae.

\section{Presentation of the method}\label{presmeth}

The method to remove the soft and collinear singularities is a modification of the subtraction method presented in \cite{Chiappetta:1996wp}. The original method was designed for $2 \rightarrow 2$ reactions at leading order where at most two hard partons fragment. It was pointlessly complicated involving some analytically unsolved one dimensional phase space integrals.\\

\noindent
In this article, the method is generalised for the case where an arbitrary number of partons in the final state fragment.
More precisely, we consider, at leading order, a partonic reaction $2 \rightarrow N -3$ where all the partons in the final state fragment\footnote{The method can be applied for a mixed case where some hard partons in the final state do not fragment. Examples will be presented in sec.~\ref{one_frag}.}. Then, at NLO approximation, we have to consider the case of a partonic reaction $2 \rightarrow N-2$ where $N-3$ hard partons fragment, the non fragmenting parton being soft and/or collinear to another one.
Furthermore, simplifications are brought in the method in such way that the phase space integrals of the subtracted terms can be performed analytically.
As already mentioned, the subtraction is performed in the hadronic centre-of-mass frame and the four-momenta are parameterised with the rapidity, the azimuthal angle and the transverse momentum. Note that, for a matter of simplicity, the subtraction terms are built for a squared matrix element summed over the colours. In the rest of the article, we introduce compact notations which are recapped in appendix \ref{notation}.
To start with, let us present the hadronic cross section at leading order.

\subsection{LO accuracy \label{secLO}}

Let us consider the inclusive hadronic reaction $H_1 + H_2 \rightarrow H_3 + H_4 + \ldots + H_{N-1} + X$ where each hadron $H_l$ has a four-momentum $K_l$. The hadronic cross section in the QCD improved parton model is given by
\begin{align}
  \sigma_H^{{\text{LO}}} &= \sum_{\seq{i}{N-1} \in S_p} \, \int \, d \bx{1} \, d \bx{2} \, \prod_{l=3}^{N-1} d \bx{l} \, d^n K_l  \, F_{i_1}^{H_1}(\bx{1},M^2) \, F_{i_2}^{H_2}(\bx{2},M^2) \, \notag \\
  &\qquad \qquad \qquad \qquad \quad {} \times \prod_{l=3}^{N-1} D_{i_l}^{H_l}(\bx{l},M_f^2) \, \hat{\sigma}_{[i]_{N-1}}\, ,
  \label{eqdefhadcrxg0}
\end{align}
where $\seq{i}{N-1} $ stands for $ i_1, \, i_2, \ldots , \, i_{N-1}$,
the summation runs over all types of partons as indicated by the use of the set $S_p = \{u,\bar{u},d,\bar{d},\ldots,g\}$\footnote{It is implicitly assumed that the partonic cross section must fulfil conservation laws, thus if the sum selects a choice of partons $i_1,\ldots,i_{N-1}$ which violates these laws the partonic cross section is set to zero.}.
The function $F_{i_k}^{H_k}(\bx{k},M^2)$ represents the partonic density of a parton $i_k$ inside a hadron $H_k$ carrying a fraction of the hadron four-momentum $\bx{k}$ at the energy scale $M$ whereas the function $D_{i_l}^{H_l}(\bx{l},M_f^2)$ represents the fragmentation function of a parton $i_l$ into a hadron $H_l$ carrying a fraction of the parton four-momentum $\bx{l}$ at the energy scale $M_f$.
The partonic cross section for the reaction $i_1 + i_2 \rightarrow i_3 + \ldots + i_{N-1}\equiv \reac{i}{N-1}$ in which each parton labelled by $i_l$ has a four-momentum $p_l$ is defined in $n=4-2\varepsilon$ dimensions as
\begin{align}
\hat{\sigma}_{\reac{i}{N-1}} &= \frac{1}{4 \, p_1 \cdot p_2} \, \frac{g_s^{2 \, (N-3)} \, \mu^{2 \, (N-3) \, \varepsilon}}{4 \, C_{i_1} \, C_{i_2}}  \int \prod_{l=3}^{N-1} \frac{d^n p_l}{(2 \, \pi)^{n-1}} \, \delta^+(p_l^2)\, \delta^n(K_l-\bx{l} \, p_l) \notag \\
&\qquad \qquad \qquad \qquad \qquad \qquad \qquad \qquad {} \times  \, (2 \, \pi)^n \, \delta^n\left({P_i-P_f}\right) \, |M^n_{\reac{i}{N-1}}|^2\, .
\label{eqdef2dNm1ph}
\end{align}
In eq.~(\ref{eqdef2dNm1ph}), $P_i \equiv p_1 + p_2$ and $P_f \equiv p_3 + \ldots + p_{N-1}$ are, respectively, the sum of initial and final state partonic momenta, $g_s$ is the QCD coupling constant and $\mu$ is the energy scale such that $g_s$ is dimensionless in a $n$-dimensional space time, $C_{i_1}$ and $C_{i_2}$ are the dimensions of the colour representations to which the partons $i_1$ and $i_2$ belong times the extra number of polarisations in a space-time of dimension $n$, namely
\begin{align}
  C_i &= \left\{ \begin{array}{cl}
    N_c & \text{for} \, i = q,\bar{q}\\
    (N_c^2-1) \, \frac{n-2}{2} & \text{for} \, i=g
\end{array} \right. ,\label{eqdefCi}
\end{align}
with $N_c$ the number of colours in the fundamental representation.
Note that, in eq.~(\ref{eqdef2dNm1ph}), $|M^n_{\reac{i}{N-1}}|^2$ represents the squared amplitude stripped from the coupling constants and the related powers of the scale $\mu$. We will keep this convention all over the article.
The integration over the four-momentum $p_l$, for each $l$, is performed to get rid of the constraint $\delta^n(K_l - \bx{l} \, p_l)$\footnote{Notice that the definition of $\hat{\sigma}_{\reac{i}{N-1}}$ given in eq. (\ref{eqdef2dNm1ph}) is not the standard definition of the partonic cross section because of the presence of $\delta^n(K_l - \bx{l} \, p_l)$, the latter one is introduced to insure the conservation of momenta.}.
Using the rapidities and the transverse momenta  for coordinates of the different four-momenta, eq.~(\ref{eqdefhadcrxg0}) reads
\begin{align}
  \sigma_H^{{\text{LO}}} &= \sum_{\seq{i}{N-1} \in S_p} \, \frac{1}{2^{N-2} \, s} \, \frac{1}{(2 \, \pi)^{(N-4) \, n - N + 3}} \, \frac{g_s^{2 \, (N-3)} \, \mu^{2 \, (N-3) \, \varepsilon}}{4 \, C_{i_1} \, C_{i_2}} \, \int d \bx{1} \, d \bx{2} \, {\PSH{\bar{x}}} \,  \notag \\
  &\quad {} \times \delta \left( (\bx{1}+\bx{2}) \, \frac{\sqrt{s}}{2} - E_+ \right) \, \delta \left( (\bx{1}-\bx{2}) \, \frac{\sqrt{s}}{2} - E_- \right) \, \notag \\
  &\quad {} \times A_{\indic{i}{N-1}}(\seq{\bar{x}}{N-1}) \,  \delta^{n-2}\left( \sum_{l=3}^{N-1} \frac{\vkt{l}}{\bx{l}} \right) |M^n_{\reac{i}{N-1}}|^2\, ,
  \label{eqdefhadcrxg2}
\end{align}
with
\begin{align}
  E_+ &= \sum_{l=3}^{N-1} \frac{\kt{l}}{\bx{l}} \, \cosh(y_l)\, , &
  E_- &= \sum_{l=3}^{N-1} \frac{\kt{l}}{\bx{l}} \, \sinh(y_l) \label{eqdefEmoinsg}\, ,
\end{align}
and $\PSH{\bar{x}}$ denoting the phase space of the hard partons which is given by
\begin{align}
  \PSH{\bar{x}} &\equiv \prod_{l=3}^{N-1} \frac{d \bx{l}}{\bx{l}^{n-2}} \, d y_l \, d^{n-2} \kt{l}\, .
\end{align}

\noindent
In eq.~(\ref{eqdefhadcrxg2}), the quantity $A_{\indic{i}{N-1}}(\seq{\bar{x}}{N-1})$ represents the combination of partonic densities and fragmentation functions for a specific partonic subprocess. We write it with the following homogeneous notation
\begin{align}
  A_{\indic{i}{N-1}}(\seq{\bar{x}}{N-1})  &= \prod_{l=1}^{N-1} D_{i_l}^{H_l}(\bx{l},M_l^2)\, ,
  \label{eqdefAijkl}
\end{align}
where the two new symbols $\indic{i}{N-1}$ and $\seq{\bar{x}}{N-1}$ stand for
\begin{align}
\indic{i}{N-1} &\equiv i_1 \,i_2 \,i_3 \, \cdots  \, i_{N-1}\, ,&
\seq{\bar{x}}{N-1} &\equiv \bx{1},\, \bx{2}, \cdots, \bx{N-1} \, .
\end{align}
Note that, in eq.~(\ref{eqdefAijkl}), not all of the quantities $D_{i_l}^{H_l}(\bx{l},M_l^2)$ have the same meaning. 
Firstly, the different energy scales $M_l$ have the following sense:
\begin{align}
    M_l \equiv \left\{
    \begin{array}{ll}
        M & \text{for $l=1,2$} \\
        M_f & \text{for $l=3, \cdots, N-1$}
    \end{array}
    \right. \, .
    \label{eqdefscaleMl}
\end{align}
Secondly, we define
\begin{align}
    D_{i_l}^{H_l}(\bx{l},M_l^2) \equiv \frac{F_{i_l}^{H_l}(\bx{l},M^2)}{\bx{l}} \quad \text{for $l=1,2$}\quad ,
    \label{eqdefDi1i2}
\end{align}
while for $l=3, \ldots, N-1$, $D_{i_l}^{H_l}(\bx{l},M_l^2)$ is the standard fragmentation function $D_{i_l}^{H_l}(\bx{l},M_f^2)$.
Note that the division by $\bx{1} \, \bx{2}$ resulting from the definition (\ref{eqdefDi1i2}), comes from the flux factor of the partonic reaction. 
The constraint on the conservation of the energy and the longitudinal momentum are eliminated by integrating on $\bx{1}$ and $\bx{2}$ leading to
\begin{align}
  \sigma_H^{{\text{LO}}} &= \sum_{\seq{i}{N-1} \in S_p} \, K^{(n) \, B}_{i_1 i_2} \, \int  \PSH{\bar{x}} \, A_{\indic{i}{N-1}}(\seq{\bar{x}}{N-1}) \, \delta^{n-2}\left( \sum_{l=3}^{N-1} \frac{\vkt{l}}{\bx{l}} \right) |M^n_{\reac{i}{N-1}}|^2\, ,
  \label{eqdefhadcrxg3}
\end{align}
with
\begin{align}
  \bx{1} &= \sum_{l=3}^{N-1} \frac{\kt{l}}{\bx{l} \, \sqrt{s}} \, e^{y_l}\, ,&
  \bx{2} &= \sum_{l=3}^{N-1} \frac{\kt{l}}{\bx{l} \, \sqrt{s}} \, e^{-y_l} \label{eqdefx2g}\, ,
\end{align}
and
\begin{align}
  K^{(n) \, B}_{i_1 i_2} &= \frac{1}{2^{N-3} \, s^2} \, \frac{1}{(2 \, \pi)^{(N-4) \, n - N + 3}} \, \frac{g_s^{2 \, (N-3)} \, \mu^{2 \, (N-3) \, \varepsilon}}{4 \, C_{i_1} \, C_{i_2}}\, .
\end{align}
Note that, although not explicitly specified, $|M^n_{\reac{i}{N-1}}|^2$ is a function of $\bx{l}$, $y_l$ and $\kt{l}$.
\\

\noindent
In order to get the structure of the collinear divergences in the initial and final state, let us recall the relations between the bare partonic densities and the renormalised ones as well as the relations between the bare fragmentation functions and the renormalised ones:
\begin{align}
  D_k^H(x,M_k^2) &= \bD_k^{H} \left( x \right) + \frac{\alpha_s}{2 \, \pi} \,  \sum_{j \in S_p} \, \left[ \calh_{kj}\left(*,\frac{\mu^2}{M_k^2}\right) \otimes \bD_j^{H}  \right]_{2} (x), \label{eqrelevolutionpdf} 
\end{align}
for $k=1,2$ and
\begin{align}
  D_k^H(x,M_k^2) &= \bD_k^{H} \left( x \right) + \frac{\alpha_s}{2 \, \pi} \,  \sum_{j \in S_p} \, \left[ \calh_{jk}\left(*,\frac{\mu^2}{M_k^2}\right) \otimes \bD_j^{H}  \right]_{1} (x).\label{eqrelevolutionff} 
\end{align}
for $k=3, \ldots, N-1$. Both in eqs.~(\ref{eqrelevolutionpdf}) and (\ref{eqrelevolutionff}), a special notation is introduced for the convolution.
To explain it, let us consider two multivariate functions $f(a_1,\ldots,a_N)$ and $g(b_1,\ldots,b_K)$. We will denote the convolution of these two functions with respect to the variables $a_k$ and $b_l$
\begin{align}
    \hspace{2em}&\hspace{-2em}\left[ f\left(a_1, \cdots, a_{k-1}, *, a_{k+1}, \cdots a_N\right) \otimes g(b_1, \cdots, b_{l-1}, * , b_{l+1}, \cdots , b_K)  \right]_{\eta} (x) \notag \\
    &\equiv \int_x^1 \frac{dz}{z^{\eta}} \, f\left(a_1, \cdots, a_{k-1}, z, a_{k+1}, \cdots a_N\right) \, g\left(b_1, \cdots, b_{l-1}, \frac{x}{z} , b_{l+1}, \cdots , b_K\right)\, .
\end{align}
Note that we also use the following convention that if a function $h$ involved in the convolution has only one argument we write in our special notation $h$ instead of $h(*)$.
In addition, the quantity $\calh_{kj}(z,\mu^2/M_k^2)$ is defined by
\begin{align}
  \calh_{kj}\left(z,\frac{\mu^2}{M_k^2}\right) &= - \frac{1}{\varepsilon} \, P^{(4)}_{kj}(z) \, \left( \frac{4 \, \pi \, \mu^2}{M_k^2} \right)^{\varepsilon} \, \frac{1}{\Gamma(1-\varepsilon)} + \text{finite terms} \, .\label{eqdefHij}
\end{align}
The quantities $P^{(4)}_{ij}(z)$ are the one-loop DGLAP kernels in four dimensions (cf. appendix \ref{APkernels}) and the finite terms are factorisation scheme  dependent and they are zero in the
$\msbar$ scheme used in this paper.
In eqs~(\ref{eqrelevolutionpdf}) and (\ref{eqrelevolutionff}), $\bD_l^{H}(x)$, are the bare partonic densities divided by $x$ for $l=1,2$ and the bare fragmentation functions for $l=3, \ldots, N-1$.\\

\noindent
Injecting eqs~(\ref{eqrelevolutionpdf}) and (\ref{eqrelevolutionff}) into eq.~(\ref{eqdefAijkl}), expanding and keeping only terms of order $\alpha_s^0$ and $\alpha_s^1$, we get
\begin{align}
  \hspace{2em}&\hspace{-2em} A_{(i)_{N-1}}(\{\bar{x}\}_{N-1}) = \bA_{(i)_{N-1}}(\{\bar{x}\}_{N-1})+ \frac{\alpha_s}{2 \, \pi} \left\{ \vphantom{\sum_{j_k \in S_p}  \left[\calh_{j_k i_k}\left(*,\frac{\mu^2}{M_f^2}\right) \otimes \bA_{\indict{i}{i_k}{j_k}{N-1}}(\seqt{\bar{x}}{\bx{k}}{*}{N-1})\right]_{1}(\bx{k})} \right. \notag \\
  &\quad\quad\quad\quad {} \sum_{l=1}^{2} \sum_{j_l \in S_p}  \left[\calh_{i_l j_l}\left(*,\frac{\mu^2}{M^2}\right) \otimes \bA_{\indict{i}{i_l}{j_l}{N-1} }(\seqt{\bar{x}}{\bx{l}}{*}{N-1})\right]_{2}(\bx{l})  \notag \\
  & \quad\quad\quad + \left. \sum_{k=3}^{N-1} \, \sum_{j_k \in S_p}  \left[\calh_{j_k i_k}\left(*,\frac{\mu^2}{M_f^2}\right) \otimes \bA_{\indict{i}{i_k}{j_k}{N-1}}(\seqt{\bar{x}}{\bx{k}}{*}{N-1})\right]_{1}(\bx{k}) \right\} \, .
  \label{eqsectotlo4}
\end{align}
In eq.~(\ref{eqsectotlo4}), we used the compact notation 
\begin{align}
\indict{i}{i_k}{j_k}{N-1} &\equiv i_1 \,i_2 \,i_3 \, \cdots i_{k-1}\, j_{k}\, i_{k+1}\cdots  \, i_{N-1}\, , \\
\seqt{\bar{x}}{\bx{k}}{*}{N-1} &\equiv \bar x_1,\, \bar x_2, \cdots, \bar x_{k-1}, \,*,\, \bar x_{k+1},\, \cdots, \bar x_{N-1} \, .
\end{align}
The quantity $\bA_{(i)_{N-1}}(\{\bar{x}\}_{N-1})$ is the combination of bare parton densities divided by their arguments and bare fragmentation functions, namely
\begin{align}
  \bA_{(i)_{N-1}}(\{\bar{x}\}_{N-1})&= \prod_{l=1}^{N-1} \bD_{i_l}^{H_l}(\bx{l}). 
  \label{eqdefA0ijkl}
\end{align}
Injecting eq.~(\ref{eqsectotlo4}) into eq.~(\ref{eqdefhadcrxg3}) and relabelling the partons yields
\begin{align}
  \sigma_H^{{\text{LO}}} &= \sum_{\seq{i}{N-1} \in S_p} \, K^{(n) \, B}_{i_1 i_2} \, \int d \text{PS}_{N-1 \,\text{h}}^{(n)}(\bar{x}) \,  \delta^{n-2}\left( \sum_{l=3}^{N-1} \frac{\vkt{l}}{\bx{l}} \right)  \left\{ \vphantom{\sum_{k=3}^{N-1} \,\int^1_{x_k} \frac{d z_k}{z_k} \, \bA_{(i)_{N-1}}\left( \bx{1},\bx{2},\cdots,\frac{\bx{k}}{z_k},\cdots \right)} \bA_{(i)_{N-1}}(\{\bar{x}\}_{N-1}) \,  |M^n_{[i]_{N-1}}|^2 \right. \label{eqdefhadcrxg4}\\
  &\quad {} +  \frac{\alpha_s}{2 \, \pi} \left(  \sum_{l=1}^{2} \sum_{j_l \in S_p} \, \frac{C_{i_l}}{C_{j_l}} \left[\calh_{i_l j_l}\left(*,\frac{\mu^2}{M^2}\right) \otimes \bA_{\indict{i}{i_l}{j_l}{N-1} }(\seqt{\bar{x}}{\bx{l}}{*}{N-1})\right]_{2}(\bx{l}) \,  |M^n_{\react{i}{i_l}{j_l}{N-1}}|^2   \right. \notag \\
  &\quad {} +  \left. \left. \sum_{k=3}^{N-1} \, \sum_{j_k \in S_p} \left[\calh_{j_k i_k}\left(*,\frac{\mu^2}{M_f^2}\right) \otimes \bA_{\indict{i}{i_k}{j_k}{N-1}}(\seqt{\bar{x}}{\bx{k}}{*}{N-1})\right]_{1}(\bx{k}) \, |M^n_{\react{i}{i_k}{j_k}{N-1}}|^2 \right) \right\} \notag\, .
\end{align}
This last equation gives the structure of the collinear divergences for the initial state (the first term in brackets) and for the final state (the second term in brackets) which are depicted in figs.~\ref{fig2} and \ref{fig1}.

\subsection{NLO accuracy}

It is well known that to reach the NLO accuracy, we have to take into account the one loop virtual corrections to the Born amplitude $M_{\reac{i}{N-1}}$ as well as the corrections originating from the phase space integration of an on-shell extra parton emitted by this Born amplitude, the so called real emission\footnote{Note that new tree level channels might open up at NLO exhibiting also collinear divergences.}. 
Since the former corrections have the same kinematics as the LO cross section, we will focus on the latter ones. 
Let us consider the same hadronic reaction but induced by a partonic reaction having $N-2$ partons in the final state $i_1 +i_2 \rightarrow i_3 + i_4 + \ldots + i_N \equiv[i]_N$. Let us denote $i_N$ the particle which can be soft or collinear to the other ones. The matrix element squared can be written as
\begin{align}
|M^{(n)}_{[i]_{N}}|^2 &= \sum_{i=1}^{N-2} \sum_{j=i+1}^{N-1} \Hn{ij}(p_N) \, E_{ij} + G^{(n)}(p_N)\, ,
\label{eqdefMatsqg}
\end{align}
where the squared eikonal factor is given by
\begin{align}
E_{ij} &\equiv \frac{p_i \cdot p_j}{p_i \cdot p_N \, p_j \cdot p_N} \notag \\
&= \frac{1}{Q^2 \, \xt{N}^2} \, \frac{p_i \cdot p_j}{p_i \cdot \hat{p}_N \, p_j \cdot \hat{p}_N} \notag \\
&\equiv \frac{1}{Q^2 \, \xt{N}^2} \, E^{\prime}_{ij}\, ,
\label{eqdefEijg}
\end{align}
with $\hat{p}_N = (\cosh(y_N), \vpthat{N}, \sinh(y_N))$ in which $\vpthat{N}$ is the unit vector in the direction of $\vpt{N}$ and $\xt{N} = \pt{N}/Q$.
The functions $\Hn{ij}(p_N)$ and $G^{(n)}(p_N)$ are regular when $p_N \rightarrow 0$ or when $i_N$ is collinear to another parton.
This decomposition is not unique but the soft and collinear limits do not depend on this ambiguity.
An arbitrary energy scale $Q$ has been introduced in order to use a dimensionless variable $\xt{N}$ for the integration on the transverse momentum of the particle $i_N$. It is obvious that the cross section for the real emission will not depend on the choice of this scale.
\\

\noindent
After having taken into account the constraint on the conservation of energy and longitudinal momentum, the hadronic cross section for the real emission reads
\begin{align}
  \sigma_H^{{\text{Real}}} &= \sum_{\seq{i}{N-1} \in S_p} \, K^{(n)}_{i_1 i_2} \, \int d \text{PS}_{N-1 \,\text{h}}^{(n)}(x) \, \int d \text{PS}_N^{(n)} \,  \delta^{n-2}\left( \sum_{l=3}^{N-1} \frac{\vkt{l}}{x_l} + \vpt{N}\right) \, \notag \\
  &\quad {} \times  \, A_{(i)_{N-1}}(\{x\}_{N-1}) \, \left[ \sum_{i=1}^{N-2} \sum_{j=i+1}^{N-1} \Hn{ij}(p_N) \, E^{\prime}_{ij} +  \xt{N}^2 \, Q^2 \, G^{(n)}(p_N) \right]\, ,
  \label{eqdefhadcrxg5}
\end{align}
where $d \text{PS}_N^{(n)}$ is the phase space of the parton $i_N$ divided by $\xt{N}^2$ which is given by
\begin{align}
  d \text{PS}_N^{(n)} &\equiv d y_N \, d \xt{N} \, \xt{N}^{n-5} \, d \phi_N \, (\sin \phi_N)^{n-4}\, .
  \label{psn}
\end{align}
The direct azimuthal angle of the vector $\vpt{N}$ with a reference vector in the transverse momentum plane is generically denoted by $\phi_N$. This reference vector will be different according to the integrands.
In eq.~(\ref{eqdefhadcrxg5}), the quantities $x_1$ and $x_2$ are given by
\begin{align}
  x_1 &= \left[ \sum_{l=3}^{N-1} \frac{\kt{l}}{\sqrt{s} \, x_l} \, e^{y_l} + \omega \, \xt{N} \, e^{y_N} \right] = \hat{x}_1 + \omega \, \xt{N} \, e^{y_N} \label{eqdefx1gp}\, ,
  \\
  x_2 &= \left[ \sum_{l=3}^{N-1} \frac{\kt{l}}{\sqrt{s} \, x_l} \, e^{-y_l} + \omega \, \xt{N} \, e^{-y_N} \right] = \hat{x}_2 + \omega \, \xt{N} \, e^{-y_N} \, ,
  \label{eqdefx2gp}
\end{align}
 where $\omega = Q/\sqrt{s}$ and
\begin{align}
  K^{(n)}_{i_1 i_2} &= \frac{1}{2^{N-2} \, s^2} \, \frac{1}{(2 \, \pi)^{(N-3) \, n - N + 2}} \, \frac{g_s^{2 \, (N-2)} \, \mu^{(N-2) \, (4-n)}}{4 \, C_{i_1} \, C_{i_2}} \,  Q^{n-4} \,V(n-2)\, .
  \label{eqdefKij}
\end{align}
In eq.~(\ref{eqdefKij}), $V(n-2)$ represents the solid angle volume of azimuthal angles in a space of dimension $n-2$, knowing that
\begin{align}
  V(n) &= \frac{2 \, \pi^{\frac{n-1}{2}}}{\Gamma\left( \frac{n-1}{2} \right)}\, .
\end{align}
A word of warning about eqs.~(\ref{eqdefx1gp}) and (\ref{eqdefx2gp}). Indeed, we can get the impression that $\hat{x}_{i} \equiv \bx{i}$ but this is not the case because the constraints on the $\kt{l}/x_l$ are different from those appearing at LO.
The requirement that $x_1$ and $x_2$ must be both less or equal to $1$ fixes the bounds on the $y_N$ integration to
\begin{align}
  \yNmax &= \ln \left( \frac{1-\hat{x}_1}{\omega \, \xt{N}} \right)\, , &
  \yNmin &= \ln \left( \frac{\omega \, \xt{N}}{1-\hat{x}_2} \right) \label{eqdefyNmin}\, .
\end{align}

\subsection{Subtraction strategy}

To start with, let us introduce some notations. We define two sets: $S_i=\{1,2\}$ which is the set of labels of initial state partons and $S_f = \{3, 4, \ldots ,N-1\}$ which is the set of labels of hard partons in the final state. 
Furthermore, with respect to this last integration (cf.~eq. (\ref{psn})), let us introduce the quantity $\calt$ as
\begin{align}
  \calt &= \int \PSN \, \sum_{i=1}^{N-2} \sum_{j=i+1}^{N-1} f_{ij}(y_N,\xt{N},\phi_N) \, E^{\prime}_{ij}\, ,
  \label{eqdefTij0!}
\end{align}
with
\begin{align}
  f_{ij}(y_N,\xt{N},\phi_N) &= \sum_{[i]_{N-1} \in S_p} \, K^{(n)}_{i_1 i_2} \, \Hn{ij}(p_N) \,  A_{(i)_{N-1}}(\{x\}_{N-1})\,  \delta^{n-2}\left( \sum_{l=3}^{N-1} \frac{\vkt{l}}{x_l} + \vpt{N}\right)\, .
  \label{eqdeffij0!}
\end{align}
Then, the sum in the right hand part of eq.~(\ref{eqdefTij0!}) is split into four parts:
\begin{align}
  \calt &= \int \PSN \, \left\{  f_{12}(y_N,\xt{N},\phi_N) \, E^{\prime}_{12} + \sum_{i=3}^{N-1} f_{1i}(y_N,\xt{N},\phi_N) \, E^{\prime}_{1i} \right. \notag \\
  &\qquad \qquad \qquad \quad {} + \left. \sum_{i=3}^{N-1} f_{2i}(y_N,\xt{N},\phi_N) \, E^{\prime}_{2i} + \sum_{i=3}^{N-2} \sum_{j=i+1}^{N-1} f_{ij}(y_N,\xt{N},\phi_N) \, E^{\prime}_{ij} \right\}\, .
  \label{eqdefTij1!}
\end{align}
The four integrands in the curly brackets of eq.~(\ref{eqdefTij1!}) are denoted respectively $T^{(1)}$, $T^{(2)}$, $T^{(3)}$ and $T^{(4)}$.
The splitting is such that the phase space integration of the first term ($T^{(1)}$) generates soft and initial state collinear divergences (ISR), the phase space integration of the second and the third term (resp. $T^{(2)}$ and $T^{(3)}$) generates soft, initial state collinear and final state collinear (FSR) divergences and the integration over the last one ($T^{(4)}$) generates soft and final state collinear divergences. 
Thus the hadronic cross section associated to the real emission can be written as
\begin{align}
  \sigma_H^{\text{Real}} &= \int \PSH{x} \, \calt + \text{finite terms}\, ,
  \label{eqsigdiv}
\end{align} 
where the finite terms are associated to the function $G^{(n)}(p_N)$ in eq. (\ref{eqdefhadcrxg5}).
\\

\noindent
In our strategy, the subtraction is performed only in some regions of phase space like in some other subtraction methods leading to more flexibility by avoiding large cancellations between positive and negative weight events. So, the phase space of the particle $i_N$ is split into two parts.

{                                    
\setlength{\leftmargini}{0.0cm}
\begin{itemize}
    \item {\bf Part I.} The momentum $\vec{p}_{N}$ is located inside a cylinder in transverse momentum around the beam axis of radius $\ptm$. In this part, by definition $\pt{N} \leq \ptm$, thus it contains the soft divergences, the initial state collinear divergences and a part of the final state collinear divergences. 
    At this level, we have to specify the integration bounds for the phase space of $i_N$. Inside the cylinder, we define
    \begin{align}
        \cTiin{k} &\equiv \int_{\text{in}} \PSN \, \Ti{k}\, ,
        \label{eqdefcTiink}
    \end{align}
    where the symbol $ \int_{\text{in}} \PSN$ is understood as
    \begin{align}
        \int_{\text{in}} \PSN &\equiv \int_0^{\xtm} d \xt{N} \, \xt{N}^{-1-2 \, \varepsilon} \, \int_0^{\pi} d \phi_N \, (\sin \phi_N)^{-2 \, \varepsilon} \, \int_{\yNmin}^{\yNmax} d y_N \, .
        \label{eqdefPSNin}
    \end{align}
    The subtraction is done, at the integrand level, by adding and subtracting a soft contribution and a collinear one. Schematically, we write
    \begin{align}
    \sigma_H^{\text{in}} &= \int \PSH{x} \, \left\{ \sum_{k=1}^{4} \left[ \cTiin{k} - \cTiins{k} - \cTiinc{k} \right] \right. \notag \\
    &\qquad \qquad \qquad \qquad \quad {} + \left.\sum_{k=1}^{4} \left[ \cTiins{k} + \cTiinc{k} \right] \right\}\, .
    \label{eqsighschem1}
    \end{align}
    Note that the eq.~(\ref{eqsighschem1}) gives the impression that the subtraction is not fully performed at the integrand level because the symbol $\cTiin{k}$ already contains an integration on $p_N$. As we will see later, to perform the analytical integration of the subtraction term, it is sometimes preferable to modify the integration bounds on $y_N$. But in this case, it is always possible to make some changes of variables in order to have a common integration for the $\cTiin{k}$ and the subtraction terms.
    Note also that the terms \textit{soft} and \textit{collinear} for the subtraction terms need some explanations. We call $\cTiins{k}$, the quantity $\cTiin{k}$ in which the variable which drives the energy of the particle $i_N$ is set to zero, it will contain the soft divergences as well as the soft-collinear ones. $\cTiinc{k}$ is the quantity $\cTiin{k}$ in which the variables which drive the rapidity and the azimuthal angle of the particle $i_N$ are set to some values at which the integrand diverges but where the soft part has been subtracted; it contains only the pure collinear divergences.
    Thus, the key point is to be able to construct for each integrand $\cTiin{k}$ (for $k=1,2,3,4$) soft and collinear contributions which have the same divergences as the original one, i.e. the first term in the curly brackets of eq.~(\ref{eqsighschem1}) is free of soft and collinear divergences and can be safely integrated numerically in four dimensions. In addition, they have to be simple enough in order that the phase space integration over $p_N$ can be performed analytically. 
    The details of this construction for these subtraction terms will be given in the next section, the way we build them will depend on the index $k$.
    This can be viewed as a loss of generality compared to ref.~\cite{Catani:1996vz} for instance but we believe that it is more efficient, especially for the soft parts, by avoiding unnecessary cancellations.
    \item  {\bf Part II.} The momentum $\vec{p}_{N}$ is located 
outside this cylinder. Then, the phase space is split into $N-2$ subparts. Namely, $N-3$ cones, denoted $\Gamma_i$ with $i=3, \ldots ,N-1,$ in rapidity and azimuthal angle, each of size $\rth$, around the different $N-3$ hard outgoing parton directions, i.e. $\Gamma_i \equiv \{\pt{N} > \ptm; d_{iN} \leq \rth\}$  and the remainder $\{\pt{N} > \ptm; d_{kN} > \rth, \, \forall \, k \in S_f\}$. The quantity $d_{km}$ represents the distance in the azimuthal angle -- rapidity plane between the partons $i_k$ and $i_m$, that is to say
$d_{km} = \sqrt{(y_k-y_m)^2 + (\phi_k - \phi_m)^2}$. So, Part II is formed by $N-3$ divergent regions containing only one type of final state collinear singularity and a region which is free of divergences corresponding to parton $i_N$ located outside the $N-3$ cones $\Gamma_i$. 
For part II, we define
\begin{align}
    \cTiout{k} &\equiv \int_{\text{out}} \PSN \, \Ti{k}\, ,
    \label{eqdefCTioutk}
\end{align}
where now the symbol $\int_{\text{out}} \PSN$ stands for
\begin{align}
    \int_{\text{out}} \PSN &\equiv  \int_0^{\pi} d \phi_N \, (\sin \phi_N)^{-2 \, \varepsilon} \, \int_{\xtm}^{\xtmax{N}}  d \xt{N} \, \xt{N}^{-1-2 \, \varepsilon} \,\int_{\yNmin}^{\yNmax} d y_N\, ,
    \label{eqdefPSNout}
\end{align}
where $\xtmax{N}$ is the maximum value taken by the variable $\xt{N}$. Note that this value depends on $\phi_N$. 
Outside the cylinder, the hadronic cross section can be written as
\begin{align}
\sigma_H^{\text{out}} &= \int \PSH{x} \, \left\{ \sum_{k=1}^{4} \, \cTiout{k} - \sum_{i=3}^{N-1} \,  \sum_{k=2}^{4} \cTioutc{k}{i} +  \sum_{i=3}^{N-1} \, \sum_{k=2}^{4} \cTioutc{k}{i} \right\}\, .
\label{eqsighschem2}
\end{align}
In this case, we need to construct subtraction terms only for the final state collinear divergences, this is the reason why the summation starts at $k=2$ in the subtraction terms.
Note that we put another exponent for the quantity $\cTioutc{k}{i}$ to indicate that it depends on the direction around which a collinear cone is drawn.
Again, these subtraction terms are such that the difference of the two first terms in the curly brackets of eq.~(\ref{eqsighschem2}) leads to a finite contribution and the integration over $p_N$ inside the cones of the subtraction terms can be performed analytically. Their constructions are postponed to the next section. 
\end{itemize}
}

\noindent
The hadronic cross section is obtained by summing the two contributions inside and outside the cylinder, that is to say
\begin{align}
   \sigma_H^{\text{Real}} =  \sigma_H^{\text{in}} + \sigma_H^{\text{out}} + \text{finite terms}\, .
   \label{eqdefsigmainout}
\end{align} 
We can split $\sigma_H^{\text{Real}}$ into two parts one  which contains no divergences and can be treated in four dimensions,
\begin{align}
    \sigma_H^{\text{finite}} &= \int \PSHQ{x} \, \left\{ \sum_{k=1}^{4} \left[ \cTiin{k} - \cTiins{k} - \cTiinc{k} \right] \right. \notag \\
    &\qquad {} + \left. \sum_{k=1}^{4} \, \cTiout{k} - \sum_{i=3}^{N-1} \, \sum_{k=2}^{4} \cTioutc{k}{i} \right\}+\text{finite terms}\, ,
\end{align}
and the other which contains the soft and collinear divergences  explicitly given as poles in the regulator $\varepsilon$
\begin{align}
    \sigma_H^{\text{div}} &= \int \PSH{x} \, \left\{ \sum_{k=1}^{4} \left[ \cTiins{k} + \cTiinc{k} \right] \right. \notag \\
    &\qquad {} + \left. \sum_{i=3}^{N-1} \, \sum_{k=2}^{4} \cTioutc{k}{i} \right\}\, .
\end{align}
We notice that the obtained results can be easily used to get the cross sections for reactions where one of the partons does not fragment, the latter one can be a photon or a jet. This case will be discussed in section 5.

\section{Detailed calculation of the subtraction terms}\label{detcalsubterm}

\subsection{Inside the cylinder}

In this section, we show how to build the different subtraction terms inside the cylinder for the different quantities $\cTiin{k}$ and perform explicitly their integration over $p_N$ analytically. 

\subsubsection{Pure FSR: both $i$ and $j$ belong to $S_f$}\label{subsecfincoll}

\noindent
\textbf{Construction of the subtraction terms}\\
Let us recall the definition of the quantity $\cTiin{4}$:
\begin{align}
  \cTiin{4} &= \int_0^{\xtm}  d \xt{N} \, \xt{N}^{-1-2 \, \varepsilon} \, \int_0^{\pi} d \phi_N \, (\sin \phi_N)^{-2 \, \varepsilon} \, \int_{\yNmin}^{\yNmax} d y_N  \, \sum_{i=3}^{N-2} \sum_{j=i+1}^{N-1} f_{ij}(y_N,\xt{N},\phi_N) \, E_{ij}^{\prime}\, .
  \label{eqdefT4ij0!}
\end{align}
The subtraction term for the soft part of $\cTiin{4}$ can be built as
\begin{align}
  \cTiins{4} &= \int_0^{\xtm}  d \xt{N} \, \xt{N}^{-1-2 \, \varepsilon} \, \int_0^{\pi} d \phi_N \, (\sin \phi_N)^{-2 \, \varepsilon} \, \int_{-\infty}^{+\infty} d y_N \, \sum_{i=3}^{N-2} \sum_{j=i+1}^{N-1} f_{ij}(y_N,0,\phi_N) \, E_{ij}^{\prime}\, .
  \label{eqdefsoftpart0!}
\end{align}
Several remarks can be pointed out. First, the quantity $f_{ij}(y_N,0,\phi_N)$ does not depend any more on either $y_N$ or $\phi_N$ because any dependence on $y_N$ or $\phi_N$ is multiplied by $\xt{N}$. Thus, this quantity can be replaced by $f_{ij}(0,0,0)$ and can be factorised out from the integral.
Second, in the limit where $\xt{N}$ goes to zero, the integration bounds on the variable $y_N$ are sent to $\infty$ (cf.~eq.~(\ref{eqdefyNmin})).
And finally, the quantity $E_{ij}^{\prime}$ does not depend on $\xt{N}$ but depends on $y_N$ and $\phi_N$. Defining the soft integral as
\begin{align}
  \Jso &= \int_0^{\pi} d \phi_N \, (\sin \phi_N)^{-2 \, \varepsilon} \, \int_{-\infty}^{+\infty} d y_N \, E_{ij}^{\prime}\, ,
  \label{eqdefJsoft120!}
\end{align}
the soft subtraction term becomes
\begin{align}
  \cTiins{4} &= \sum_{i=3}^{N-2} \sum_{j=i+1}^{N-1} f_{ij}(0,0,0) \, \Jso \, \int_0^{\xtm}  d \xt{N} \, \xt{N}^{-1-2 \, \varepsilon}\, .
  \label{eqdefsoftpart1!}
\end{align}
\\

\begin{figure}[ht]
  \begin{center}
  \includegraphics[scale=0.7]{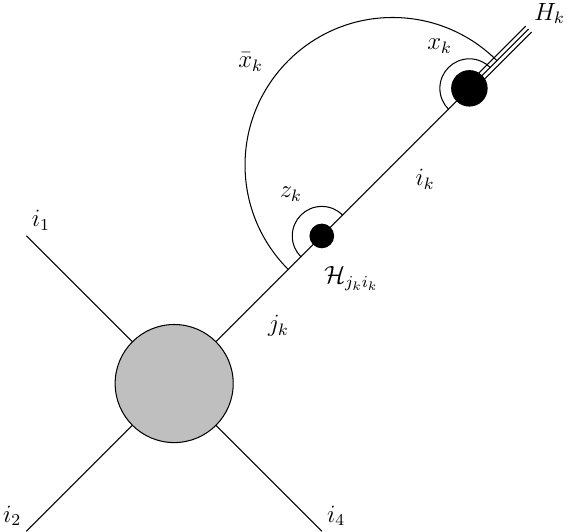}
  \end{center}
\caption{\footnotesize
Labelling of parton radiation in the final state.
The variables $x_k, z_k$ and $\bx{k}$ are defined as follows:
$K_k = x_k p_k$, $p_k = z_k q_k$ and $K_k = \bx{k} q_k$, where
$p_k$, $q_k$, and $K_k$ are the four-momenta of the partons $i_k$, $j_k$ and the hadron $H_k$,
respectively.
}
\label{fig2}
\end{figure}

\noindent
However, the quantity $\cTiin{4}-\cTiins{4}$ is still divergent in the collinear regions $p_N // p_i$ (for $i=3, \ldots, N-1$) and $\xt{N} \neq 0$. 
To get rid of these divergences, we have to introduce a new subtraction term $\cTiinc{4}$.
To build it, let us restart from eq.~(\ref{eqdefT4ij0!}) and write $E^{\prime}_{ij}$ as
\begin{align}
  E^{\prime}_{ij} &=  g_{ij}(y_N,\phi_N) \, \left( \frac{1}{p_i \cdot \hat{p}_N} + \frac{1}{p_j \cdot \hat{p}_N} \right), 
  \label{eqdefEprimeijp!}
\end{align}
with
\begin{equation}
  g_{ij}(y_N,\phi_N) = \frac{p_i \cdot p_j}{(p_i+p_j) \cdot \hp_N} = g_{ji}(y_N,\phi_N).
  \label{eqdefgij0!}
\end{equation}
Inserting eq.~(\ref{eqdefEprimeijp!}) into eq.~(\ref{eqdefT4ij0!}) leads to
\begin{align}
  \cTiin{4}-\cTiins{4} &= \int_0^{\xtm}  d \xt{N} \, \xt{N}^{-1-2 \, \varepsilon} \, \int_0^{\pi} d \phi_N \, (\sin \phi_N)^{-2 \, \varepsilon} \, \int_{\yNmin}^{\yNmax} d y_N \, \notag \\
  &\quad {} \times \sum_{i=3}^{N-1} \frac{1}{\hp_i \cdot \hp_N} \,\left[ L_{i}(y_N,\xt{N},\phi_N) - \hat{L}_{i}(y_N,\phi_N) \right]\, ,
  \label{eqdefT4ij1!}
\end{align}
where
\begin{align}
  L_{i}(y_N,\xt{N},\phi_N) &= \left[ \sum_{j=i+1}^{N-1} f_{ij}(y_N,\xt{N},\phi_N) g_{ij}(y_N,\phi_N) \right. \notag \\
  &\quad {} + \left. \sum_{j=3}^{i-1} f_{ji}(y_N,\xt{N},\phi_N) \, g_{ji}(y_N,\phi_N) \right] \, \frac{1}{\pt{i}},
  \label{eqdefLi0!}
\end{align}
and
\begin{align}
  \hat{L}_{i}(y_N,\phi_N) &= \left[ \sum_{j=i+1}^{N-1} f_{ij}(0,0,0) g_{ij}(y_N,\phi_N) + \sum_{j=3}^{i-1} f_{ji}(0,0,0) \, g_{ji}(y_N,\phi_N) \right] \,\frac{1}{\pt{i}}\, .
  \label{eqdefLhi0!}
\end{align}
In eqs.~(\ref{eqdefLi0!}) and (\ref{eqdefLhi0!}), we took the convention that if the lower bound of the sum is greater than the upper bound, then the sum gives zero.
Furthermore, for each term of the summation over the subscript $i$, a collinear approximation $p_N = (1-z_i)/z_i \, p_i$ is done in the term enclosed by squared brackets in eq.~(\ref{eqdefT4ij1!}).
The variable $z_i$ represents the ratio of the energy of the parton after the emission of $i_N$ over the energy before the emission, cf.\ fig. \ref{fig2},
yielding the collinear subtraction term
\begin{align}
  \cTiinc{4} &= \sum_{i=3}^{N-1} \xt{i}^{-2 \, \varepsilon}  \, J^{\text{coll}} \, \int_{\zm{i}}^{1}  \frac{d z_i}{z_i} \, z_i^{2 \, \varepsilon} \, (1-z_i)^{-1-2 \, \varepsilon} \, \left[ L_{i}\left(y_i,\frac{1-z_i}{z_i} \, \xt{i},0\right) - \hat{L}_{i}(y_i,0) \right] \, .
  \label{eqdefT4collij0!}
\end{align}
The collinear integral $J^{\text{coll}}$, appearing in eq.~(\ref{eqdefT4collij0!}), is defined as
\begin{align}
  J^{\text{coll}} &\equiv \int_0^{\pi} d \phi_N \, (\sin \phi_N)^{-2 \, \varepsilon} \, \int_{-\infty}^{+\infty} d y_N \,  \frac{\cos \phi_N}{\hp_i \cdot \hp_N}\, .
  \label{eqdefJcoll!}
\end{align}
The details of the computation of $J^{\text{coll}}$ are given is appendix \ref{collint}.
Concerning eq.~(\ref{eqdefJcoll!}), two remarks are in order. First, to simplify the analytic computation of $J^{\text{coll}}$ the bounds of the $y_N$ integration are sent to infinity. This is justified by the fact that the collinear divergence is at $y_N = y_i$ and not at the boundary of the integration. The choice of these bounds washes out the dependence on $i$ in the final result of $J^{\text{coll}}$.
Second, in the collinear subtraction term, we choose to multiply the integrand by a factor $\cos \phi_N$, it does not change the divergence which is located at $\phi_N = 0$.
If this factor is not present, once the analytical computation over $\phi_N$ and $y_N$ has been performed, a global factor $\Gamma^2(1-\varepsilon)/\Gamma(1 - 2 \, \varepsilon)$ appears, which is different from the global factor found in the soft case or if $i$ or/and $j$ belong to set $S_i$ (see below), namely $\Gamma^2(1/2-\varepsilon)/\Gamma(1 - 2 \, \varepsilon)$. To factorise out the latter, the former global factor has to be expressed in terms of the latter.
It gives some spurious factors which blur uselessly the formulae obtained at the end.
Note that, in the collinear approximation, $L_{i}(y_i,(1-z_i)/z_i \, \xt{i},0)$ and $\hat{L}_{i}(y_i,0)$ become simply
\begin{align}
  L_{i}\left(y_i,\frac{1-z_i}{z_i} \, \xt{i},0\right) &= \sum_{j=i+1}^{N-1} f_{ij}\left(y_i,\frac{1-z_i}{z_i} \, \xt{i},0 \right)+ \sum_{j=3}^{i-1} f_{ji}\left(y_i,\frac{1-z_i}{z_i} \, \xt{i},0 \right) \label{eqdefLi1!}\, , \\
  \hat{L}_{i}(y_i,0) &= \sum_{j=i+1}^{N-1} f_{ij}\left(0,0,0 \right)+ \sum_{j=3}^{i-1} f_{ji}\left(0,0,0 \right)\, .
  \label{eqdefLhi1!}
\end{align}

\noindent
\textbf{Analytical integration of the subtraction terms}\\
The gory details of the computation of the two integrals $\Jso$ and $J^{\text{coll}}$ are given in, respectively, the appendix \ref{softint} and \ref{collint}. Let us mention the final result:
\begin{align}
  \Jso &= 2^{-2 \, \varepsilon} \, \frac{\Gamma^2 \left( \frac{1}{2} - \varepsilon \right)}{\Gamma(1 - 2 \, \varepsilon)} \, \left\{ -\frac{2}{\varepsilon} + 2 \, \ln \left( 2 \, \bd_{ij} \right) - \varepsilon \, \ln^2 \left( 2 \, \bd_{ij} \right) + 4 \, \varepsilon \, (\ystij)^2  \right\}\, ,
  \label{eqdefJsoft12!} \\
  J^{\text{coll}} &= \frac{2^{-2 \, \varepsilon}}{-\varepsilon} \, \frac{\Gamma^2\left(\frac{1}{2}-\varepsilon\right)}{\Gamma(1-2 \, \varepsilon)}\, ,
  \label{eqdefJcolla!}
\end{align}
where $\bd_{ij} = \cosh(y_i - y_j) - \cos(\phi_{i} - \phi_j)$ and $\ystij = (y_i-y_j)/2$.
With these results, the analytical integration over $p_N$ of the soft subtraction term can be performed easily yielding
\begin{align}
  \cTiins{4} &= 2^{-1-2 \, \varepsilon} \, \frac{\Gamma^2 \left( \frac{1}{2} - \varepsilon \right)}{\Gamma(1 - 2 \, \varepsilon)} \, \xtm^{-2 \, \varepsilon} \, \sum_{i=3}^{N-2} \sum_{j=i+1}^{N-1} f_{ij}(0,0,0) \, \notag \\
  &\qquad {} \times \left\{ \frac{2}{\varepsilon^2} - \frac{2}{\varepsilon} \, \ln \left( 2 \, \bd_{ij} \right) + \ln^2 \left( 2 \, \bd_{ij} \right) - 4 \, (\ystij)^2  \right\}\, ,
  \label{eqdefsoftpart2!}
\end{align}
while the analytical integration of the collinear subtraction term leads to
\begin{align}
  \cTiinc{4} &= \frac{2^{-2 \, \varepsilon}}{-\varepsilon} \, \frac{\Gamma^2\left(\frac{1}{2}-\varepsilon\right)}{\Gamma(1-2 \, \varepsilon)} \,  \sum_{i=3}^{N-1} \xt{i}^{-2 \, \varepsilon} \, \int_{\zm{i}}^{1}  \frac{d z_i}{z_i} \, z_i^{2 \, \varepsilon} \, (1-z_i)^{-1-2 \, \varepsilon} \, \notag \\
  &\qquad \qquad \qquad \qquad \qquad \qquad {} \times \left[ L_{i}\left(y_i,\frac{1-z_i}{z_i} \, \xt{i},0\right) - \hat{L}_{i}(y_i,0) \right]\, ,
  \label{eqdefT4collij1}
\end{align}
where $\zm{i} = \xt{i}/(\xt{i} + \xtm)$.

\vspace{0.5cm}

\noindent
It remains to express the coefficients of the collinear divergences when $p_N // p_i$ in terms of the "plus" distributions. For that, note that the structure in $z_i$ of eq.~(\ref{eqdefT4collij1}) is of the type
\begin{equation}
  \cala_1 \equiv \xt{i}^{- 2 \, \varepsilon} \, \int^1_{\zm{i}} \frac{d z_i}{z_i \, (1-z_i)} \, \left( \frac{1-z_i}{z_i} \right)^{- 2 \, \varepsilon} \, \left[ F(z_i) - F(1) \right]\, .
\end{equation}
It can be further re-written as
\begin{equation}
  \cala_1 = \xt{i}^{- 2 \, \varepsilon} \, \int^1_{\zm{i}} \frac{d z_i}{z_i} \, z_i^{2 \, \varepsilon} \, \left[ \frac{1}{1-z_i} - 2 \, \varepsilon \, \frac{\ln(1-z_i)}{1-z_i} \right] \, \left[ F(z_i) - F(1) \right]\, .
\end{equation}
Note that, as will be clear later, the term $z_i^{2 \, \varepsilon}$ does not need to be expanded around $\varepsilon=0$ because 
it drops out after a change of variables to recover the collinear structure, cf.~appendix \ref{gutsdivterm}.
$\cala_1$ will be written in terms of the "plus" distributions which is defined as 
\begin{align}
  \int_0^1 dx \, (g(x))_{+} \, F(x) &\equiv \int_0^1 dx \, g(x) \, (F(x) - F(1))\, ,
\end{align}
where $g(x)$ is a function singular at $x=1$ such that $(1-x) \, g(x)$ is integrable and $F(x)$ is a regular one at the same point.\\

\noindent
Using that
\begin{align}
  \hspace{2em}&\hspace{-2em} \int_{\zm{i}}^1 \frac{d z_i}{z_i} \, z_i^{2 \, \varepsilon} \, \frac{F(z_i)}{(1-z_i)_{+}} \notag \\
  &= \int_{\zm{i}}^1  \frac{d z_i}{z_i} \, z_i^{2 \, \varepsilon} \, \frac{F(z_i) - F(1)}{1-z_i} + F(1) \, \left\{ -\ln\left( \frac{\zm{i}}{1-\zm{i}} \right) \right. \notag \\
  &\qquad \qquad {} + \left. 2 \, \varepsilon \, \left[ -\frac{1}{2} \, \ln^2(\zm{i}) + \dilog(\zm{i}) - \frac{\pi^2}{6} + \ln(1-\zm{i}) \, \ln(\zm{i}) \right] \right\},
  \label{eqdistribplus0}
\end{align}
and
\begin{align}
  \hspace{2em}&\hspace{-2em} \int_{\zm{i}}^1 \frac{d z_i}{z_i} \, z_i^{2 \, \varepsilon} \, \left( \frac{\ln(1-z_i)}{1-z_i} \right)_{+} \, F(z_i) \notag \\
  &= \int_{\zm{i}}^1  \frac{d z_i}{z_i} \, z_i^{2 \, \varepsilon} \, \frac{F(z_i) - F(1)}{1-z_i} \, \ln(1-z_i) + F(1) \, \left\{ \frac{1}{2} \, \ln^2(1-\zm{i}) + \dilog(\zm{i}) - \frac{\pi^2}{6} \right\}\, ,
  \label{eqdistriblogplus0}
\end{align}
the $\cala_1$ term becomes
\begin{align}
  \cala_1 &= \xt{i}^{- 2 \, \varepsilon} \, \left\{ \int_{\zm{i}}^1 \frac{d z_i}{z_i} \, z_i^{2 \, \varepsilon} \, \frac{F(z_i)}{(1-z_i)_{+}} - 2 \, \varepsilon \, \int_{\zm{i}}^1 \frac{d z_i}{z_i} \, z_i^{2 \, \varepsilon} \, \left( \frac{\ln(1-z_i)}{1-z_i} \right)_{+} \, F(z_i) \right. \notag \\
  &\quad + \left.  F(1) \left[ \ln\left( \frac{\zm{i}}{1-\zm{i}} \right) + \varepsilon \, \ln^2\left( \frac{\zm{i}}{1-\zm{i}} \right) \right] \right\}\, .
\end{align}
Making the appearance of the "plus" distributions generates some soft terms. Therefore, we will group the results of the analytical integration on $\xt{N}$, $\phi_N$ and $y_N$ of the soft and collinear subtraction terms into the quantity $\cTiindiv{4} \equiv \cTiins{4} + \cTiinc{4}$ yielding
\begin{align}
\hspace{2em}&\hspace{-2em} \cTiindiv{4} = 2^{-2 \, \varepsilon} \, \frac{\Gamma^2\left(\frac{1}{2}-\varepsilon\right)}{\Gamma(1-2 \, \varepsilon)}  \notag \\
&\quad {} \times\left\{ -\frac{1}{\varepsilon} \, \sum_{i=3}^{N-1} \xt{i}^{- 2 \, \varepsilon} \, \, \left[ \int_{\zm{i}}^1 \frac{d z_i}{z_i} \, \frac{z_i^{2 \, \varepsilon}}{(1-z_i)_{+}} \, \calf_{i}\left(y_i,\frac{1-z_i}{z_i} \, \xt{i},0\right) \right. \right. \notag \\
&\qquad \qquad \qquad \qquad \quad {} - \left. 2 \, \varepsilon \, \int_{\zm{i}}^1 \frac{d z_i}{z_i} \, z_i^{2 \, \varepsilon} \, \left( \frac{\ln(1-z_i)}{1-z_i} \right)_{+} \, \calf_{i}\left(y_i,\frac{1-z_i}{z_i} \, \xt{i},0\right) \right] \notag \\
&\qquad \quad {} +  \sum_{i=3}^{N-2} \sum_{j=i+1}^{N-1} f_{ij}(0,0,0) \, \left[ \frac{1}{\varepsilon^2} - \frac{1}{\varepsilon} \, \ln(2 \, \xt{i} \, \xt{j} \, d_{ij}) + \ln^2(\xt{i}) +  \ln^2(\xt{j}) \right. \notag \\
&\qquad \qquad \qquad \qquad \qquad \qquad {} + \left. \left. 2 \, \ln(\xtm) \, \ln(2 \, \bd_{ij}) + \frac{1}{2} \, \ln^2(2 \, \bd_{ij}) - 2 \, \left( \ystij \right)^2  \vphantom{\frac{\Gamma^2(1-\varepsilon)}{\Gamma(1 - 2 \, \varepsilon)}} \right]  \vphantom{\frac{f_{34} \left( y_3, \frac{1-z_3}{z_3} \, \pt{3}, 0 \right)}{(1-z_3)_{+}}} \right\},
\label{eqdefI3480}
\end{align}
where, to lighten the notations, the following quantity has been introduced
\begin{align}
  \calf_{i}(y_N,\xt{N},\phi_N) &= \sum_{j=i+1}^{N-1} f_{ij}(y_N,\xt{N},\phi_N)+ \sum_{j=3}^{i-1} f_{ji}(y_N,\xt{N},\phi_N)\, .
  \label{eqdefFcal0}
\end{align}

\subsubsection{Mixed terms ISR and FSR: $i$ belongs to $S_i$ and $j$ belongs to $S_f$}

This case is a bit more complicated due to the appearance of initial and final state collinear divergences. Let us treat in detail the case where $i=1$. The case where $i=2$ can be obtained from the former one by changing the label 1 into the label 2 and the sign of the rapidities.\\

\noindent
\textbf{Construction of the subtraction terms}\\
Let us remind the definition of $\cTiin{2}$:
\begin{align}
  \cTiin{2} &\equiv \int_0^{\xtm}  d \xt{N} \, \xt{N}^{-1-2 \, \varepsilon} \, \int_0^{\pi} d \phi_N \, (\sin \phi_N)^{-2 \, \varepsilon} \, \int_{\yNmin}^{\yNmax} d y_N \, \sum_{j=3}^{N-1} f_{1j}(y_N,\xt{N},\phi_N) \, E_{1j}^{\prime}\, .
  \label{eqdefT2ij0!}
\end{align}
The $y_N$ integration range is split in two parts\footnote{For the sake of simplicity, it is assumed that $y_j$ belongs to the range $[\yNmin,\yNmax]$, that is to say that $\Delta y \geq 0$. If it is not the case, no collinear divergence shows up.}:
\begin{align}
  \int_{\yNmin}^{\yNmax} d y_N  &= \int_{y_j}^{\yNmax} d y_N + \int_{\yNmin}^{y_j} d y_N\, ,
  \label{eqsplitrangeyN1!}
\end{align}
and the change of variable $\Delta y = y_N - y_j$ in the first (respectively $\Delta y = y_j-y_N$ in the second) integral of the right hand side of eq.~(\ref{eqsplitrangeyN1!}) is performed. This leads to
\begin{align}
  \cTiin{2} &= \int_0^{\xtm}  d \xt{N} \, \xt{N}^{-1-2 \, \varepsilon} \, \int_0^{\pi} d \phi_N \, (\sin \phi_N)^{-2 \, \varepsilon} \, \notag \\
  &\quad {} \times \left[  \sum_{j=3}^{N-1} \int_{0}^{\dyM} d \Delta y \, f_{1j}(y_j+\Delta y,\xt{N},\phi_N) \, \frac{e^{\Delta y}}{\cosh(\Delta y)-\cos \phi_N} \right. \notag \\
  &\qquad {} + \left. \sum_{j=3}^{N-1} \int^{\dym}_{0} d \Delta y \, f_{1j}(y_j-\Delta y,\xt{N},\phi_N) \, \frac{e^{-\Delta y}}{\cosh(\Delta y)-\cos \phi_N}\right]\, ,
  \label{eqdefT2ij1!}
\end{align}
where $\dyM = \yNmax-y_j$ and $\dym = y_j-\yNmin$. In the limit $\xt{N} \rightarrow 0$, the two bounds $\dyM$ and $\dym$ are sent to $+\infty$ (see eqs.~(\ref{eqdefyNmin})) and, because of the factor $e^{\Delta y}$ in its integrand, the first integral in the square brackets diverges in this region in addition to the final state collinear divergence at $\Delta y = 0$ and $\phi_N = 0$. The divergence at $\Delta y = +\infty$ originates from the collinear divergence when the parton $i_N$ flies along the beam direction. To disentangle these two divergences, we use a partial fraction decomposition to write $\cTiin{2}$ in the following form:
\begin{align}
  \cTiin{2} &= \int_0^{\xtm}  d \xt{N} \, \xt{N}^{-1-2 \, \varepsilon} \, \int_0^{\pi} d \phi_N \, (\sin \phi_N)^{-2 \, \varepsilon} \, \notag \\
  &\quad {} \times \left[  2 \, \sum_{j=3}^{N-1} \int_{0}^{\dyM} d \Delta y \, f_{1j}(y_j+\Delta y,\xt{N},\phi_N)  \right. \notag \\
  &\qquad \quad {} - \sum_{j=3}^{N-1}\int_{0}^{\dyM} d \Delta y \, f_{1j}(y_j+\Delta y,\xt{N},\phi_N) \, \frac{e^{-\Delta y}-2 \, \cos \phi_N}{\cosh(\Delta y)-\cos \phi_N} \notag \\
  &\qquad \quad {} + \left. \sum_{j=3}^{N-1} \int^{\dym}_{0} d \Delta y \, f_{1j}(y_j-\Delta y,\xt{N},\phi_N) \, \frac{e^{-\Delta y}}{\cosh(\Delta y)-\cos \phi_N}\right]\, .
  \label{eqdefT2ij2!}
\end{align}
Then, introducing the change of variables $\Delta y = \ln(1/t)$ leads to
\begin{align}
  \cTiin{2} &= 2 \, \int_0^{\xtm}  d \xt{N} \, \xt{N}^{-1-2 \, \varepsilon} \, \int_0^{\pi} d \phi_N \, (\sin \phi_N)^{-2 \, \varepsilon} \, \notag \\
  &\quad {} \times \left[  \sum_{j=3}^{N-1} \int^{1}_{e^{-\dyM}} \frac{dt}{t} \, f_{1j}\left(y_j+\ln\left( \frac{1}{t} \right),\xt{N},\phi_N\right)  \right. \notag \\
  &\qquad \quad {} - \sum_{j=3}^{N-1} \int^{1}_{e^{-\dyM}} d t \, f_{1j}\left(y_j+\ln\left( \frac{1}{t} \right),\xt{N},\phi_N\right) \, \frac{t-2 \, \cos \phi_N}{t^2+1-2 \, t \, \cos \phi_N} \notag \\
  &\qquad \quad {} + \left. \sum_{j=3}^{N-1} \int^{1}_{e^{-\dym}} d t \, f_{1j}\left(y_j-\ln\left( \frac{1}{t} \right),\xt{N},\phi_N\right) \, \frac{t}{t^2+1-2 \, t \, \cos \phi_N}\right] \, .
  \label{eqdefT2ij3!}
\end{align}
The subtraction terms for the last two terms of eq.~(\ref{eqdefT2ij3!}) will be constructed, simply, by sending $\dym$ and $\dyM$ to infinity and taking the function $f_{1j}$ at $\Delta y=0$ and $\phi_N=0$. Let us now focus on the first term  of eq.~(\ref{eqdefT2ij3!}). The key point is that $\exp(-\dyM)$ depends on $\xt{N}$ in a complicated way. It will thus be replaced in the construction of the subtraction term by the expression $\exp(-\dytM) = \xt{N} \, \omega \, \exp(y_j)/(1-\bx{1})$ which goes to zero at the same speed as $\exp(-\dyM)$ when $\xt{N} \rightarrow 0$ but is a linear function of $\xt{N}$. Let us consider the quantity
\begin{align}
  \calt_{\text{in}}^{(2) \, \prime} &= 2 \,  \sum_{j=3}^{N-1} \,  \int_0^{\xtm}  d \xt{N} \, \xt{N}^{-1-2 \, \varepsilon} \, \int_0^{\pi} d \phi_N \, (\sin \phi_N)^{-2 \, \varepsilon} \, \notag \\
  &\quad {} \times \int^{1}_{\frac{\xt{N} \, \omega \, e^{y_j}}{1-\bx{1}}} \frac{dt}{t} \, f_{1j}\left(y_j+\ln\left( \frac{1}{t} \right),\xt{N},\phi_N\right) \, .
  \label{eqdefT2pij1!}
\end{align}
The order of integration over the variables $\xt{N}$ and the $t$ will be exchanged in eq.~(\ref{eqdefT2pij1!}). The structure of $\calt_{\text{in}}^{(2) \, \prime}$ with respect to these two variables is of the type
\begin{align}
  \text{Structure of $\calt_{\text{in}}^{(2) \, \prime}$} &= \int_0^{\xtm}  d \xt{N} \, \xt{N}^{-1-2 \, \varepsilon} \, \int^{1}_{\xt{N} \, \beta_{1j}} \frac{dt}{t} \, F(t,\xt{N}) \, ,
  \label{eqstrucT2p0!}
\end{align}
where $\beta_{1j} = \omega \, \exp(y_j)/(1-\bx{1})$.
This change of the integration order leads to a dichotomy of cases.
This can be understood by remembering that when the fraction of 4-momentum $x_1$ or $x_2$ goes to 1, there is almost no room to emit a soft gluon, such that the limit on $\pt{N}$ due to kinematics becomes smaller than the size of the cylinder inducing this dichotomy.\\

\noindent
{\bf 1)} $\xtm \leq 1/\beta_{1j}$\\
This case yields two terms and the structure of $\calt_{\text{in}}^{(2) \, \prime}$ after a change of variable $\xt{N} = z \, t$ in one of them, becomes
\begin{align}
  \text{Structure of $\calt_{\text{in}}^{(2) \, \prime}$} &= \int^{\beta_{1j} \, \xtm}_{0} d t \, t^{-1-2 \, \varepsilon} \, \int^{1/\beta_{1j}}_{0} d z \, z^{-1-2 \, \varepsilon} \, F(t,z \, t) \notag \\
  &\quad {} + \int^{1}_{\beta_{1j} \, \xtm} \frac{dt}{t} \, \int^{\xtm}_{0} d \xt{N} \, \xt{N}^{-1 - 2 \, \varepsilon} \, F(t,\xt{N})\, .
  \label{eqstrucT2p1}
\end{align}

\noindent
{\bf 2)} $\xtm > 1/\beta_{1j}$\\
This case generates only one term and by changing the variable $\xt{N} = z \, t$, the structure of $\calt_{\text{in}}^{2 \, \prime}$ becomes
\begin{align}
  \text{Structure of $\calt_{\text{in}}^{(2) \, \prime}$} &= \int^{1}_{0} d t \, t^{-1-2 \, \varepsilon} \, \int^{1/\beta_{1j}}_{0} d z \, z^{-1-2 \, \varepsilon} \, F(t,z \, t) \, .
  \label{eqstrucT2p2}
\end{align}
\\

\begin{figure}[ht]
  \begin{center}
  \includegraphics[scale=0.7]{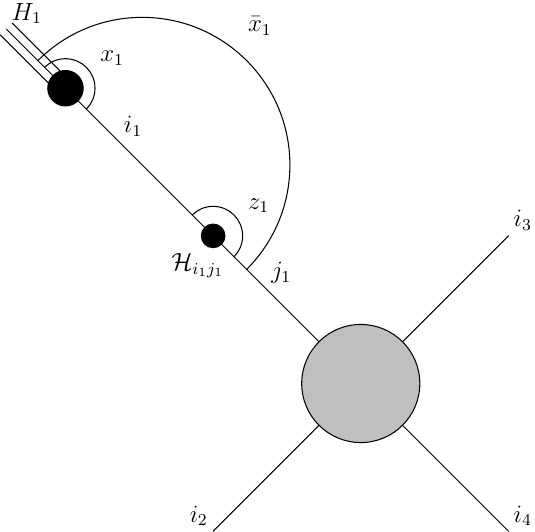}
  \end{center}
\caption{\footnotesize
Labelling of parton radiation in the initial state. 
The variables $x_1, z_1$ and $\bx{1}$ are defined as follows:
$p_1 = x_1 K_1$, $q_1 = z_1 p_1$, and $q_1 = \bx{1} K_1$ where
$p_1$, $q_1$, and $K_1$ are the four-momenta of the partons $i_1$, $j_1$ and the hadron $H_1$, 
respectively.
}
\label{fig1}
\end{figure}

\noindent
Note that expressing the components of the four-momentum $p_N$ in terms of the variables $z$ and $t$ leads to
\begin{align}
  p_N &= Q \, z \left( \frac{1}{2} \, \left( e^{y_j} + t^2 \, e^{-y_j} \right), t \, \vpthat{N}, \frac{1}{2} \, \left( e^{y_j} - t^2 \, e^{-y_j} \right) \right) \, .
  \label{eqcomppN}
\end{align}
By inspecting eq.~(\ref{eqcomppN}), it is easy to realise that
the vanishing of the variable $z$ leads to the soft limit while the vanishing of the variable $t$ leads to the collinear limit $p_{N} // p_1$.
Indeed, in the limit $t \rightarrow 0$, the four-momentum $p_N$ becomes
\begin{equation}
  p_N = z \, e^{y_j} \, \frac{Q}{2} \, \left(1, \vec{0}, 1\right)\, ,
  \label{eqdefpNcoll}
\end{equation}
such that $p_N$ is collinear to $K_1$. The four-momentum $p_1$ of the parton $i_1$ is $p_1=x_1 \, K_1$, the four-momentum of the parton $j_1$ is $z_1 \, x_1 \, K_1 \equiv \bx{1} \, K_1$, cf.~fig.~\ref{fig1}. Thus, the momentum $p_N$ is equal to $(1 - z_1) \, x_1 \, K_1 = (x_1 - \bx{1}) \, K_1$ and from eq.~(\ref{eqdefpNcoll}), $x_1$ reads $x_1 = \bx{1} + z \, \omega \, e^{y_j}$.

\vspace{0.5cm}

\noindent
All this preliminary discussion yields the construction of the subtraction terms. The one for the initial state collinear divergence will be constructed from eq.~(\ref{eqstrucT2p1}) or eq.~(\ref{eqstrucT2p2}) as
\begin{align}
    \cTiinci{2} &= 2 \, \sum_{j=3}^{N-1} \, \int_0^{\pi} d \phi_N \, (\sin \phi_N)^{-2 \, \varepsilon} \,  \int_{0}^{1/\beta_{1j}} dz \, z^{-1-2\varepsilon} \, \left( f_{1j}^c(z) - f_{1j}^c(0) \right) \notag \\
    &\quad {} \times \left[  \Theta\left( \frac{1}{\beta_{1j}} - \xtm\right) \,   \int_{0}^{\beta_{1j} \, \xtm} dt \, t^{-1 - 2 \, \varepsilon}  + \Theta\left( \xtm - \frac{1}{\beta_{1j}} \right) \,   \int_{0}^{1} dt \, t^{-1 - 2 \, \varepsilon} \,  \right]\, .
    \label{eqdefT2ijsub0!}
\end{align}
In eq.~(\ref{eqdefT2ijsub0!}), $f_{1j}^{c}(z)$ corresponds to the function  $f_{1j}(y_N,\xt{N},\phi_N)$  in which all the scalar products containing $p_N$ are evaluated in the configuration where $p_N // p_1$, that is to say by reading the components of $p_N$ in eq.~(\ref{eqdefpNcoll}). It is easy to realise that, with respect to the variables describing the phase space of the particle  $i_N$, it is thus a function of $z$ only. Note also that $f_{1j}^{c}(0)$ corresponds to $f_{1j}(0,0,0)$.
For the final state collinear divergence, the subtraction term can be read directly from eq.~(\ref{eqdefT2ij3!}):
\begin{align}
    \cTiincf{2} &= 2 \, \sum_{j=3}^{N-1} \,  \int_{0}^{\xtm} d \xt{N} \, \xt{N}^{-1-2 \, \varepsilon} \, \left[ f_{1j}(y_j,\xt{N},0) - f_{1j}(0,0,0) \right] \, \notag \\
    &\quad {} \times \int_0^{\pi} d \phi_N \, (\sin \phi_N)^{-2 \, \varepsilon} \,  \int_{0}^{1} dt \, \frac{2 \, \cos \phi_N}{t^2 + 1 - 2 \, t \, \cos \phi_N}   \notag \\
    &= 2 \, \sum_{j=3}^{N-1} \, J^{\text{coll}} \, \int_{0}^{\xtm} d \xt{N} \, \xt{N}^{-1-2 \, \varepsilon} \, \left[ f_{1j}(y_j,\xt{N},0) - f_{1j}(0,0,0) \right]\, .
    \label{eqdefT2ijsub1!}
\end{align}
The subtraction term for the soft divergence can be constructed by looking at eqs.~(\ref{eqdefT2ij3!}) and (\ref{eqstrucT2p1}) (or (\ref{eqstrucT2p2})) as
\begin{align}
    \cTiins{2} &= 2\sum_{j=3}^{N-1} \, f_{1j}(0,0,0) \int_0^{\pi} d \phi_N \, (\sin \phi_N)^{-2 \, \varepsilon} \left\{ \int_{0}^{\xtm} d \xt{N} \, \xt{N}^{-1-2 \, \varepsilon} \int_{0}^{1} dt \, \frac{2 \, \cos \phi_N}{t^2 + 1 - 2 \, t \, \cos \phi_N} \right. \notag \\
    &\qquad \qquad {} + \Theta\left( \frac{1}{\beta_{1j}} - \xtm\right) \, \left[ \int_{0}^{\beta_{1j} \, \xtm} dt \, t^{-1 - 2 \, \varepsilon} \, \int_{0}^{1/\beta_{1j}} dz \, z^{-1-2\varepsilon} \right. \notag \\
    &\qquad \qquad \qquad \qquad \qquad \qquad \qquad \qquad {} + \left. \int_{\beta_{1j} \, \xtm}^1 \frac{dt}{t} \, \int_{0}^{\xtm} dz \, z^{-1-2\varepsilon} \right] \notag \\
    &\qquad \qquad {} + \left.  \Theta\left( \xtm - \frac{1}{\beta_{1j}} \right) \, \int_{0}^{1} dt \, t^{-1 - 2 \, \varepsilon} \, \int_{0}^{1/\beta_{1j}} dz \, z^{-1-2\varepsilon} \right\} \, .
    \label{eqdefcTiins2}
\end{align}
\\

\noindent
\textbf{Analytical integration of the subtraction terms}\\
The analytical integration of $\cTiins{2}$, $\cTiinci{2}$ and $\cTiincf{2}$ are easy to perform and the results are
\begin{align}
    \cTiins{2} &=  2^{-2 \, \varepsilon} \, \frac{\Gamma^2\left( \frac{1}{2} - \varepsilon\right)}{\Gamma(1 - 2 \, \varepsilon)}\, \sum_{j=3}^{N-1} \, f_{1j}(0,0,0) \, \xtm^{-2 \, \varepsilon} \, \notag \\
    &\quad {} \times \left\{ \frac{1}{\varepsilon^2} + \frac{1}{\varepsilon} \, \ln(\beta_{1j} \, \xtm) + \Theta\left( \xtm - \frac{1}{\beta_{1j}} \right) \, \, \ln^2(\beta_{1j} \, \xtm) \right\}\, ,
    \label{eqresintcTiins2} \\
    \cTiinci{2} &= - 2^{-2 \, \varepsilon} \, \frac{\Gamma^2\left( \frac{1}{2} - \varepsilon\right)}{\Gamma(1 - 2 \, \varepsilon)} \, \frac{1}{\varepsilon} \, \sum_{j=3}^{N-1} \, \int_{0}^{1/\beta_{1j}} dz \, z^{-1-2\varepsilon} \, \left[ f_{1j}^c(z) - f_{1j}^c(0) \right] \notag \\
    &\qquad {} \times \left[  \Theta\left( \frac{1}{\beta_{1j}} - \xtm\right) \,(\beta_{1j} \, \xtm)^{-2 \, \varepsilon} + \Theta\left( \xtm - \frac{1}{\beta_{1j}} \right)\right] \, ,
    \label{eqrescTiinci2} \\
    \cTiincf{2} &= - 2^{-2 \, \varepsilon} \, \frac{\Gamma^2\left( \frac{1}{2} - \varepsilon\right)}{\Gamma(1 - 2 \, \varepsilon)} \, \frac{1}{\varepsilon} \, \sum_{j=3}^{N-1} \, \int_{0}^{\xtm} d \xt{N} \, \xt{N}^{-1-2 \, \varepsilon} \, \left[ f_{1j}(y_j,\xt{N},0) - f_{1j}(0,0,0) \right]\, .
    \label{eqrescTiincf2}
\end{align}

\noindent
In order to make the "plus" distributions appear explicitly, the following changes of variable $z = (\bx{1}/z_1 - \bx{1}) \, \exp(-y_j)/\omega$, respectively  $\xt{N} = (1-z_j)/z_j \, \xt{j}$, are applied on the terms containing an integral over $z$
in eq.~(\ref{eqrescTiinci2}), respectively over $\xt{N}$ in eq.~(\ref{eqrescTiincf2}). 
Let us discuss in detail the first change of variable. The terms containing an integral over $z$ in eq.~(\ref{eqrescTiinci2}) are of the type
\begin{equation}
  \int_0^{(1-\bx{1}) \, e^{-y_j}/\omega} dz \, z^{-1-2 \, \varepsilon} \, \left[ G(z) - G(0) \right] \equiv \cala_2 \, .
  \label{eqdefT20}
\end{equation}
After the first change of variable, it becomes
\begin{equation}
  \cala_2 = \left( \frac{\bx{1} \, e^{-y_j}}{{\omega}} \right)^{-2 \, \varepsilon} \, \int_{\bx{1}}^1 d z_1 \, z_1^{-1+2 \, \varepsilon} \, (1-z_1)^{-1-2 \, \varepsilon} \, \left( F(z_1) - F(1) \right),
  \label{eqdefT21}
\end{equation}
with $F(z_1) = G(\bx{1} \, e^{-y_j}/\omega \, (1-z_1)/z_1)$.
Making explicit the appearance of the "plus" distributions, with the help of eqs.~(\ref{eqdistribplus0}) and (\ref{eqdistriblogplus0}), leads to
\begin{align}
  \cala_2 &= \left( \frac{\bx{1} \, e^{-y_j}}{\omega} \right)^{-2 \, \varepsilon} \, \left\{ \int_{\bx{1}}^1  \frac{d z_1}{z_1} \, z_1^{2 \, \varepsilon} \,
   \frac{F(z_1)}{(1-z_1)_{+}}
   - 2 \, \varepsilon  \, \int_{\bx{1}}^1  \frac{d z_1}{z_1} \, z_1^{2 \, \varepsilon} \,  \left( \frac{\ln(1-z_1)}{1-z_1} \right)_{+} \, F(z_1) \right. \notag \\
    &\qquad \qquad \qquad \qquad {} + \left.  F(1) \, \left[  \ln \left( \frac{\bx{1}}{1-\bx{1}} \right) + \varepsilon \, \ln^2 \left( \frac{\bx{1}}{1-\bx{1}} \right) \right] \right\}\, .
    \label{eqdistribplus1}
\end{align}
As in the preceding case, we will group the divergent parts because the evaluation of the "plus" distributions generates some soft terms. For that purpose we define $\cTiindiv{2} = \cTiins{2} + \cTiinci{2} + \cTiincf{2}$.
Thus, using the result given by eq.~(\ref{eqdistribplus1}) and the one associated to the change of variable  $\xt{N} = (1-z_i)/z_i \, \xt{i}$ (cf. eqs~(\ref{eqdistribplus0}) and (\ref{eqdistriblogplus0})) and neglecting terms which vanish when $\varepsilon \rightarrow 0$, the divergent part reads
\begin{align}
  \cTiindiv{2} &= 2^{-2 \, \varepsilon} \, \frac{\Gamma^2 \left( \frac{1}{2}-\varepsilon \right)}{\Gamma(1 - 2 \, \varepsilon)} \, \sum_{j=3}^{N-1} \left\{  -\frac{\chi(\bx{1},y_j)^{- 2 \, \varepsilon}}{\varepsilon} \, \int_{\bx{1}}^{1} \frac{dz_1}{z_1} \, z_1^{2 \, \varepsilon} \, \frac{f_{1j}^c \left( \frac{\bx{1} \, e^{-y_j}}{\omega} \, \frac{1-z_1}{z_1} \right)}{(1-z_1)_{+}} \right. \notag \\
  &\qquad \qquad \qquad \qquad \quad {}+ 2 \,  \int_{\bx{1}}^{1} \frac{d z_1}{z_1} \, \left(\frac{\ln(1-z_1)}{1-z_1}\right)_{+} \, f_{1j}^c \left( \frac{\bx{1} \, e^{-y_j}}{\omega} \, \frac{1-z_1}{z_1} \right) \notag \\
  &\qquad \qquad \qquad \qquad \quad {} - \frac{1}{\varepsilon} \, \xt{j}^{- 2 \, \varepsilon} \, \int_{\zm{j}}^{1} \frac{d z_j}{z_j} \, z_j^{2 \, \varepsilon} \, \frac{f_{1j}\left(y_j,\frac{1-z_j}{z_j} \, \xt{j},0\right)}{(1-z_j)_{+}} \, \label{eqdefI1id20} \\
  &\qquad \qquad \qquad \qquad \quad {}+ 2 \, \xt{j}^{- 2 \, \varepsilon} \, \int_{\zm{j}}^{1} \frac{d z_j}{z_j} \, z_j^{2 \, \varepsilon} \, \left(\frac{\ln(1-z_j)}{1-z_j}\right)_{+} \, f_{1j}\left(y_j,\frac{1-z_j}{z_j} \, \xt{j},0\right) \notag \\
  &\qquad \qquad \qquad \qquad \quad {} - f_{1j}^c(0)  \, \frac{1}{\varepsilon} \, \left[  - \frac{1}{\varepsilon} + \ln\left( \frac{\bx{1} \, e^{-y_j} \, \xtm}{\omega} \right)  \right. 
  -  \varepsilon \, \left( \Upsilon(\bx{1},y_j) + \ln^2(\xtm) \right) \notag \\
  &\qquad \qquad \qquad \qquad \qquad \qquad \qquad \quad  {} + \left. \left. \xt{j}^{- 2 \, \varepsilon} \, \left\{ \ln   \left( \frac{\xt{j}}{\xtm} \right) + \varepsilon \, \ln^2 \left( \frac{\xt{j}}{\xtm} \right) \right\}  \right] \vphantom{\frac{f_{1j}^c \left( \frac{\bx{1} \, e^{-y_j}}{\omega} \, \frac{1-z_1}{z_1} \right)}{(1-z_1)_{+}}}  \right\}, \notag 
\end{align}
with $\zm{j} = \xt{j}/(\xt{j}+\xtm)$,
\begin{align}
  \chi(\bx{1},y_j) &= \left\{
  \begin{array}{cc}
    \frac{\bx{1} \, \xtm}{(1-\bx{1})} & \text{if} \quad \xtm \leq \frac{(1-\bx{1}) \, e^{-y_j}}{\omega} \\
    \frac{\bx{1} \, e^{-y_j}}{\omega} & \text{if} \quad \xtm > \frac{(1-\bx{1}) \, e^{-y_j}}{\omega}
  \end{array}
  \right. \, ,
\end{align}
and
\begin{align}
  \Upsilon(\bx{1},y_j) &= \left\{
  \begin{array}{cc}
    \ln^2 \left( \frac{\bx{1}}{1-\bx{1}} \right) +2 \, \ln(\xtm) \, \ln\left( \frac{\bx{1} \, e^{-y_j}}{\omega} \right) - \ln^2(\xtm) & \text{if} \quad \xtm \leq \frac{(1-\bx{1}) \, e^{-y_j}}{\omega} \\
    \ln^2 \left( \frac{\bx{1} \, e^{-y_j}}{\omega} \right) & \text{if} \quad \xtm > \frac{(1-\bx{1}) \, e^{-y_j}}{\omega}
  \end{array}
  \right.\, .
\end{align}

\noindent
Let us finish this part by the following remark. The dichotomy of cases yields the two conditions $\xtm \leq (1-\bx{1}) \, e^{-y_j}/\omega$ and $\xtm > (1-\bx{1}) \, e^{-y_j}/\omega$. For the case $\cTiin{3}$ the conditions would have been $\xtm \leq (1-\bx{2}) \, e^{y_j}/\omega$ and $\xtm > (1-\bx{2}) \, e^{y_j}/\omega$. 
For practical applications, in order to avoid numerous cases, the value of $\xtm$ can be adjusted in such a way that the condition $\xtm$ less than (or greater than) is always true irrespective of the index $j$.
Since all the final state hadrons are detected, their rapidity $y_j$ for $j=3, \ldots, N-1$ must be in the range $y_{\text{min}} \leq y_j \leq y_{\text{max}}$ where $y_{\text{min}}$ and $y_{\text{max}}$ are determined by experiments. Then, defining the rapidity $y_M$ as $\max(|y_{\text{min}}|,|y_{\text{max}}|)$, we have that
\begin{equation}
  e^{-y_M} \leq e^{\pm y_j} \quad \forall j \in \{3, 4, \ldots, N-1\} \, . 
  \label{eqineq0}
\end{equation}
Thus, if on one hand, we demand that $\xtm$ is chosen as
\begin{equation}
  \xtm \equiv \lambda \, \frac{1-\max(\bx{1},\bx{2})}{\omega} \, e^{-y_M},
  \label{eqdefxtm0}
\end{equation}
with $0 \leq \lambda \leq 1$, then the inequality
\begin{equation}
  \xtm \leq \min \left( \frac{(1-\bx{1})e^{-y_j}}{\omega}, \, \frac{(1-\bx{2})e^{y_j}}{\omega} \right) \quad \forall j \in \{3, 4, \ldots, N-1\},
  \label{eqineq1}
\end{equation}
is always true. 
On the other hand, we could have chosen for $\xtm$
\begin{equation}
  \xtm \equiv \lambda \, \frac{1-\min(\bx{1},\bx{2})}{\omega} \, e^{y_M},
  \label{eqdefxtm1}
\end{equation}
with $\lambda > 1$ which would have led to the inequality
\begin{equation}
  \xtm \geq \max \left( \frac{(1-\bx{1})e^{-y_j}}{\omega}, \, \frac{(1-\bx{2})e^{y_j}}{\omega} \right) \quad \forall j \in \{3, 4, \ldots, N-1\}\, .
  \label{eqineq2}
\end{equation}
These two definitions of $\xtm$ (eqs.~(\ref{eqdefxtm0}) or (\ref{eqdefxtm1})) will clearly reduce the number of cases to deal with.

\subsubsection{Pure ISR: both $i$ and $j$ belong to $S_i$}

\noindent
\textbf{Construction of the subtraction terms}\\
Let us remind the quantity $\cTiin{1}$
\begin{align}
  \cTiin{1} &\equiv \int_0^{\xtm}  d \xt{N} \, \xt{N}^{-1-2 \, \varepsilon} \, \int_0^{\pi} d \phi_N \, (\sin \phi_N)^{-2 \, \varepsilon} \, \int_{\yNmin}^{\yNmax} d y_N \,f_{12}(y_N,\xt{N},\phi_N) \, E_{12}^{\prime}.
  \label{eqdefT1120}
\end{align}
The squared eikonal factor appearing in eq.~(\ref{eqdefT1120}) is particularly simple since, indeed, $E_{12}^{\prime}=2$.
The $y_N$ integration range is split in two parts
\begin{align}
  \int_{\yNmin}^{\yNmax} d y_N  &= \int_{y_0}^{\yNmax} d y_N + \int_{\yNmin}^{y_0} d y_N,
  \label{eqsplitrangeyN11}
\end{align}
where $y_0$ is arbitrary (and can be chosen as $0$ or any of the $y_i$). The change of variable $\Delta y = y_N - y_0$ (respectively $\Delta y = y_0-y_N$) in the first (respectively the second) integral of the right hand side of eq.~(\ref{eqsplitrangeyN11}) is performed leading to
\begin{align}
  \cTiin{1} &= 2 \, \int_0^{\xtm}  d \xt{N} \, \xt{N}^{-1-2 \, \varepsilon} \, \int_0^{\pi} d \phi_N \, (\sin \phi_N)^{-2 \, \varepsilon} \, \notag \\
  &\quad {} \times \left[  \int_{0}^{\dyM} d \Delta y \, f_{12}(y_0+\Delta y,\xt{N},\phi_N)  +  \int^{\dym}_{0} d \Delta y \, f_{12}(y_0-\Delta y,\xt{N},\phi_N) \right].
  \label{eqdefT1121}
\end{align}
The two terms in the squared brackets in eq.~(\ref{eqdefT1121}) have the same form as the first term of eq.~(\ref{eqdefT2ij2!}), thus the way to construct the subtraction terms will proceed in the same way. Since,
in this case, there are only divergences when $p_N // p_1$, $p_N // p_2$ or $p_N = 0$, the subtracted term can be built as
\begin{align}
    \cTiinci{1} &= 2 \, \sum_{l=1}^{2} \, \int_0^{\pi} d \phi_N \, (\sin \phi_N)^{-2 \, \varepsilon} \,  \int_{0}^{1/\beta_{l0}} dz \, z^{-1-2\varepsilon} \, \left( f_{12}^{(l) \, c}(z) - f_{12}^{(l) \, c}(0) \right) \notag \\
    &\quad {} \times \left[  \Theta\left( \frac{1}{\beta_{l0}} - \xtm\right) \,   \int_{0}^{\beta_{l0} \, \xtm} dt \, t^{-1 - 2 \, \varepsilon}  +   \Theta\left( \xtm - \frac{1}{\beta_{l0}} \right) \,   \int_{0}^{1} dt \, t^{-1 - 2 \, \varepsilon} \,  \right],
    \label{eqdefcTiinci1}
\end{align}

\begin{align}
    \cTiins{1} &= 2 \, f_{12}(0,0,0)\, \sum_{l=1}^{2} \, \int_0^{\pi} d \phi_N \, (\sin \phi_N)^{-2 \, \varepsilon} \left\{   \Theta\left( \frac{1}{\beta_{l0}} - \xtm\right) \, \right. \notag \\
    &\qquad \qquad {} \times \left[ \int_{0}^{\beta_{l0} \, \xtm} dt \, t^{-1 - 2 \, \varepsilon} \, \int_{0}^{1/\beta_{lj}} dz \, z^{-1-2\varepsilon} + \int_{\beta_{l0} \, \xtm}^1 \frac{dt}{t} \, \int_{0}^{\xtm} dz \, z^{-1-2\varepsilon} \right] \notag \\
    &\qquad \qquad \qquad  {} + \left.  \Theta\left( \xtm - \frac{1}{\beta_{l0}} \right) \, \int_{0}^{1} dt \, t^{-1 - 2 \, \varepsilon} \, \int_{0}^{1/\beta_{l0}} dz \, z^{-1-2\varepsilon} \right\},
    \label{eqdefcTiins1}
\end{align}

\noindent
where the function $f_{12}^{(1) \, c}$, respectively $f_{12}^{(2) \, c}$, corresponds to $f_{12}(+\infty,0,\phi_N)$, respectively
$f_{12}(-\infty,0,\phi_N)$, that is to say to the function $f_{12}(y_N,\xt{N},\phi_N)$ in which all the scalar products containing the four-momentum $p_N$ are evaluated using $p_N = z \, e^{y_0} \, Q/2 \, (1, \vec{0}, 1)$, resp. $p_N = z \, e^{-y_0} \, Q/2 \, (1, \vec{0}, -1)$.
In addition, $\beta_{20}$ is defined as $\beta_{20} = \omega \, \exp(-y_0)/(1 - \bx{2})$.\\

\noindent
\textbf{Analytical integration of the subtraction terms}\\
The integration over the variables $z$, $\phi_N$ and $t$ can be easily performed.
But again, we want to express the divergent part in terms of the "plus" distributions. We thus introduce the changes of variable $z=(\bx{1}/z_1 - \bx{1}) \, \exp(-y_0)/\omega$ in the first term of the sum in eq.~(\ref{eqdefcTiinci1}) and  $z=(\bx{2}/z_2 - \bx{2}) \, \exp(y_0)/\omega$ in the second one. Introducing $\cTiindiv{1} = \cTiins{1} + \cTiinci{1}$ leads to
\begin{align}
  \cTiindiv{1} &= 2^{-2 \, \varepsilon} \, \frac{\Gamma^2 \left( \frac{1}{2}-\varepsilon \right)}{\Gamma(1 - 2 \, \varepsilon)} \, \left\{  -\frac{\chi(\bx{1},y_0)^{- 2 \, \varepsilon}}{\varepsilon} \, \int_{\bx{1}}^{1} \frac{dz_1}{z_1} \, z_1^{2 \, \varepsilon} \, \frac{f_{12}^{(1) \, c} \left( \frac{\bx{1} \, e^{-y_0}}{\omega} \, \frac{1-z_1}{z_1} \right)}{(1-z_1)_{+}} \right. \notag \\
  &\qquad \qquad \qquad \qquad \quad {}+ 2 \,  \int_{\bx{1}}^{1} \frac{d z_1}{z_1} \, \left(\frac{\ln(1-z_1)}{1-z_1}\right)_{+} \, f_{12}^{(1) \, c} \left( \frac{\bx{1} \, e^{-y_0}}{\omega} \, \frac{1-z_1}{z_1} \right) \notag \\
  &\qquad \qquad \qquad \qquad \quad {} -\frac{\chi(\bx{2},-y_0)^{- 2 \, \varepsilon}}{\varepsilon} \, \int_{\bx{2}}^{1} \frac{dz_2}{z_2} \, z_2^{2 \, \varepsilon} \, \frac{f_{12}^{(2) \, c} \left( \frac{\bx{2} \, e^{y_0}}{\omega} \, \frac{1-z_2}{z_2} \right)}{(1-z_2)_{+}} \,  \label{eqdefI12d10} \\
  &\qquad \qquad \qquad \qquad \quad {}+ 2 \,  \int_{\bx{2}}^{1} \frac{d z_2}{z_2} \, \left(\frac{\ln(1-z_2)}{1-z_2}\right)_{+} \, f_{12}^{(2) \, c} \left( \frac{\bx{2} \, e^{y_0}}{\omega} \, \frac{1-z_2}{z_2} \right) \notag  \\
  &\qquad \qquad \qquad \qquad \quad {} - f_{12}(0,0,0) \, \frac{1}{\varepsilon} \, \left[  - \frac{1}{\varepsilon} + \ln\left( \frac{\bx{1} \, \bx{2}}{\omega^2} \right)  \right. \notag \\
  &\qquad \qquad \qquad \qquad \qquad \qquad \qquad \qquad \quad {} -  \left. \left. \varepsilon \, \left( \Upsilon(\bx{1},y_0) +  \Upsilon(\bx{2},-y_0) \right) \vphantom{\frac{1}{\varepsilon}}\right] \vphantom{\frac{f_{1j}^c \left( \frac{\bx{1} \, e^{-y_j}}{\omega} \, \frac{1-z_1}{z_1} \right)}{(1-z_1)_{+}}}  \right\}.
 \notag
\end{align}

\subsection{Outside the cylinder}

In this case, $\xt{N}$ cannot reach zero ($\xt{N} \geq \xtm$) such that only collinear divergences remain when the parton $i_N$ is collinear to the final state parton $i_j$. The subtractions are performed inside the cones of size $\rth$ in rapidity -- azimuthal angle drawn around the direction of each hard parton of the final state.

\subsubsection{Pure FSR: both $i$ and $j$ belong to $S_f$}

\textbf{Construction of the subtraction terms}\\
The starting point is the following formula
\begin{align}
  \cTiout{4} &= \int_0^{\pi} d \phi_N \, (\sin \phi_N)^{-2 \, \varepsilon} \, \int_{\xtm}^{\xtmax{N}}  d \xt{N} \, \xt{N}^{-1-2 \, \varepsilon} \, \int_{\yNmin}^{\yNmax} d y_N \, \notag \\
  &\quad {} \times \sum_{i=3}^{N-1} \frac{1}{\hp_i \cdot \hp_N} \, \calf_{i}(y_N,\xt{N},\phi_N) \, \frac{p_i \cdot p_j}{\pt{i} \, (p_i+p_j) \cdot \hp_N}.
  \label{eqdefT4ijc0}
\end{align}
where the function $\calf_{i}$ is defined in eq.~(\ref{eqdefFcal0}).
Inside each cone $\Gamma_i$ around the 3-vector $\vec{p}_i$, the subtraction part is given by
\begin{align}
  \cTioutc{4}{i} &= \int_{\xtm}^{\xtmaxp{N}}  d \xt{N} \, \xt{N}^{-1-2 \, \varepsilon} \, \calf_{i}(y_i,\xt{N},0) \, \iint_{\Gamma_i} d \phi_N \, d y_N \, (\phi_N)^{-2 \, \varepsilon} \, \frac{2}{(y_i-y_N)^2 + \phi_{N}^2}\, ,
  \label{eqdefT4ijc1}
\end{align}
where $\xtmaxp{N} = \xtmax{N}|_{\phi_N=0}$\footnote{Note that the exact value of $\xtmax{N}$ (or $\xtmaxp{N}$) depends on the kinematics of the outgoing hadrons and thus is process dependent. It will not be given explicitly.}.
The constraint  $d_{iN} = \sqrt{(y_i-y_N)^2 + \phi_N^2} \leq \rth$ complicates the analytical computation of a collinear integral of the type $J^{\text{coll}}$. This is for this reason that, in eq.~(\ref{eqdefT4ijc1}), the denominator $\cosh(y_i-y_N) - \cos \phi_N$ appearing in the integrand,  as well as the measure $(\sin \phi_N)^{-2 \, \varepsilon}$ are replaced by the first non-vanishing term of their Taylor expansion when $y_N \rightarrow y_i$ and $\phi_N \rightarrow 0$.
\\ 

\noindent
\textbf{Analytical integration of the subtraction terms}\\
The details for the integration over $\phi_{N}$ and $y_N$ are given in appendix \ref{intir}.
Since there is no soft divergence, we set $\cTioutdiv{k} = \sum_{i=3}^{N-1} \cTioutc{k}{i}$ in order to keep the same notation as in the "inside the cylinder" case.
Note that, as explained in appendix \ref{intir}, the size of the cone $\rth$ is not a fixed value but can be squeezed by the kinematics. Using the result of this appendix, the divergent part coming from the integration of the subtracted term is
\begin{align}
  \cTioutdiv{4} &= 2^{-2 \, \varepsilon} \, \frac{\Gamma^2\left( \frac{1}{2} - \varepsilon \right)}{\Gamma(1 - 2 \, \varepsilon)} \; \frac{\rth^{-2 \, \varepsilon}}{-\varepsilon} \, \sum_{i=3}^{N-1} \int_{\xtm}^{\xtmaxp{N}}  d \xt{N} \, \xt{N}^{-1-2 \, \varepsilon}  \, \calf_{i}(y_i,\xt{N},0).
  \label{eqdefT4ijcdG0}
\end{align}
Then, making the change of variable $\xt{N} = (1-z_i)/z_i \, \xt{i}$, eq.~(\ref{eqdefT4ijcdG0}) becomes
\begin{align}
\cTioutdiv{4} &= 2^{-2 \, \varepsilon} \, \frac{\Gamma^2\left( \frac{1}{2} - \varepsilon \right)}{\Gamma(1 - 2 \, \varepsilon)} \, \frac{\rth^{-2 \, \varepsilon}}{-\varepsilon}\, \sum_{i=3}^{N-1} \, \xt{i}^{- 2 \, \varepsilon} \,  \int_{\zmin{i}}^{\zm{i}} \frac{d z_i}{z_i} \, z_i^{2 \, \varepsilon} \, \left[ \frac{1}{1-z_i} - 2 \, \varepsilon \, \frac{\ln(1-z_i)}{1-z_i} \right] \,  \notag \\
&\quad {} \times   \, \calf_{i}\left(y_i,\frac{1-z_i}{z_i} \, \xt{i},0\right),
\label{eqdefT4ijcdG1}
\end{align}
with $\zmin{i}=\xt{i}/(\xt{i}+\xtmaxp{N})$.
The determination of the lower bound on the $z_i$ integration follows from the fact that $x_1$ and $x_2$ must be less or equal to one.
While the upper bound of the $z_i$ integration comes from the fact that $\xt{N} \geq \xtm$, that is to say $\zm{i} = \xt{i}/(\xt{i}+\xtm)$.
Since the integration variable $z_i$ runs between $\zmin{i}$ and $\zm{i}$ which  never reaches 1, eq.~(\ref{eqdefT4ijcdG1}) can be written as
\begin{align}
\cTioutdiv{4} &= -2^{-2 \, \varepsilon} \, \frac{\Gamma^2\left( \frac{1}{2} - \varepsilon \right)}{\Gamma(1 - 2 \, \varepsilon)} \, \frac{\rth^{-2 \, \varepsilon}}{\varepsilon}\, \sum_{i=3}^{N-1} \, \xt{i}^{- 2 \, \varepsilon} \, \int_{\zmin{i}}^{\zm{i}} \frac{d z_i}{z_i} \, z_i^{2 \, \varepsilon} \,   \notag \\
&\quad {} \times  \left[ \frac{1}{(1-z_i)_{+}} - 2 \, \varepsilon \, \left(\frac{\ln(1-z_i)}{1-z_i}\right)_{+} \right] \,\calf_{i}\left(y_i,\frac{1-z_i}{z_i} \, \xt{i},0\right).
\label{eqdefI3413}
\end{align}

\subsubsection{Mixed terms ISR and FSR: $i$ belongs to $S_i$ and $j$ belongs to $S_f$}

\textbf{Construction of the subtraction terms}\\
In this case also, we treat in detail the case where $i=1$ and we define
\begin{align}
  \cTiout{2} &\equiv \int_0^{\pi} d \phi_N \, (\sin \phi_N)^{-2 \, \varepsilon} \, \int_{\xtm}^{\xmax{N}}  d \xt{N} \, \xt{N}^{-1-2 \, \varepsilon} \, \int_{\yNmin}^{\yNmax} d y_N \, \sum_{j=3}^{N-1} f_{1j}(y_N,\xt{N},\phi_N) \, E_{1j}^{\prime}.
  \label{eqdefT2ijout0}
\end{align}
Since $\xt{N}$ cannot reach zero, only subtraction terms for final state collinearity are necessary. Thus the collinear subtraction term will have the same structure as in the "pure FSR" case, the only difference will be the coefficient in front. The required subtracted term is
\begin{align}
  \cTioutc{2}{i} &= \int_{\xtm}^{\xtmaxp{N}}  d \xt{N} \, \xt{N}^{-1-2 \, \varepsilon} \, \sum_{j=3}^{N-1} f_{1j}(y_j,\xt{N},0) \, \iint_{\Gamma_j} d \phi_N \, d y_N \, (\phi_N)^{-2 \, \varepsilon} \, \frac{2}{(y_j-y_N)^2 + \phi_{N}^2}.
  \label{eqdefT2ijout1}
\end{align}

\noindent
\textbf{Analytical integration of the subtraction terms}\\
From the eq.~(\ref{eqdefI3413}), we immediately get that
\begin{align}
  \cTioutdiv{2} &= -2^{-2 \, \varepsilon} \, \frac{\Gamma^2\left( \frac{1}{2} - \varepsilon \right)}{\Gamma(1 - 2 \, \varepsilon)} \, \frac{\rth^{-2 \, \varepsilon}}{\varepsilon}\,\sum_{j=3}^{N-1} \xt{j}^{- 2 \, \varepsilon} \, \int_{\zmin{j}}^{\zm{j}} \frac{d z_j}{z_j} \, z_j^{2 \, \varepsilon} \, \notag \\
  &\quad {} \times \left[ \frac{1}{(1-z_j)_{+}} - 2 \, \varepsilon \, \left(\frac{\ln(1-z_j)}{1-z_j}\right)_{+} \right] \,   f_{1j}\left(y_j,\frac{1-z_j}{z_j} \, \xt{j},0\right).
  \label{eqdefT2ijout2}
\end{align}

\subsubsection{Pure ISR: both $i$ and $j$ belong to $S_i$}

Let us define the quantity $\cTiout{1}$
\begin{align}
  \cTiout{1} &\equiv \int_0^{\pi} d \phi_N \, (\sin \phi_N)^{-2 \, \varepsilon} \, \int_{\xtm}^{\xmax{N}}  d \xt{N} \, \xt{N}^{-1-2 \, \varepsilon} \, \int_{\yNmin}^{\yNmax} d y_N \, f_{12}(y_N,\xt{N},\phi_N) \, E_{12}^{\prime}.
  \label{eqdefT112out0}
\end{align}
The integration variable $\xt{N}$ cannot vanish thus the right hand side of eq.~(\ref{eqdefT112out0}) does not diverge and there is nothing to subtract.

\section{The divergent terms}\label{divterm}

After having collected all the divergent terms from the analytical phase space integration on $p_N$ of the different counter terms, we have to show that the ones of collinear origin are  absorbed into the redefinitions of the PDFs and the FFs and the ones of soft origin cancel against the divergences coming from the virtual contribution. This implies that the coefficients $H_{ij}$ have to verify some conditions in the collinear and the soft limits. Let us present in this section these equations as well as the finite pieces associated to the divergent terms once the poles in $\varepsilon$ have been cancelled. We will give only the results and relegate all the details to appendix \ref{gutsdivterm}.\\

\noindent
Let us discuss first the part associated to the initial state collinear divergences. 
The comparison of the structure of collinear divergences coming from the initial state, derived in sec. \ref{secLO} eq.~(\ref{eqdefhadcrxg4}), and of the results obtained after  integration over the phase space of the soft/collinear parton of the subtracted terms leads to the following conditions required to absorb the collinear divergences into a redefinition of the partonic density functions:
\begin{align}
  z_1 \, \left[ \Hn{12}((1-z_1) \, p_1) + \sum_{l=3}^{N-1} \Hn{1l}((1-z_1) \, p_1) \right]  &= a^{(n)}_{j_1 i_1}(z_1) \, \frac{C_{i_1}}{C_{j_1}} \, |M^n_{\react{i}{i_1}{j_1}{N-1}}|^2, \label{eqrelini1} \\
  z_2 \, \left[ \Hn{12}((1-z_2) \, p_2) +  \sum_{l=3}^{N-1} \Hn{2l}((1-z_2) \, p_2) \right] &= a^{(n)}_{j_2 i_2}(z_2) \, \frac{C_{i_2}}{C_{j_2}} \, |M^n_{\react{i}{i_2}{j_2}{N-1}}|^2. \label{eqrelini2}
\end{align}
The functions $a^{(n)}_{ij}(z)$ are the coefficient of the distribution $1/(1-z)_{+}$ in the one-loop DGLAP kernels in $n$ dimensions, cf. appendix~\ref{APkernels}.
The finite terms associated to the initial state collinear divergences are given by\footnote{Keeping in mind that only the case where $\xtm$ fulfills the condition (\ref{eqineq1}) is shown.}
\begin{align}
  \hspace{2em}&\hspace{-2em} \sigma^{\text{ini. coll.}}_{H} = \sum_{\seq{i}{N-1} \in S_p}  K^{(4) \, B}_{i_1 i_2} \, \frac{\alpha_s}{2 \, \pi} \, \int d \text{PS}_{N-1 \,\text{h}}^{(n)}(\bar{x}) \,  \delta^{2}\left( \sum_{l=3}^{N-1} \frac{\vkt{l}}{\bx{l}}\right)   \nonumber\\
  &\quad {} \times   \, \left\{ \sum_{j_1\in S_p}\int_{\bx{1}}^{1} \frac{dz_1}{z_1^2} \, A_{(i)_{N-1}}\left(\seqt{\bar{x}}{\bx{1}}{\frac{\bx{1}}{z_1}}{N-1}\right) \frac{C_{i_1}}{C_{j_1}}  \right. 
  \left[ \frac{a^{(n-4)}_{j_1 i_1}(z_1)}{(1-z_1)_{+}} + \frac{a^{(4)}_{j_1 i_1}(z_1)}{(1-z_1)_{+}} \ln \left(\Lambda_1\right) \right. \notag \\
  &\qquad \qquad \qquad \qquad \qquad {} +\left. 2 \, a^{(4)}_{j_1 i_1}(z_1) \, \left( \frac{\ln(1-z_1)}{1-z_1} \right)_{+} \right] \, |M^4_{\react{i}{i_1}{j_1}{N-1}}|^2 \left. \vphantom{\sum_{j_1\in S_p}\int_{\bx{1}}^{1} \frac{dz_1}{z_1^2} \, A_{(i)_{N-1}}\left(\seqt{\bar{x}}{\bx{1}}{\frac{\bx{1}}{z_1}}{N-1}\right) \frac{C_{i_1}}{C_{j_1}}  
  \left[ \frac{a^{(n-4)}_{j_1 i_1}(z_1)}{(1-z_1)_{+}} + \frac{a^{(4)}_{j_1 i_1}(z_1)}{(1-z_1)_{+}} \ln \left(\Lambda_1\right)\right]} + 1 \longleftrightarrow 2 \right\}\, ,
\label{eqdefhadcrxinicoll3} 
\end{align}
with $\Lambda_i=\frac{Q^2 \, \xtm^2 \, \bx{i}^2}{z_i^2 \, M^2 \, (1-\bx{i})^2}$ for $i=1, 2$.
As expected, the structure of eq.~(\ref{eqdefhadcrxinicoll3}) is the convolution of a PDF times a product of a one-loop DGLAP kernel and a partonic amplitude squared. It gives the dependence of the NLO partonic cross section on the factorisation scale for initial state.
\newline

\noindent
The collinear divergences originating from the final state have to be absorbed into a redefinition of the fragmentation functions. To fulfil this requirement, the collinear limit of the coefficient $H_{ij}$ must obey to
\begin{align}
  z_k \,\Xi_k\left((1-z_k) \, \frac{\kt{k}}{\bx{k}}\right) = a^{(n)}_{i_k j_k}(z_k) \, |M^n_{\react{i}{i_k}{j_k}{N-1}}|^2 &\qquad \text{for each $k$ in $S_f$}, \label{eqrelfink}
\end{align}
with
\begin{align}
  \Xi_k\left(p_N\right) &= \sum_{j=k+1}^{N-1} \Hn{kj}\left(p_N\right) + \sum_{j=1}^{k-1} \Hn{jk}\left(p_N\right).
  \label{eqdefXi0}
\end{align}

\noindent
The finite parts associated to the final state collinear divergences are given by
\begin{align}
  \hspace{2em}&\hspace{-2em} \sigma^{\text{fin. coll.}}_{H \, k} = \sum_{\seq{i}{N-1} \in S_p}  K^{(4) \, B}_{i_1 i_2} \, \frac{\alpha_s}{2 \, \pi} \, \int d \text{PS}_{N-1 \,\text{h}}^{(4)}(\bar{x}) \,  \delta^{2}\left( \sum_{l=3}^{N-1} \frac{\vkt{l}}{\bx{l}}\right) \nonumber\\
  &\quad {} \times  \left\{ \int_{\bx{k}}^{1} \frac{dz_k}{z_k} A_{(i)_{N-1}}\left(\seqt{\bar{x}}{\bx{k}}{\frac{\bx{k}}{z_k}}{N-1}\right)  \right. \left[ \frac{a^{(n-4)}_{i_k j_k}(z_k)}{(1-z_k)_{+}}+ \frac{a^{(4)}_{i_k j_k}(z_k)}{(1-z_k)_{+}} \ln \left( \frac{z_k^2 \, \Xt{k}^2 \, Q^2}{\bx{k}^2 \, M_f^2} \right) \right. \notag \\
  &\qquad \qquad \qquad \qquad \qquad {} + \left. 2 \, a^{(4)}_{i_k j_k}(z_k) \, \left( \frac{\ln(1-z_k)}{1-z_k} \right)_{+}\right] \, |M^{(4)}_{\react{i}{i_k}{j_k}{N-1}}|^{2} \notag \\
  &\quad {} + \left. \ln \left( \rth^2 \right) \, \int_{\bx{k}}^{\czm{k}} \frac{dz_k}{z_k} \, A_{(i)_{N-1}}\left(\seqt{\bar{x}}{\bx{k}}{\frac{\bx{k}}{z_k}}{N-1}\right)  \,   \, \frac{a^{(4)}_{i_k j_k}(z_k)}{(1-z_k)_{+}} \, |M^{(4)}_{\react{i}{i_k}{j_k}{N-1}}|^{2}  \right\}.
  \label{eqdefhadcrxfincoll3} 
\end{align}
Also in this case, the structure of the terms in the curly brackets of eq.~(\ref{eqdefhadcrxfincoll3}) is a convolution of a FF times the product of a one-loop DGLAP kernel and a partonic amplitude squared. Note that the first term, depending on the factorisation scale, receives contributions from inside and outside the cylinder while the last term, depending on the cone size, receives contributions only from outside the cylinder. 
This is why the upper bound of the $z_k$ integration, 
given by\footnote{The bounds on the $z_k$ integration are different from the ones given in eq.~(\ref{eqdefT4ijcdG1}) because a change of variables has been performed to recover the structure of eq.~(\ref{eqdefhadcrxg4}), cf. appendix \ref{gutsdivterm}.} 
\begin{align}
  \czm{k} &= \frac{\Xt{k}-\bx{k} \, \xtm}{\Xt{k}}\, ,
  \label{eqdefczm}
\end{align}
does not reach 1.

\noindent
The cancellation of the soft divergences between the real emission and the virtual one will also yield some conditions that the coefficients $H_{ij}$ have to satisfy in the soft limit. Before giving them, let us recap the structure of the virtual contribution:
\begin{align}
  \sigma_H^{\text{virt}} &\equiv \sum_{\seq{i}{N-1} \in S_p} \, K^{(n) \, B}_{i_1 i_2} \, \left(\frac{4 \, \pi \, \mu^2}{Q^2}\right)^{\varepsilon} \, \frac{\alpha_s}{2 \, \pi} \, \frac{1}{\Gamma(1-\varepsilon)} \, \int d \text{PS}_{N-1 \,\text{h}}^{(n)}(\bar{x}) \,\notag \\
  &\quad {} \times A_{(i)_{N-1}}(\{\bar{x}\}_{N-1}) \, \left\{  \left[ \frac{\cala^{(n)}}{\varepsilon^2} + \frac{\calb^{(n)}}{\varepsilon} \right] \, |M^{(n)}_{[i]_{N-1}}|^{2} \right. \notag \\
  &\quad {} + \left. \frac{1}{\varepsilon} \, \left[ \sum_{i=1}^{N-2} \sum_{j=i+1}^{N-1} \calc^{(n)}_{ij} \, \ln \left( \frac{2 \, p_i \cdot p_j}{Q^2} \right)  \right] + \calf^{(n)}(Q^2) \vphantom{\frac{\cala}{\varepsilon^2}} \right\}.
  \label{eqvirtstruct}
\end{align}
The energy scale $Q$ appearing in eq.~(\ref{eqvirtstruct}) is the same as the one appearing in eq.~(\ref{eqdefEijg}). As in the case of the real emission, the virtual cross section is independent of this scale by construction.
The function $\calf^{(n)}(Q^2)$ is finite when $\varepsilon \rightarrow 0$. After having collected the divergences of soft origin, as well as the finite pieces associated, coming from the analytical integration over $p_N$ of the different subtraction terms and having compared them to the virtual term leads to the following relations valid in $n$ dimensions:
\begin{align}
  \cala^{(n)} \, |M^{(n)}_{[i]_{N-1}}|^{2} &= - \sum_{i=1}^{N-2} \sum_{j=i+1}^{N-1} \Hn{ij}(0), &
  \calb^{(n)} &= - \sum_{k=1}^{N-1} b_{i_k i_k}, &
  \calc^{(n)}_{ij} &= \Hn{ij}(0),
  \label{eqrels3}
\end{align}
where the $b_{ij}$ are the coefficients of the $\delta(1-z)$ of the one-loop DGLAP kernels (cf. appendix~\ref{APkernels}).
The finite part associated to the soft divergences is given by

\begin{align}
  \hspace{2em}&\hspace{-2em} \sigma_H^{\text{soft}} = \sum_{\seq{i}{N-1}\in S_p} \, K^{(4) \, B}_{i_1 i_2} \, \frac{\alpha_s}{2 \, \pi} \, \int d \text{PS}_{N-1 \,\text{h}}^{(n)}(\bar{x}) \, \delta^{2}\left( \sum_{l=3}^{N-1} \frac{\vkt{l}}{\bx{l}}\right)\, A_{(i)_{N-1}}(\{\bar{x}\}_{N-1}) \notag \\
  &\times   \, \left\{ \left[ \sum_{k=3}^{N-1} \, \ln^2\left( \frac{\xt{k}}{\xtm} \right) \,  a^{(4)}_{i_k i_k}(1)  -  \ln^2(\xtm) \, \sum_{k=1}^{N-1} a^{(4)}_{i_k i_k}(1) \right. \right.\notag \\
  &+ \left. \ln^2\left( \frac{\bx{1}}{1 - \bx{1}} \, \right) \, a^{(4)}_{i_1 i_1}(1)  +  \ln^2\left( \frac{\bx{2}}{1 - \bx{2}} \, \right) \, a^{(4)}_{i_2 i_2}(1)-\sum_{k=1}^{2} b_{i_k i_k} \, \ln\left( \frac{M^2}{Q^2} \right)\right.\nonumber \\
  &-\left.\sum_{k=3}^{N-1} b_{i_k i_k} \,  \ln\left( \frac{M_f^2}{Q^2} \right)\right]|M^{(4)}_{[i]_{N-1}}|^{2} + 2 \, \ln(\xtm) \, \sum_{i=1}^{N-2} \sum_{j=i+1}^{N-1} \Hq{ij}(0) \, \ln \left( \frac{2 \, p_i \cdot p_j}{Q^2} \right) \notag \\
  & + \left. \sum_{i=3}^{N-2} \sum_{j=i+1}^{N-1} \Hq{ij}(0) \, \left[ \frac{1}{2} \, \ln^2(2 \, \bd_{ij}) - 2 \, \left( \ystij \right)^2 \right] + \calf^{(4)}(Q^2)
  \right\}. 
  \label{eqdefhadcrxsoft3}
\end{align}

\noindent
Some terms are proportional to the coefficient in front of the plus distribution of the diagonal one-loop DGLAP kernel taken at $z=1$ times a partonic amplitude squared and others are not. This is related to the well-known fact that in QCD, since the gluons carry colour charges,
the amplitude of real emission in the soft limit is proportional to the colour connected Born amplitudes. Squaring the latter does not 
always lead to the Born amplitude squared. However, this will not prevent us from having cancellation/absorption of divergences. 
A non trivial example is given in appendix \ref{qqbgg}.
Note also that eqs.~(\ref{eqdefhadcrxinicoll3}) and (\ref{eqdefhadcrxsoft3}) mirror the dependence on $\xtm$ of the subtraction terms. 
They vanish logarithmically as $\xtm \rightarrow 0$ as expected.

\section{Cases with non fragmenting partons \label{one_frag}}

We have to treat the case where one or several partons, say $i_{l_1}, \, i_{l_2}, \,  \ldots $ do not fragment, this is typically the case if these partons are photons or they initiate jets. Let us discuss these two cases in more detail. 
For simplicity, we will discuss the case where only one parton does not fragment. It is easy to extend the results obtained in this section to the case where several partons do not fragment. The non-fragmenting parton will be denoted $i_{N-1}$.

\subsection{Parton $i_{N-1}$ is a photon}

It is well known that a high-$p_T$ photon can be produced by two mechanisms: either it comes directly from the partonic sub-process or it is emitted collinearly by a parton produced at large transverse momentum.
The latter case is described by a fragmentation function of the parton into a photon and thus the results of the preceding section can be used. Note that since the photon is observed, its four-momentum cannot be soft nor collinear to the beams, the photon plays the same role as any other hard parton.\\ 

\noindent
In this subsection, we will see that the direct production can also be described by the general formula given in sec.~\ref{secLO} at the cost of introducing a technical fragmentation function of a photon parton into a photon. At lowest order 
in the electromagnetic coupling at which we are working, 
this fragmentation function is merely a Dirac distribution which should be integrated for practical implementation. Nevertheless, for the uniformity of the presentation, it is interesting to keep this constraint unsatisfied.\\

\noindent
Let us first discuss the fragmentation of a parton (including a photon parton) into a photon. As in the hadronic case, the renormalised fragmentation function is written in terms of the bare one\footnote{Since we are interested only by the direct contribution, we consider only the inhomogeneous term in the evolution equation, cf. ref.~\cite{Bourhis:1997yu} for example.} 
\begin{align}
  D_k^{\gamma}(x,M_f^2) &= \bD_k^{\gamma} \left( x \right) + \frac{\alpha}{2 \, \pi} \,  \sum_{l \in S^{\prime}_p} \, \left[ \calh_{lk}\left(*,\frac{\mu^2}{M_f^2}\right) \otimes \bD_l^{\gamma}  \right]_{1}(x), \label{eqrelevolutionffgamma} 
\end{align}
with $S^{\prime}_p = S_p \cup \{\gamma\}$. Since we consider only point-like interactions, the bare fragmentation $\bD_l^{\gamma}(z)$ is given by
\begin{align}
  \bD_l^{\gamma}(z) &= \delta(1-z) \, \delta_{\gamma l}.
\end{align}
Injecting this result into eq.~(\ref{eqrelevolutionffgamma}) gives
\begin{align}
  D_k^{\gamma}(x,M_f^2) &= \delta(1-x) \, \delta_{k \gamma} + \frac{\alpha}{2 \, \pi} \, \calh_{\gamma k}\left(x,\frac{\mu^2}{M_f^2}\right).  \label{eqrelevolutionffgamma1}
\end{align}
Note that, at NLO QCD approximation and lowest order in QED (neglecting QED radiative corrections), we have 
\begin{align}
    \calh_{\gamma \gamma}\left( x , \frac{\mu^2}{M_f^2} \right) &= \calh_{\gamma g}\left(x,\frac{\mu^2}{M_f^2}\right) = 0,
    \label{eqdefcalhgammagamma}
\end{align}
and 
\begin{align}
    \calh_{\gamma q}\left(x,\frac{\mu^2}{M_f^2}\right) = Q_q^2 \, \frac{(1 + (1-x)^2)}{x} \, \left( \frac{4 \, \pi \, \mu^2}{M_f^2} \right)^{\varepsilon},
    \label{eqdefcalhgammaq}
\end{align}
where $Q_q$ is the fractional electric charge of the quark $q$, i.e., $Q_u =2/3$ for an up-type quark and $Q_d = -1/3$ for a down-type quark.
\\

\noindent
The LO approximation for the reaction $H_1 + H_2 \rightarrow H_3 + \ldots + H_{N-2} + \gamma$ can thus be described by eq.~(\ref{eqdefhadcrxg0}) using only the first term of the right hand side of eq.~(\ref{eqrelevolutionffgamma1}) for the fragmentation function of a parton into a photon.
We then get
\begin{align}
  \sigma_H^{\text{LO}} &= \sum_{\seq{i}{N-2} \in S_p} \, \mK^{(n) \, B}_{i_1 i_2} \, \int \PSH{\bar{x}} \, A_{(i)_{N-1}}(\{\bar{x}\}_{N-1}) \, \delta^{n-2}\left( \sum_{l=3}^{N-1} \frac{\vkt{l}}{\bx{l}} \right) |M^n_{[i]_{N-1}}|^2,
  \label{eqdefhadcrxgo3}
\end{align}
with $\bx{1}$ and $\bx{2}$ given by eq.~(\ref{eqdefx2g}). Note that the first sum concerns only the partons $i_1, \ldots, i_{N-2}$, because at this level $i_{N-1} \equiv \gamma$.
The only difference, compared to eq.~(\ref{eqdefhadcrxg0}), is that a factor $g_s^2$ is transformed into a $e^2$ in the overall normalisation factor
\begin{align}
  \mK^{(n) \, B}_{i_1 i_2} &= \frac{1}{2^{N-2} \, s^2} \, \frac{1}{(2 \, \pi)^{(N-4) \, n - N + 3}} \, \frac{g_s^{2 \, (N-4)} \, e^2 \, \mu^{2 \, (N-3) \, \varepsilon}}{4 \, C_{i_1} \, C_{i_2}}.
\end{align}

\noindent
When an extra parton is emitted, the structure of the collinear emission contains terms similar to those appearing in the general case (with the constraint $\delta(1-\bx{N-1})$) plus a term describing the collinear emission of a photon by a parton.
The term of order $\alpha^0$ in eq.~(\ref{eqrelevolutionffgamma1}) gives eq.~(\ref{eqdefhadcrxgo3}) from eq.~(\ref{eqdefhadcrxg3}) while the term of order $\alpha$ is used to build the structure of the collinear divergences which is given by
\begin{align}
  \sigma_H^{{\text{LO}}} &= \sum_{\seq{i}{N-2} \in S_p} \, \mK^{(n) \, B}_{i_1 i_2} \, \int d \text{PS}_{N-1 \,\text{h}}^{(n)}(\bar{x}) \,  \delta^{n-2}\left( \sum_{l=3}^{N-1} \frac{\vkt{l}}{\bx{l}} \right)  \left\{ \vphantom{\sum_{k=3}^{N-1} \,\int^1_{x_k} \frac{d z_k}{z_k} \, \bA_{(i)_{N-1}}\left( \bx{1},\bx{2},\cdots,\frac{\bx{k}}{z_k},\cdots \right)} \bA_{(i)_{N-1}}(\{\bar{x}\}_{N-1}) \,  |M^n_{[i]_{N-1}}|^2 \right. \notag \\
  & +  \frac{\alpha_s}{2 \, \pi} \left(  \sum_{l=1}^{2} \sum_{j_l \in S_p} \, \frac{C_{i_l}}{C_{j_l}} \left[\calh_{i_l j_l}\left(*,\frac{\mu^2}{M^2}\right) \otimes \bA_{\indict{i}{i_l}{j_l}{N-1} }(\seqt{\bar{x}}{\bx{l}}{*}{N-1})\right]_{2}(\bx{l}) \,  |M^n_{\react{i}{i_l}{j_l}{N-1}}|^2   \right. \notag \\
  & +  \sum_{k=3}^{N-2} \, \sum_{j_k \in S_p} \left[\calh_{j_k i_k}\left(*,\frac{\mu^2}{M_f^2}\right) \otimes \bA_{\indict{i}{i_k}{j_k}{N-1}}(\seqt{\bar{x}}{\bx{k}}{*}{N-1})\right]_{1}(\bx{k}) \, |M^n_{\react{i}{i_k}{j_k}{N-1}}|^2 \notag \\
 &+\left. \left. \sum_{j_{N-1} \in S_p}\, \calh_{\gamma j_{N-1}}\left(\bx{N-1},\frac{\mu^2}{M_f^2}\right) \, 
   \bA_{(i)_{N-2}}\left(\{\bar{x}\}_{N-2} \right) \, |M^n_{\react{i}{i_{N-1}}{j_{N-1}}{N-1}}|^2 \right) \right\}.
  \label{eqdefhadcrxgo4}
\end{align}
Note that, in eq.(\ref{eqdefhadcrxgo4}), for compactness reasons, a normalisation factor $\mK^{(n) \, B}_{i_1 i_2}$ containing a factor $e^2$ has been factored out, thus the term in the last line is multiplied by $\alpha_s/(2 \, \pi)$ instead of $\alpha/(2 \, \pi)$ as suggested by eq.~(\ref{eqrelevolutionffgamma1}). In addition, in this term the extra constraint which reads $\delta(1-\bx{N-1}/z_{N-1})$ has been taken into account hence the missing integration over $z_{N-1}$.
\\

\noindent
At NLO approximation, the introduction of a technical fragmentation function of a photon parton into a photon (first term of eq.~(\ref{eqrelevolutionffgamma1})) enables the use of the formula (\ref{eqdefhadcrxg5}) to describe the cross section for the real emission up to a different overall normalisation factor, that is to say
\begin{align}
  \sigma_H^{\text{Real}} &= \sum_{\seq{i}{N-1} \in S_p} \, \mK^{(n)}_{i_1 i_2} \, \int \PSH{x} \, \int \PSN \, A_{(i)_{N-1}}(\{x\}_{N-1}) \,  \delta^{n-2}\left( \sum_{l=3}^{N-1} \frac{\vkt{l}}{x_l} +  \vpt{N}\right) \, \notag \\
  &\qquad \qquad \qquad {} \times \left[ \sum_{i=1}^{N-2} \sum_{j=i+1}^{N-1} \Hn{ij}(p_N) \, E^{\prime}_{ij} + \xt{N}^2 \; Q^2 \, G^{(n)}(p_N) \right],
  \label{eqdefhadcrxgo5}
\end{align}
where the quantities $x_1$ and $x_2$ are given by eqs.~(\ref{eqdefx1gp}) and (\ref{eqdefx2gp}), and the overall normalisation factor reads
\begin{align}
  \mK^{(n)}_{i_1 i_2} &= \frac{1}{2^{N-2} \, s^2} \, \frac{1}{(2 \, \pi)^{(N-3) \, n - N + 2}} \, \frac{g_s^{2 \, (N-3)} \, e^2 \, \mu^{2 \, (N-2) \, \varepsilon}}{4 \, C_{i_1} \, C_{i_2}} \,  Q^{-2 \, \varepsilon} \,V(n-2).
  \label{eqdefmKij}
\end{align}

\noindent
The strategy for the subtraction is exactly the same as in the case with $N-1$ fragmenting partons. The subtracted terms can be analytically integrated over the phase space of the parton $i_N$. The noticeable difference is a new term for final state divergences, describing the collinear splitting of the parton $j_{N-1}$ into a photon and the parton $i_N$, which reads
\begin{align}
  \hspace{2em}&\hspace{-2em} \sigma^{\text{fin. coll.}}_{H \, N-1} = \sum_{\seq{i}{N-1} \in S_p}  \mK^{(4) \, B}_{i_1 i_2} \, \frac{\alpha_s}{2 \, \pi} \, \int \PSHQP{N-2}{\bar{x}} \, \delta^{2}\left( \sum_{l=3}^{N-1} \frac{\vkt{l}}{\bx{l}} \right) \notag \\
&\qquad {} \times  \left\{  A_{(i)_{N-2}}\left(\left\{\bar{x}\right\}_{N-2}\right) \,   \left[  \frac{a^{(4)}_{\gamma j_{N-1}}(\bx{N-1})}{(1-\bx{N-1})_{+}} \, \ln \left( \frac{\Xt{N-1}^2 \, Q^2}{M_f^2} \right) \,  \right. \right.  \notag \\
  &\qquad  {}  + \left. \frac{a^{(n-4)}_{\gamma j_{N-1}}(\bx{N-1})}{(1-\bx{N-1})_{+}} + 2 \, a^{(4)}_{\gamma j_{N-1}}(\bx{N-1}) \, \left( \frac{\ln(1-\bx{N-1})}{1-\bx{N-1}} \right)_{+}\right] \, |M^{(4)}_{\react{i}{i_{N-1}}{j_{N-1}}{N-1}}|^{2} \notag \\
  &\qquad  {} + \left. \ln \left( \rth^2 \right) \, \Theta\left( \zm{N-1} - \bx{N-1}\right) \,  \frac{a^{(4)}_{\gamma j_{N-1}}(\bx{N-1})}{(1-\bx{N-1})_{+}} \, |M^{(4)}_{\react{i}{i_{N-1}}{j_{N-1}}{N-1}}|^{2}  \right\}.
  \label{eqdefhadcrxfincoll3g}
\end{align}
Note that this corresponds to the eq.~(\ref{eqdefhadcrxfincoll3}) with the constraint $\delta(1-\bx{N-1}/z_{N-1})$.
Furthermore, this constraint translates into a new upper bound over the $\bx{N-1}$ integration for the last term in curly brackets, with respect to the one appearing in eq.~(\ref{eqdefhadcrxfincoll3}),  given by $\zm{N-1} = \Xt{N-1}/(\Xt{N-1}+\xtm)$.\\

\noindent
The case where the photon is in the initial state can be obviously treated by this method. Nevertheless, it is more complicated to find a way to present the results without introducing numerous new formulae. Thus, in order to reduce the size of the article, we choose not to present this case here.

\subsection{Case of jets}

In this subsection we look at the case where some partons do not fragment and are combined to form jets. In order to lighten this article, we will treat the case where only the parton $i_{N-1}$ does not fragment. At LO accuracy, the formula is the same as for the photon case, the parton $i_{N-1}$ forms the jet and $p_{N-1} = p_{\text{jet}}$, but at NLO, what is fixed is the momentum of the jet which can be formed by either the parton $i_{N-1}$ or the parton $i_N$ or by both partons $i_{N-1}$ and $i_N$.
Thus, the parton $i_{N-1}$ can also be soft and/or collinear. The phase space is then sliced in two parts $\pt{N-1} \geq \pt{N}$ and $\pt{N-1} \leq \pt{N}$. Each part has a collinear divergence and the sum of the two vanishes due to the Kinoshita-Lee-Nauenberg (KLN) theorem. Let us sketch this cancellation. Starting from eq.~(\ref{eqdefhadcrxfincoll1})
by putting $\bD_{i_{N-1}}^{i_{N-1}}(x_{N-1}/z_{N-1}) = \delta(1-x_{N-1}/z_{N-1})$ and neglecting terms of order  ${\cal O}(\alpha_s^2)$, the 
non-cancelled collinear divergence carried by $\sigma^{\text{fin. coll.}}_{H \, N-1}$ reads\footnote{Hereafter, to keep the formulae as compact as possible, the variable $\bx{N-1}$ is named only $z$.}
\begin{align}
  \hspace{2em}&\hspace{-2em} \sigma^{\text{fin. coll.}}_{H \, N-1} = \sum_{\seq{i}{N-2} \in S_p}  K^{(n) \, B}_{i_1 i_2} \, \frac{\alpha_s}{2 \, \pi} \, \frac{1}{\Gamma(1-\varepsilon)} \, \int \PSHP{N-2}{\bar{x}} \, d y_{N-1} \, d^{n-2} \kt{N-1} \notag \\
  &\qquad {} \times  \int_{1/2}^{1} \frac{dz}{z^{n-2}} \, \delta^{n-2}\left( \sum_{l=3}^{N-2} \frac{\vkt{l}}{\bar{x}_l} + \frac{\vkt{N-1}}{z} \right) \,   A_{(i)_{N-2}}\left(\{\bar{x}\}_{N-2}\right) \,  \notag \\
  &\qquad {} \times \left\{ \left[  \calh_{i_{N-1} j_{N-1}}\left(z,\frac{\mu^2}{M_f^2}\right) + \frac{a^{(n-4)}_{i_{N-1} j_{N-1}}(z)}{(1-z)_{+}} + \frac{a^{(4)}(z)_{i_{N-1} j_{N-1}}}{(1-z)_{+}} \, \ln \left( \frac{\Xt{N-1}^2 \, Q^2}{M_f^2} \right)  \right. \right.  \notag \\
  &\qquad \qquad \qquad {}  +  \left. 2 \, a^{(4)}_{i_{N-1} j_{N-1}}(z) \, \left( \frac{\ln(1-z)}{1-z} \right)_{+}\right] \, |M^{(n)}_{\react{i}{i_{N-1}}{j_{N-1}}{N-1}}|^{2} \notag \\
  &\qquad \qquad \quad + \ln \left( \rth^2 \right) \, \Theta(\zm{N-1} - z)  \,  \left. \frac{a^{(4)}_{i_{N-1} j_{N-1}}(z)}{(1-z)_{+}} \, |M^{(n)}_{\react{i}{i_{N-1}}{j_{N-1}}{N-1}}|^{2}  \right\},
  \label{eqdefhadcrxfincoll3j}
\end{align}
with $\zm{N-1} = \Xt{N-1}/(\Xt{N-1}+\xtm)$.
The interesting quantity is the four momentum of the jet $K_{\text{jet}}$ which is, in the collinear approximation, $K_{\text{jet}} = K_{N-1}/z$. Thus, changing $K_{N-1}$ against $K_{\text{jet}}$ leads to
\begin{align}
  \hspace{2em}&\hspace{-2em} \sigma^{\text{fin. coll.}}_{H \, N-1} = \sum_{\seq{i}{N-2} \in S_p}  K^{(n) \, B}_{i_1 i_2} \, \frac{\alpha_s}{2 \, \pi} \, \frac{1}{\Gamma(1-\varepsilon)} \, \int \PSHP{N-2}{\bar{x}} \, d \yj \, d^{n-2} \ktj \notag \\
  &\qquad {} \times  \delta^{n-2}\left( \sum_{l=3}^{N-2} \frac{\vkt{l}}{\bar{x}_l} + \vktj \right) \, A_{(i)_{N-2}}\left(\{\bar{x}\}_{N-2}\right) \,  |M^{(n)}_{\react{i}{i_{N-1}}{j_{N-1}}{N-1}}|^{2} \notag \\
  &\qquad {} \times \left\{  \int_{1/2}^{1} dz\left[  \calh_{i_{N-1} j_{N-1}}\left(z,\frac{\mu^2}{M_f^2}\right) + \frac{a^{(n-4)}_{i_{N-1} j_{N-1}}(z)}{(1-z)_{+}} + \frac{a^{(4)}(z)_{i_{N-1} j_{N-1}}}{(1-z)_{+}} \, \ln \left( \frac{z^2 \, \Xtj^2 \, Q^2}{M_f^2} \right)  \right. \right.  \notag \\
  &\qquad  {}  +  \left. 2 \, a^{(4)}_{i_{N-1} j_{N-1}}(z) \, \left( \frac{\ln(1-z)}{1-z} \right)_{+}\right] \, 
  + \left. \ln \left( \rth^2 \right) \, \int_{1/2}^{\zmj} dz  \, \frac{a^{(4)}_{i_{N-1} j_{N-1}}(z)}{(1-z)_{+}}  \right\},
  \label{eqdefhadcrxfincoll4}
\end{align}
with $\zmj = (\Xtj - \xtm)/\Xtj$.
The integrals over $z$ can be performed. Nevertheless, the term $\calh_{i_{N-1} j_{N-1}}(z,\mu^2/M_f^2)$ in square brackets in eq.~(\ref{eqdefhadcrxfincoll4}) contains a collinear divergence and the $z$ integration does not remove it. 
However,  in eq.~(\ref{eqdefhadcrxfincoll4}), there is only the contribution where $i_N$ is soft and/or collinear, and one has to add the contribution where $i_{N-1}$ is soft and/or collinear. Thus in general, we get the following result
\begin{align}
  \hspace{2em}&\hspace{-2em} \sigma^{\text{fin. coll.}}_{H \, \text{jet}} = \sum_{\seq{i}{N-2} \in S_p}  K^{(n) \, B}_{i_1 i_2} \, \frac{\alpha_s}{2 \, \pi} \, \frac{1}{\Gamma(1-\varepsilon)} \, \int \PSHP{N-2}{\bar{x}} \, d \yj \, d^{n-2} \ktj \notag \\
  &\qquad {} \times  \delta^{n-2}\left( \sum_{l=3}^{N-2} \frac{\vkt{l}}{\bar{x}_l} + \vktj \right) \, A_{(i)_{N-2}}\left(\{\bar{x}\}_{N-2}\right) \,  \sum_{j_{N-1}} |M^{(n)}_{\react{i}{i_{N-1}}{j_{N-1}}{N-1}}|^{2} \, \notag \\
  &\qquad {} \times \calj_{j_{N-1}}\left(\Xtj,\frac{Q^2}{M_f^2},\xtm\right).
  \label{eqdefhadcrxfincoll5}
\end{align}
Let us discuss the dependence of the above mentioned divergences on the type of parton which initiates the jet. 
Let us assume first that $j_{N-1}$, the parton which initiates the jet, is a quark (or an anti-quark), then $i_N$ can be a gluon and $i_{N-1}$ a quark (or an anti-quark) or vice versa. Summing the two contributions $i_N$ soft and/or collinear and $i_{N-1}$ soft and/or collinear, leads to
\begin{align}
  \calj_{q}\left(\Xtj,\frac{Q^2}{M_f^2},\xtm\right) &= \int_{1/2}^{1} dz \left\{  \calh_{qq}\left(z,\frac{\mu^2}{M_f^2}\right) + \calh_{gq}\left(z,\frac{\mu^2}{M_f^2}\right) + \frac{a^{(n-4)}_{qq}(z)}{(1-z)_{+}} + \frac{a^{(n-4)}_{gq}(z)}{(1-z)_{+}}  \right. \notag \\
  &\qquad \qquad {} + \left[\frac{a^{(4)}_{qq}(z)}{(1-z)_{+}}+\frac{a^{(4)}_{gq}(z)}{(1-z)_{+}}\right] \, \ln \left( \frac{z^2 \, \Xtj^2 \, Q^2}{M_f^2} \right)   \notag \\
  &\qquad \qquad {} + 2 \, \left[a^{(4)}_{qq}(z)+ a^{(4)}_{gq}(z) \right] \, \left( \frac{\ln(1-z)}{1-z} \right)_{+} \notag \\
  &\qquad \qquad {} + \left. \ln \left( \rth^2 \right) \, \int_{1/2}^{\zmj} dz  \, \left[\frac{a^{(4)}_{qq}(z)}{(1-z)_{+}}+\frac{a^{(4)}_{gq}(z)}{(1-z)_{+}}\right] \right\}.
  \label{defcalq0}
\end{align}
The collinear divergence presents in eq.~(\ref{defcalq0}) vanishes because the coefficient in front the divergence vanishes, indeed
\begin{align}
  \int_{1/2}^1 dz \, \left[\calh_{qq}\left(z,\frac{\mu^2}{M_f^2}\right) + \calh_{gq}\left(z,\frac{\mu^2}{M_f^2}\right)\right] &= - \frac{1}{\varepsilon} \, \left( \frac{4 \, \pi \, \mu^2}{M_f^2} \right)^{\varepsilon} \, \frac{1}{\Gamma(1-\varepsilon)} \,  \notag \\
  &\qquad{} \times  \int_{1/2}^1 dz \, \left[P^{(4)}_{qq}(z) + P^{(4)}_{gq}(z)\right]. \label{eqdivcolljetq}
\end{align}
But the quantity $\int_{1/2}^1 dz \, \left[P^{(4)}_{qq}(z) + P^{(4)}_{gq}(z)\right]$ sums to zero as it should be.\\

\noindent
Let us assume now that the jet has been initiated by a gluon $j_{N-1} = g$, then $i_{N}$ and $i_{N-1}$ can be a pair of quark -- anti-quark of a certain flavour or $i_{N}$ and $i_{N-1}$ are gluons. Thus summing the different contributions leads to
\begin{align}
  \calj_{g}\left(\Xtj,\frac{Q^2}{M_f^2},\xtm\right) &= \int_{1/2}^{1} dz \left\{  2 \, N_F \, \calh_{qg}\left(z,\frac{\mu^2}{M_f^2}\right) + \calh_{gg}\left(z,\frac{\mu^2}{M_f^2}\right) \right. \notag \\
  &\qquad \qquad {} + 2 \, N_F \, \frac{a^{(n-4)}_{qg}(z)}{(1-z)_{+}} + \frac{a^{(n-4)}_{gg}(z)}{(1-z)_{+}} \notag \\
  &\qquad \qquad {} + \left[2 \, N_F \, \frac{a^{(4)}_{qg}(z)}{(1-z)_{+}}+\frac{a^{(4)}_{gg}(z)}{(1-z)_{+}}\right] \, \ln \left( \frac{z^2 \, \Xtj^2 \, Q^2}{M_f^2} \right)   \notag \\
  &\qquad \qquad {} + 2 \, \left[2 \, N_F \, a^{(4)}_{qg}(z)+ a^{(4)}_{gg}(z) \right] \, \left( \frac{\ln(1-z)}{1-z} \right)_{+} \notag \\
  &\qquad \qquad {} + \left. \ln \left( \rth^2 \right) \, \int_{1/2}^{\zmj} dz  \, \left[2 \, N_F \, \frac{a^{(4)}_{qg}(z)}{(1-z)_{+}}+\frac{a^{(4)}_{gg}(z)}{(1-z)_{+}}\right] \right\}.
  \label{defcalg0}
\end{align}
The collinear divergence presents in eq.~(\ref{defcalg0}) also vanishes because the coefficient in front of the divergence vanishes, indeed
\begin{align}
  \int_{1/2}^1 dz \, \left[ 2 \, N_F \, \calh_{qg}\left(z,\frac{\mu^2}{M_f^2}\right) + \calh_{gg}\left(z,\frac{\mu^2}{M_f^2}\right)\right] &= - \frac{1}{\varepsilon} \, \left( \frac{4 \, \pi \, \mu^2}{M_f^2} \right)^{\varepsilon} \, \frac{1}{\Gamma(1-\varepsilon)} \, \notag \\
  & \times \int_{1/2}^1 dz \, \left[ 2 \, N_F \, P^{(4)}_{qg}(z) + P^{(4)}_{gg}(z)\right],
  \label{eqdivcolljetg}
\end{align}
where $N_F$ is the number of active flavours. Again, the quantity $\int_{1/2}^1 dz \, \left[ 2 \, N_F \, P^{(4)}_{qg}(z) + P^{(4)}_{gg}(z)\right]$ sums to zero in agreement with the KLN  theorem which states that degenerate states like a jet are free of collinear divergences.\\

\noindent
The "jet functions" introduced in eqs.~(\ref{defcalq0}) and (\ref{defcalg0}) have some similarities with the ones used in ref.~\cite{Mukherjee:2012uz}. Note however, that the cone of size $\rth$ is not a jet cone in the sense that our cone is centred on the direction of the hardest parton which is the jet direction only in the collinear limit. Despite that, the merging rule to build the jet is close to the so called $k_t$ algorithm \cite{Catani:1993hr,Ellis:1993tq} which, for a jet made of at most two partons, reduces to $d_{N-1 N} \leq \rth$; this is verified in our case.
The integral over $z$  in eqs.~(\ref{defcalq0}) and (\ref{defcalg0}) can be performed analytically and we get
\begin{align}
    \calj_{q}\left(\Xtj,\frac{Q^2}{M_f^2},\xtm\right) &= C_F \, \left[ \frac{13}{2}- \frac{2 \, \pi^2}{3}- \frac{3}{2} \, \ln \left( \frac{\Xtj^2 \, Q^2}{M_f^2} \right) \right] \notag \\
    &\quad {} + \ln(\rth^2) \, C_F \, \left[ \frac{3}{2} - 3 \, \zmj + 2 \, \ln \left( \frac{\zmj}{1 - \zmj}\right)\right],
    \label{eqrescaljq}
\end{align}
and 
\begin{align}
    \calj_{g}\left(\Xtj,\frac{Q^2}{M_f^2},\xtm\right) &= - \frac{23 \, N_F}{18} + N_c \, \left[ \frac{67}{9}- \frac{2 \, \pi^2}{3} \right]- \frac{11 \, N_c - 2 \, N_F}{6} \, \ln \left( \frac{\Xtj^2 \, Q^2}{M_f^2} \right) \notag \\
    &\quad {} + \ln(\rth^2) \, \left[2 \, N \, \ln \left( \frac{\zmj}{1-\zmj} \right) - \frac{N_F}{3} + \frac{11}{6} \, N_c + (N_F - 4 \, N_c) \, \zmj \right. \notag \\
    &\qquad \qquad \qquad {} + \left.  (N_c - N_F) \, \zmj^2 + \frac{2}{3} \, (N_F - N_c) \, \zmj^3\right].
    \label{eqrescaljg}
\end{align}
Note that a dependence on the scale $M_f^2$ is still present in eqs~(\ref{eqrescaljq}) and (\ref{eqrescaljg}). It is cancelled by terms coming from the soft part, cf.~eq~(\ref{eqdefhadcrxsoft3}). Indeed, from this equation, the coefficient $b_{j_{N-1} j_{N-1}}$ in front of  $\ln(Q^2/M_f^2)$ will be either a $b_{qq}$ or a $b_{gg}$  (coefficients in front of the log in (\ref{eqrescaljq}) and (\ref{eqrescaljg}))  depending on the flavour of the parton $j_{N-1}$ which is the jet at LO and which initiates it at NLO.

\section{Summary and prospects}

In this article, we have presented a novel general method for subtracting collinear and soft divergences at NLO accuracy, specifically designed for processes involving an arbitrary number of fragmentation functions. While several general subtraction methods exist, the one discussed in this article introduces several new features:
\begin{enumerate}
    \item Analytical integration of the subtraction terms is performed in the hadronic centre-of-mass frame.
    \item Longitudinal Lorentz boost invariant variables are employed to describe the phase space.
\end{enumerate}
We have explicitly addressed scenarios where all hard partons fragment, providing recipes for constructing the various subtraction terms and analytically integrating them over the phase space of the parton which may be soft or collinear with respect to others. 
As anticipated, collinear divergences can be absorbed into a redefinition of the PDFs or FFs, while the soft divergences cancel out 
when the virtual contribution is added.
Additionally, we have investigated situations where one hard parton in the final state does not fragment. Our results demonstrate that the subtraction method remains effective in such cases, including scenarios where the unfragmented hard parton is a photon or contributes to a jet. Notably, our method imposes no restrictions on the number of hard partons that do not fragment, although for the sake of brevity, 
we have focused on the case of a single unfragmented hard parton in this article.
\\

\noindent
An immediate application of this method will involve the revision of the DiPhox/JetPhox numerical codes, which currently employ phase space slicing techniques to address soft and collinear divergences. Despite their age, these codes are still utilized by experimental collaborations, particularly those focusing on characterising the quark-gluon plasma. These collaborations study various correlation variables between particles that easily escape the plasma (typically photons) and those strongly interacting with it (such as jets or hadrons). The existing codes are well-suited for analysing these observables.
Furthermore, while codes incorporating NNLO corrections to di-photon production exist, none of them integrate the two fragmentation components by their own. We plan to address this gap in a forthcoming practical article dedicated to rewriting these legacy codes of the Phox family. 
In the present article, we have provided comprehensive results to address more complex processes, such as NLO corrections to di-photon plus jets including the fragmentation components and photon + $k$ jets ($k > 1$) with fragmentation.
Regarding applications to reactions containing heavy quarks, the current method is limited to scenarios where the typical energy scale, such as transverse momentum or invariant mass, is significantly larger than the mass of the heavy quark. 
However, this method can be extended to handle cases involving massive hard partons, thereby enabling the description 
of the full kinematic range of reactions involving heavy quarks.

\section*{Acknowledgments}
We would like to dedicate this article to Eric Pilon.
When this work started, Eric already suffered seriously from his cancer. Nevertheless, he was always available for discussions on Physics up to the last months before he passed away.
This article aims to be a modest tribute to him after thirty years of fruitful collaborations.

\noindent
This work is supported by the French National Research Agency in the framework of the "Investissements d'avenir" program (ANR-15-IDEX-02).

\clearpage
\appendix

\section{One-loop DGLAP kernels}\label{APkernels}

\noindent
In this appendix, we provide the expressions of the one-loop DGLAP kernels and define various functions related to them, which are utilized in the main text. The one-loop DGLAP kernels in $n=4-2 \varepsilon$ dimensions split in the form
\begin{align}
  P^{(n)}_{ij}(z) &= \frac{a^{(n)}_{ij}(z)}{(1-z)_{+}} + b_{ij} \, \delta(1-z) \notag \\
  &= \frac{a^{(4)}_{ij}(z)}{(1-z)_{+}} - \varepsilon \, \frac{a^{(n-4)}_{ij}(z)}{(1-z)_{+}} + b_{ij} \, \delta(1-z) \notag \\
  &= P^{(4)}_{ij}(z) - \varepsilon \, \frac{a^{(n-4)}_{ij}(z)}{(1-z)_{+}},
  \label{eqdefapndim0}
\end{align}
plus eventually some terms of order $O(\varepsilon^2)$ which play no role in a NLO computation.
The expressions of the functions $a^{(4)}_{ij}(z)$  are given by
\begin{align}
  a^{(4)}_{gg}(z) &= 2 \, N_c \, \left[ z + \frac{(1-z)^ 2}{z} + z \, (1-z)^2 \right], \label{eqdefa4gg} \\
  a^{(4)}_{qq}(z) &= C_F \, \left[ 1 + z^2 \right], \label{eqdefa4qq} \\
  a^{(4)}_{gq}(z) &= C_F \, \left[ \frac{1+(1-z)^ 2}{z} \right] \, (1-z), \label{eqdefa4gq} \\
  a^{(4)}_{qg}(z) &= T_F \, \left[ z^2 + (1-z)^2 \right] \, (1-z), \label{eqdefa4qg}
\end{align}
where $N_c$ is the numbers of colours, $C_F = (N^2_c-1)/(2 \, N_c)$ and $T_F = 1/2$. Note that at the order of accuracy used (NLO), the flavours of the quarks do not need to be specified, that is to say $a^{(4)}_{q_i q_i}(z) = a^{(4)}_{q_k q_k}(z)$ (where $q_i$ and $q_k$ are quarks of different flavours).
The extra parts needed to get the one-loop DGLAP kernels in $n$ dimensions are given by
\begin{align}
  a^{(n-4)}_{gg}(z) &= 0, \label{eqdefanm4gg} \\
  a^{(n-4)}_{qq}(z) &= C_F \, (1-z)^2, \label{eqdefanm4qq} \\
  a^{(n-4)}_{gq}(z) &= C_F \, z \, (1-z), \label{eqdefanm4gq} \\
  a^{(n-4)}_{qg}(z) &= T_F \, 2 \, z \, (1-z)^2. \label{eqdefanm4qg}
\end{align}
The coefficients $b_{ij}$ read
\begin{align}
  b_{ij} &= 0 \quad \quad (\text{if} \quad i \neq j), \\
  b_{gg} &= \frac{11 \, N_c - 2 \, N_F}{6}, \label{eqdefbgg} \\
  b_{qq} &= \frac{3}{2} \, C_F. \label{eqdefbqq}
\end{align}

\noindent
To be complete, one has to add the case where $i$ or $j$ is a photon.
\begin{align}
  a^{(4)}_{\gamma q}(z) &= Q_f^2 \, \left[ \frac{1+(1-z)^ 2}{z} \right] \, (1-z), \label{eqdefa4gaq} \\
  a^{(4)}_{q \gamma}(z) &= Q_f^2 \,\left[ z^2 + (1-z)^2 \right] \, (1-z), \label{eqdefa4qga} \\
  a^{(n-4)}_{\gamma q}(z) &= Q_f^2 \, z \, (1-z), \label{eqdefanm4gaq} \\
  a^{(n-4)}_{q \gamma}(z) &= Q_f^2 \, 2 \, N_c \, z \, (1-z)^2, \label{eqdefanm4qga}
\end{align}
where $Q_f$ represents the charge of the quark in units of $e$.

\section{The soft integral}\label{softint}

Let us introduce a "Feynman" parameter, denoted $u$, in order to write the eikonal factor $E_{ij}^{\prime}$ as
\begin{align}
  E_{ij}^{\prime} &= \int_0^1 du \, \frac{p_i \cdot p_j}{((u \, p_i + (1-u) \, p_j) \cdot \hat{p}_N)^2}.
  \label{eqnewEij0}
\end{align}
We set
\begin{align}
  \piij &\equiv u \, p_i + (1-u) \, p_j \notag \\
  &= (\mtij \, \cosh(y_{ij}), \vpitij, \mtij \, \sinh(y_{ij}) ).
  \label{eqdefpiij}
\end{align}
Since $p_i$ and $p_j$ are lightlike, $\piij^2 = 2 \, u \, (1-u) \, p_i \cdot p_j$, and the transverse momentum $\pitij$\footnote{We denote $\pitij$ the length of the vector $\vpitij$.} is given by
\begin{align}
  \pitij^2 &= u^2 \, \pt{i}^2 + (1-u)^2 \, \pt{j}^2 + 2 \, u \, (1-u) \, \pt{i} \, \pt{j} \, \cos(\phi_{ij}),
  \label{eqdefpitijs}
\end{align}
where $\phi_{ij}$ is the azimuthal angle between the two vectors $\vpt{i}$ and $\vpt{j}$ and
the transverse mass $\mtij$ is defined as $\mtij^2 = \piij^2 + \pitij^2$.
The rapidity $y_{ij}$ can be extracted from one of these two equations:
\begin{align}
  \mtij \, \cosh(y_{ij}) &= u \, \pt{i} \, \cosh(y_i) + (1-u) \, \pt{j} \, \cosh(y_j), \label{eqdef1yij} \\
  \mtij \, \sinh(y_{ij}) &= u \, \pt{i} \, \sinh(y_i) + (1-u) \, \pt{j} \, \sinh(y_j). \label{eqdef2yij}
\end{align}
Furthermore, the scalar product between the four momenta $\piij$ and $\hat{p}_N$ is
\begin{align}
  \piij \cdot \hat{p}_N &= \mtij \, \cosh(y_{ij} - y_N) - \pitij \, \cos(\phi_N), \label{eqscaprodpiijpn}
\end{align}
where $\phi_N$ is the direct azimuthal angle between the vectors $\vpitij$ and $\vpthat{N}$.\\

\noindent
The soft integral, defined in eq.(\ref{eqdefJsoft120!}), becomes then
\begin{align}
  \Jso &= p_i \cdot p_j \, \int_0^1 du \, \int_0^{\pi} d \phi_N \, \sin(\phi_N)^{-2 \, \varepsilon} \, \notag \\
  &\quad {} \times \int^{+\infty}_{-\infty} d y_N \, \frac{1}{(\mtij \, \cosh(y_{ij} - y_N) - \pitij \, \cos(\phi_N))^2}.
  \label{eqdefJsoft0}
\end{align}
A priori, this integral seems more complicated to compute than the equivalent one using the polar angle as variable because in this case, an extra integration has to be performed\footnote{There is one integral over the azimuthal angle and one over the rapidity instead of only one over the polar angle between $\vec{p}_{N}$ and $\vec{\Pi}_{ij}$.}. As we will see, this is not a problem and the calculation proceeds in the same way as in the standard case.\\

\noindent
To start with, let us split the range of the $y_N$ integration in two parts
\begin{align}
  \int_{-\infty}^{+\infty} d y_N \, F(y_N) &=\int_{y_{ij}}^{+\infty} d y_N  \, F(y_N) + \int_{-\infty}^{y_{ij}} d y_N \, F(y_N). \label{eqsplitrangeyN}
\end{align}
We make the changes of variable $\Delta y = y_N-y_{ij}$ in the first integral of the right hand side of eq.~(\ref{eqsplitrangeyN}) and $\Delta y = y_{ij} - y_N$ in the second one
\begin{align}
  \int_{-\infty}^{+\infty} d y_N \, F(y_N) &= \int_{0}^{+\infty} d \Delta y  \, F(y_{ij}+\Delta y) + \int^{+\infty}_{0} d \Delta y \, F(y_{ij}-\Delta y). \label{eqsplitrangeyNp}
\end{align}
The integrands $F(y_{ij} \pm \Delta y)$ depend on $\Delta y$ only through $\cosh(\pm \Delta y)$ such that the two integrals are equal and the soft integral becomes simply
\begin{align}
  \Jso &= 2 \, p_i \cdot p_j \, \int_0^1 du \, \int_0^{\pi} d \phi_N \, \sin(\phi_N)^{-2 \, \varepsilon} \, \int^{+\infty}_{0} d \Delta y \, \frac{1}{(\mtij \, \cosh(\Delta y) - \pitij \, \cos(\phi_N))^2}.
  \label{eqdefJsoft1}
\end{align}
Then, the change of variable $\cos(\phi_N) = 2 \, x - 1$ is performed leading to
\begin{align}
  \Jso &= 2^{1-2 \, \varepsilon} \, p_i \cdot p_j \, \int_0^1 du \, \int^{+\infty}_{0} d \Delta y \,  \left( \mtij \, \cosh(\Delta y) + \pitij \right)^{-2} \, \notag \\
  &\quad {} \times  \int_0^{1} d x \, x^{-1/2-\varepsilon} \, (1-x)^{-1/2-\varepsilon} \, \left( 1 - \frac{2 \, \pitij}{\mtij \, \cosh(\Delta y) + \pitij} \, x \right)^{-2} \notag \\
  &=  2^{1-2 \, \varepsilon} \, p_i \cdot p_j \, \int_0^1 du \, \int^{+\infty}_{0} d \Delta y \,  \left( \mtij \, \cosh(\Delta y) + \pitij \right)^{-2} \, \notag \\
  &\quad {} \times \frac{\Gamma^2 \left( \frac{1}{2} - \varepsilon \right)}{\Gamma(1 - 2 \, \varepsilon)} \; \Hyp \! \left( 2,\frac{1}{2}-\varepsilon;1 - 2 \, \varepsilon;z \right),
  \label{eqdefJsoft2}
\end{align}
where $\Hyp$ is the Gauss hypergeometric function (cf. ref. \cite{abramowitz+stegun}) and
\begin{equation}
  z = \frac{2 \, \pitij}{\mtij \, \cosh(\Delta y) + \pitij}.
\end{equation}
Using the quadratic transformation given by eq.~(15.3.16) of ref. \cite{abramowitz+stegun}, yields
\begin{align}
  \Jso &= 2^{1-2 \, \varepsilon} \, p_i \cdot p_j \,  \frac{\Gamma^2 \left( \frac{1}{2} - \varepsilon \right)}{\Gamma(1 - 2 \, \varepsilon)} \, \int_0^1 du \, \int^{+\infty}_{0} d \Delta y \,  \left( \mtij \, \cosh(\Delta y) \right)^{-2} \, \notag \\
  &\quad {} \times\; \Hyp \! \left( 1,\frac{3}{2};1 - \varepsilon;z^{\prime \, 2} \right),
  \label{eqdefJsoft3}
\end{align}
with
\begin{equation}
  z^{\prime} = \frac{\pitij}{\mtij \, \cosh(\Delta y)}.
\end{equation}
The four-momentum $\piij$ is time like because $p_i \cdot p_j \geq 0$ and $z^{\prime} \leq 1$, such that the hypergeometric function 
can be expanded in a series\footnote{We disregard the case where $p_i \cdot p_j =0$ which is not relevant for NLO computations. Furthermore, we assume that $u \neq 0,1$ in such way that $z^{\prime} < 1$ to justify the expansion. Once the integration over $\Delta y$ has been performed, the result will be a regular function of $u$ for these values.}
\begin{align}
  \Jso &= 2^{1-2 \, \varepsilon} \, p_i \cdot p_j \,  \frac{\Gamma^2 \left( \frac{1}{2} - \varepsilon \right)}{\Gamma(1 - 2 \, \varepsilon)} \, \int_0^1 du \left( \mtij \right)^{-2} \, \notag \\
  &\quad {} \times \frac{\Gamma(1-\varepsilon)}{\Gamma \left( \frac{3}{2} \right)} \, \sum_{n=0}^{\infty} \frac{\Gamma(1+n) \, \Gamma \left( \frac{3}{2} + n \right)}{\Gamma(1-\varepsilon+n)}\, \frac{1}{n!} \, \left( \frac{\pitij}{\mtij} \right)^{2 \, n} \,  \int^{+\infty}_{0} d \Delta y \, \cosh^{-2 \, (n+1)}(\Delta y).
  \label{eqdefJsoft4}
\end{align}
 In order to perform the $\Delta y$ integration term by term, let us introduce
 \begin{align}
 U(n) &\equiv \int^{+\infty}_{0} d \Delta y \, \frac{1}{\cosh^{2 \, (n+1)}(\Delta y)}. \label{eqdefUn0}
 \end{align}
 It is easy to get that
 \begin{align}
   U(0) &= \int^{+\infty}_{0} d \Delta y \, \frac{1}{\cosh^{2}(\Delta y)} \notag \\
   &= \left[ \tanh(\Delta y) \right]_0^{+\infty} \notag \\
   &= 1.
   \label{eqdefU0}
 \end{align}
 Let us establish a recurrence relation for $U(n)$. In order to do that, we use
 the following relation, valid for $m>1$ and easily established by an integration by parts:
 \begin{align}
   \int \frac{dz}{\cosh^m(z)} &= \frac{1}{m-1} \, \frac{\sinh(z)}{\cosh^{m-1}(z)} + \frac{m-2}{m-1} \, \int \frac{dz}{\cosh^{m-2}(z)} \quad \text{for} \; m\neq 1.
   \label{eqdefreccosh}
 \end{align}
 The first term on the right hand side of eq.~(\ref{eqdefreccosh}) will always vanish when the integration bounds are $0$ and $+\infty$ for $m > 2$. But this condition is satisfied if $n \geq 1$, thus we get that
 \begin{align}
   U(n) &= \frac{2 \, n}{2 \, n + 1} \, U(n-1) \notag \\
   &= \frac{2 \, n \; 2 \, (n-1)}{2 \, n + 1 \; 2 \, n - 1} \, U(n-2) \notag \\
   &= \frac{2 \, n \; 2 \, (n-1) \cdots 2}{(2 \, n + 1) \; (2 \, n - 1) \cdots 3} \, U(0).
   \label{eqdefrecU}
 \end{align}
 Using the result of eq.~(\ref{eqdefU0}) leads to
 \begin{align}
   U(n) &=  \frac{2^{2 \, n} \, (n!)^2}{(2 \, n + 1)!} 
   = 2^{2 \, n} \,\frac{\Gamma^2(n+1)}{\Gamma(2 \, (n+1))}.
   \label{eqresUn}
 \end{align}
Applying the duplication formula for the gamma function, we then obtain 
\begin{align}
  U(n) &=  \frac{\sqrt{\pi}}{2} \, \frac{\Gamma(n+1)}{\Gamma \left( n + \frac{3}{2} \right)}.
  \label{eqresUn1}
\end{align}
Putting the result of eq.~(\ref{eqresUn1}) into eq.~(\ref{eqdefJsoft4}) yields
\begin{align}
  \Jso &= 2^{1-2 \, \varepsilon} \, p_i \cdot p_j \,  \frac{\Gamma^2 \left( \frac{1}{2} - \varepsilon \right)}{\Gamma(1 - 2 \, \varepsilon)} \, \int_0^1 du \left( \mtij \right)^{-2} \, \notag \\
  &\quad {} \times \frac{\Gamma(1-\varepsilon)}{\Gamma \left( \frac{3}{2} \right)} \,  \frac{\sqrt{\pi}}{2} \, \sum_{n=0}^{\infty} \frac{\Gamma^2 \left( 1 + n \right)}{\Gamma(1-\varepsilon+n)}\, \frac{1}{n!} \, \left( \frac{\pitij}{\mtij} \right)^{2 \, n}.
  \label{eqdefJsoft5}
\end{align}
The series in the right hand side of eq.~(\ref{eqdefJsoft5}) can be summed up into a new hypergeometric function
\begin{align}
  \Jso &= 2^{1-2 \, \varepsilon} \, p_i \cdot p_j \,  \frac{\Gamma^2 \left( \frac{1}{2} - \varepsilon \right)}{\Gamma(1 - 2 \, \varepsilon)} \, \int_0^1 du \left( \mtij \right)^{-2} \, \Hyp \! \left( 1,1;1-\varepsilon;\frac{\pitij^2}{\mtij^2} \right).
  \label{eqdefJsoft6}
\end{align}
Using the linear transformation eq.~(15.3.6) of ref.~\cite{abramowitz+stegun}, eq.~(\ref{eqdefJsoft6}) becomes
\begin{align}
  \Jso &= 2^{1-2 \, \varepsilon} \, p_i \cdot p_j \,  \frac{\Gamma^2 \left( \frac{1}{2} - \varepsilon \right)}{\Gamma(1 - 2 \, \varepsilon)} \, \int_0^1 du \left( \mtij \right)^{-2} \notag \\
  &\quad {} \times \left\{ \frac{\Gamma(1-\varepsilon) \, \Gamma(-1-\varepsilon)}{\Gamma(-\varepsilon) \, \Gamma(- \varepsilon)}   \, \Hyp \! \left( 1,1;2+\varepsilon;\frac{\piij^2}{\mtij^2} \right) \right. \notag \\
  &\qquad \quad {}+ \left. \left( \frac{\piij^2}{\mtij^2} \right)^{-1-\varepsilon} \, \Gamma(1-\varepsilon) \, \Gamma(1+\varepsilon) \; \Hyp \! \left( -\varepsilon,-\varepsilon;-\varepsilon;\frac{\piij^2}{\mtij^2} \right)\right\}.
  \label{eqdefJsoft7}
\end{align}
Remembering that $\Hyp(a,b;b;z) = (1-z)^{-a}$, eq.~(\ref{eqdefJsoft7}) can be rearranged in the following way
\begin{align}
  \Jso &= 2^{1-2 \, \varepsilon} \, p_i \cdot p_j \,  \frac{\Gamma^2 \left( \frac{1}{2} - \varepsilon \right)}{\Gamma(1 - 2 \, \varepsilon)} \, \int_0^1 du \left( \mtij \right)^{-2} \notag \\
  &\quad {} \times \left\{ \frac{\varepsilon}{1+\varepsilon}   \; \Hyp \! \left( 1,1;2+\varepsilon;\frac{\piij^2}{\mtij^2} \right) \right. \notag \\
  &\qquad \quad {}+ \left.  \Gamma(1-\varepsilon) \, \Gamma(1+\varepsilon) \, \left( \piij^2 \right)^{-1-\varepsilon} \, \mtij^2 \, \left( \pitij^2 \right)^{\varepsilon} \vphantom{\Hyp \! \left( 1,1;2+\varepsilon;\frac{\piij^2}{\mtij^2} \right)} \right\}.
  \label{eqdefJsoft8}
\end{align}
Note that since the remaining hypergeometric function is finite when $\varepsilon=0$ and is multiplied by $\varepsilon$, the arguments of the latter can be taken at $\varepsilon=0$. Thus, using that $\Hyp(1,1;2;z) = - \ln(1-z)/z$ and dropping some terms which vanish as $\varepsilon=0$, eq.~(\ref{eqdefJsoft8}) can be re-written as
\begin{align}
  \Jso &= 2^{1-2 \, \varepsilon} \, p_i \cdot p_j \,  \frac{\Gamma^2 \left( \frac{1}{2} - \varepsilon \right)}{\Gamma(1 - 2 \, \varepsilon)} \, \int_0^1 du  \notag \\
&\quad {} \times \left\{ -\frac{\varepsilon}{\piij^2}  \, \ln \left(\frac{\pitij^2}{\mtij^2}\right) + \Gamma(1-\varepsilon) \, \Gamma(1+\varepsilon) \, \left( \piij^2 \right)^{-1-\varepsilon} \, \left( \pitij^2 \right)^{\varepsilon} \right\}.
  \label{eqdefJsoft9}
\end{align}
\\

\noindent
Let us focus on the first term inside the curly brackets of eq.~(\ref{eqdefJsoft9}). The quantity $B_1$ which is defined to be
\begin{align}
  B_1 &\equiv \int_0^1 du \, \frac{1}{\piij^2}  \, \ln \left(\frac{\pitij^2}{\mtij^2}\right),
  \label{eqdefB10}
\end{align}
can be written as
\begin{align}
  B_1 &= \frac{1}{2 \, p_i \cdot p_j} \, \left\{ \int_0^1 \frac{du}{u} \, \left[ \ln(\pitij^2) - \ln(\mtij^2) \right] + \int_0^1 \frac{du}{1-u} \, \left[ \ln(\pitij^2) - \ln(\mtij^2) \right] \right\}.
  \label{eqdefB11}
\end{align}
The quantities $\pitij^2$ and $\mtij^2$ are second order polynomials in the variable $u$ and can be written as
\begin{align}
  \pitij^2 &= \pt{j}^2 \, \left( 1 - \frac{u}{\tup} \right) \, \left( 1 - \frac{u}{\tum} \right), \label{eqwrtpitij1} \\
  \mtij^2 &= \pt{j}^2 \, \left( 1 - \frac{u}{\bup} \right) \, \left( 1 - \frac{u}{\bum} \right), \label{eqwrtmtij1}
\end{align}
where
\begin{align}
  \tupm &= \frac{\pt{j}}{\pt{j} - \pt{i} \, e^{\mp i \, \phi_{ij}}}, \label{eqdefbupm} \\
  \bupm &= \frac{\pt{j}}{\pt{j} - \pt{i} \, e^{\mp (y_i-y_j)}}. \label{eqdeftupm}
\end{align}
In eqs.~(\ref{eqwrtpitij1}) and (\ref{eqwrtmtij1}), the way of writing a second order polynomial in terms of its roots is not standard but it will simplify the computation of the two integrals of eq.~(\ref{eqdefB11}). Indeed, we can
notice that $\tup$ and $\tum$ are complex conjugate while $\bup$ and $\bum$ are real but do not belong to the range $[0,1]$ 
as it can be inferred from eq.~(\ref{eqdeftupm})
and that $1-u/\bup$ and $1-u/\bum$ are positive when $u$ is in $[0,1]$. Thus the logarithms of $\pitij^2$ and $\mtij^2$ can be safely split, namely
\begin{align}
  \ln(\pitij^2) &= \ln(\pt{j}^2) + \ln\left( 1 - \frac{u}{\tup} \right) + \ln\left( 1 - \frac{u}{\tum} \right), \label{eqwrtlnpitij1} \\
  \ln(\mtij^2) &= \ln(\pt{j}^2) + \ln\left( 1 - \frac{u}{\bup} \right) +  \ln\left( 1 - \frac{u}{\bum} \right). \label{eqwrtlnmtij1}
\end{align}
The first integral in the curly brackets of eq.~(\ref{eqdefB11}) can be performed easily leading to
\begin{align}
  \int_0^1 \frac{du}{u} \, \left[ \ln(\pitij^2) - \ln(\mtij^2) \right] &= -\dilog \left( \frac{1}{\tup} \right)-\dilog \left( \frac{1}{\tum} \right)+\dilog \left( \frac{1}{\bup} \right)+\dilog \left( \frac{1}{\bum} \right).
  \label{eqdeffirstint}
\end{align}
For the second integral, after a change of variable $t=1-u$, the quantities $\pitij^2$ and $\mtij^2$ can be written as
\begin{align}
  \pitij^2 &= \pt{i}^2 \, \left( 1 - \frac{t}{1-\tup} \right) \, \left( 1 - \frac{t}{1-\tum} \right), \label{eqwrtpitij2} \\
  \mtij^2 &= \pt{i}^2 \, \left( 1 - \frac{t}{1-\bup} \right) \, \left( 1 - \frac{t}{1-\bum} \right). \label{eqwrtmtij2}
\end{align}
For the same reasons, the logarithm of these quantities can be safely split leading for this second integral to
\begin{align}
  \int_0^1 \frac{du}{1-u} \, \left[ \ln(\pitij^2) - \ln(\mtij^2) \right] &= -\dilog \left( \frac{1}{1-\tup} \right)-\dilog \left( \frac{1}{1-\tum} \right) \notag \\
  &\quad {} + \dilog \left( \frac{1}{1-\bup} \right)+\dilog \left( \frac{1}{1-\bum} \right).
  \label{eqdefsecondint}
\end{align}
Using the following property of the dilogarithms, namely
\begin{align}
  \dilog \left( \frac{1}{z} \right) + \dilog \left( \frac{1}{1-z} \right) &= -\frac{1}{2} \, \ln^2 \left( 1 - \frac{1}{z} \right) \quad \text{for $z \notin [0,1]$ },\label{eqpropdilog1}
\end{align}
the quantity $B_1$ becomes
\begin{align}
  B_1 &= \frac{1}{4 \, p_i \cdot p_j} \, \left\{ \ln^2 \left( 1 - \frac{1}{\tup} \right) + \ln^2 \left( 1 - \frac{1}{\tum} \right) -  \ln^2 \left( 1 - \frac{1}{\bup} \right) - \ln^2 \left( 1 - \frac{1}{\bum} \right) \right\}.
  \label{eqdefB12}
\end{align}
\\

\noindent
For the second term in the curly brackets of eq.~(\ref{eqdefJsoft9}), let us introduce
\begin{align}
  B_2 &= \int_0^1 du \, \left( \piij^2 \right)^{-1-\varepsilon} \, \left( \pitij^2 \right)^{\varepsilon},
  \label{eqdefB20}
\end{align}
which can be written as
\begin{align}
  B_2 &= (2 \, p_i \cdot p_j)^{-1-\varepsilon} \, \left[ \int_0^1 du \, u^{-1-\varepsilon} \, (1-u)^{-\varepsilon} \, \left( \pitij^2 \right)^{\varepsilon} + \int_0^1 du \, u^{-\varepsilon} \, (1-u)^{-1-\varepsilon} \, \left( \pitij^2 \right)^{\varepsilon} \right].
  \label{eqdefB21}
\end{align}
Let us focus on the first integral of eq.~(\ref{eqdefB21}). The distribution $u^{-1-\varepsilon}$ can be replaced by
\begin{equation}
  u^{-1-\varepsilon} \simeq -\frac{1}{\varepsilon} \, \delta(u) + \frac{1}{(u)_{+}} - \varepsilon \, \left( \frac{\ln(u)}{u} \right)_{+} + O(\varepsilon^2),
  \label{eqdefdistribu}
\end{equation}
and the rest of the integrand is expanded around $\varepsilon=0$ yielding
\begin{align}
  \int_0^1 du \, u^{-1-\varepsilon} \, (1-u)^{-\varepsilon} \, \left( \pitij^2 \right)^{\varepsilon} &= -\frac{1}{\varepsilon} \, \left[ 1 + \varepsilon \, \ln(\pt{j}^2) + \frac{\varepsilon^2}{2} \, \ln^2(\pt{j}^2) \right] + \int_0^1 \frac{du}{(u)_{+}} \notag \\
  &\quad {} + \varepsilon \, \left[ \int_0^1 \frac{du}{u} \left( \ln \left( 1 - \frac{u}{\tup} \right) + \ln \left( 1 - \frac{u}{\tum} \right) \right) \right. \notag \\
  &\qquad {} - \left. \int_0^1 \frac{du}{u} \, \ln(1-u) \right] - \varepsilon \, \int_0^1 du \, \left( \frac{\ln(u)}{u} \right)_{+} \notag \\
  &=  -\frac{1}{\varepsilon} \, \left[ 1 + \varepsilon \, \ln(\pt{j}^2) + \frac{\varepsilon^2}{2} \, \ln^2(\pt{j}^2) \right] \notag \\
  &\quad {}- \varepsilon \left[ \dilog \left( \frac{1}{\tup} \right) + \dilog \left( \frac{1}{\tum} \right) - \frac{\pi^2}{6} \right].
  \label{eqdefintone}
\end{align}
The second integral of eq.~(\ref{eqdefB21}), with a change of variable $t=1-u$, gives
\begin{align}
  \int_0^1 du \, u^{-\varepsilon} \, (1-u)^{-1-\varepsilon} \, \left( \pitij^2 \right)^{\varepsilon} &=  -\frac{1}{\varepsilon} \, \left[ 1 + \varepsilon \, \ln(\pt{i}^2) + \frac{\varepsilon^2}{2} \, \ln^2(\pt{i}^2) \right] \notag \\
  &\quad {}- \varepsilon \left[ \dilog \left( \frac{1}{1-\tup} \right) + \dilog \left( \frac{1}{1-\tum} \right) - \frac{\pi^2}{6} \right].
  \label{eqdefinttwo}
\end{align}
Plugging the results of eqs.~(\ref{eqdefintone}) and (\ref{eqdefinttwo}) into eq.~(\ref{eqdefB21}) leads to
\begin{align}
  B_2 &= \frac{1}{2 \, p_i \cdot p_j} \, \left\{  -\frac{1}{\varepsilon} \, \left[ 2 - 2 \, \varepsilon \, \ln \left( \frac{2 \, p_i \cdot p_j}{\pt{i} \, \pt{j}} \right) + \frac{\varepsilon^2}{2} \, \left[ \ln^2 \left( \frac{\pt{i}^2}{2 \, p_i \cdot p_j} \right) + \ln^2 \left( \frac{\pt{j}^2}{2 \, p_i \cdot p_j} \right)\right] \right] \right. \notag \\
  &\quad {} + \left. \varepsilon \, \left[ \frac{1}{2} \, \ln^2 \left( 1 - \frac{1}{\tup} \right) + \frac{1}{2} \, \ln^2 \left( 1 - \frac{1}{\tum} \right) + \frac{\pi^2}{3} \right] \right\}.
  \label{eqdefB22}
\end{align}
\\

\noindent
Using the results of eqs.~(\ref{eqdefB12}) and (\ref{eqdefB22}) and the fact that $\Gamma(1-\varepsilon) \, \Gamma(1+\varepsilon) \simeq 1 + \varepsilon^2 \, \pi^2/6$, eq.~(\ref{eqdefJsoft9}) becomes
\begin{align}
  \Jso &= 2^{-2 \, \varepsilon} \,\frac{\Gamma^2 \left( \frac{1}{2} - \varepsilon \right)}{\Gamma(1 - 2 \, \varepsilon)} \, \left\{ -\frac{2}{\varepsilon} + 2 \, \ln \left( \frac{2 \, p_i \cdot p_j}{\pt{i} \, \pt{j}} \right) \right. \notag \\
  &\quad {}- \frac{\varepsilon}{2} \,  \left[ \ln^2 \left( \frac{\pt{i}^2}{2 \, p_i \cdot p_j} \right) + \ln^2 \left( \frac{\pt{j}^2}{2 \, p_i \cdot p_j} \right)\right] \notag \\
  &\quad {}+ \left. \frac{\varepsilon}{2} \, \left[ \ln^2 \left( 1 - \frac{1}{\bup} \right) + \ln^2 \left( 1 - \frac{1}{\bum} \right) \right]  \right\}.
  \label{eqdefJsoft10}
\end{align}
Finally, using the definition of the two roots $\bup$ and $\bum$ as well as introducing $\bd_{ij} = \cosh(2 \, \ystij ) - \cos(\phi_{ij})$ with $\ystij = (y_i-y_j)/2$, we end up with
\begin{align}
  \Jso &= 2^{-2 \, \varepsilon} \,   \frac{\Gamma^2 \left( \frac{1}{2} - \varepsilon \right)}{\Gamma(1 - 2 \, \varepsilon)} \, \left\{ -\frac{2}{\varepsilon} + 2 \, \ln \left( 2 \, \bd_{ij} \right) - \varepsilon \, \ln^2 \left( 2 \, \bd_{ij} \right) + 4 \, \varepsilon \, (\ystij)^2  \right\}.
  \label{eqdefJsoft11}
\end{align}
Note that the definition of $\bd_{ij}$ is equivalent to the one given in subsection \ref{subsecfincoll}.
Let us notice that the result of eq.~(\ref{eqdefJsoft11}) is particularly simple, the dilogarithms combine to logarithms squared and the dependence on the azimuthal angle is through $\bar d_{ij}$. In addition, this result is explicitly invariant under boosts along the beam direction as it depends only on difference of rapidities and azimuthal angles.

\section{The collinear integral inside the cylinder}\label{collint}

Let us recall the integral to compute (see eq. (\ref{eqdefJcoll!})):
\begin{align}
  J^{\text{coll}} &= \int_0^{\pi} d \phi_N \, (\sin \phi_N)^{-2 \, \varepsilon} \, \int_{-\infty}^{+\infty} d y_N \,  \frac{\cos \phi_N}{\cosh(y_N-y_i) - \cos \phi_N}.
  \label{eqdefJcoll0}
\end{align}
As mentioned earlier, the result of this integral will not depend on $y_i$ because the range of integration in $y_N$ extends to infinity.
As in the appendix \ref{softint}, the $y_N$ integration range is divided into two parts
\begin{align}
  \int_{-\infty}^{+\infty} d y_N \, F(y_N) &= \int_{y_{i}}^{+\infty} d y_N  \, F(y_N) + \int_{-\infty}^{y_{i}} d y_N \, F(y_N). \label{eqsplitrangeyNc}
\end{align}
We again make the changes of variable $\Delta y = y_N-y_{i}$ in the first integral on the right hand side of eq.~(\ref{eqsplitrangeyNc}) and $\Delta y = y_{i} - y_N$ in the second one
\begin{align}
  \int_{-\infty}^{+\infty} d y_N \, F(y_N) &= \int_{0}^{+\infty} d \Delta y  \, F(y_{i}+\Delta y) + \int^{+\infty}_{0} d \Delta y \, F(y_{i}-\Delta y). \label{eqsplitrangeyNpc}
\end{align}
The integrand depends on $\Delta y$ only through $\cosh(\pm \Delta y)$ such that the two integrals are equal and thus
\begin{align}
  J^{\text{coll}} &= 2 \, \int_0^{\pi} d \phi_N \, \left( \sin \phi_N \right)^{-2 \, \varepsilon} \, \int_0^{\infty} d \Delta y \, \frac{\cos \phi_N}{\cosh(\Delta y) - \cos \phi_N},
  \label{eqdefhatJcoll0}
\end{align}
which can be written as
\begin{align}
  J^{\text{coll}} &= 2 \, \int_0^{\infty} d \Delta y \, \int_0^{\pi} d \phi_N \, \left( \sin \phi_N \right)^{-2 \, \varepsilon} \, \left[  -1 + \frac{\cosh(\Delta y)}{\cosh(\Delta y) - \cos \phi_N}\right].
  \label{eqdefhatJcoll1}
\end{align}
Setting $\cos \phi_N = 2 \, x -1$ leads to
\begin{align}
  J^{\text{coll}} &= 2^{1-2 \, \varepsilon} \, \int_0^{\infty} d \Delta y \, \int_0^{1} d x \, x^{-1/2-\varepsilon} \, (1-x)^{-1/2-\varepsilon} \, \left[  -1 + \frac{\cosh(\Delta y)}{\cosh(\Delta y)+1} \, (1 - z \, x)^{-1}\right],
  \label{eqdefhatJcoll10}
\end{align}
where $z=2/(1+\cosh(\Delta y))$. The $x$ integration gives a Gauss hypergeometric function
\begin{align}
  J^{\text{coll}} &= 2^{1-2 \, \varepsilon} \, \frac{\Gamma^2 \left( \frac{1}{2} - \varepsilon \right)}{\Gamma(1- 2 \, \varepsilon)} \,  \int_0^{\infty} d \Delta y \, \left[  -1 + \frac{\cosh(\Delta y)}{\cosh(\Delta y)+1} \, \Hyp \left( 1, \frac{1}{2}-\varepsilon;1-2 \, \varepsilon;z \right) \right].
  \label{eqdefhatJcoll11}
\end{align}
Using the quadratic transformation, eq.~(15.3.16) of ref. \cite{abramowitz+stegun}, yields
\begin{align}
  J^{\text{coll}} &= 2^{1-2 \, \varepsilon} \, \frac{\Gamma^2 \left( \frac{1}{2} - \varepsilon \right)}{\Gamma(1- 2 \, \varepsilon)} \, \int_0^{\infty} d \Delta y \, \left[  -1 + \Hyp \left( \frac{1}{2},1;1-\varepsilon;z^{\prime \, 2} \right) \right],
  \label{eqdefhatJcoll2}
\end{align}
with $z^{\prime} = 1/\cosh(\Delta y)$. As in the preceding case, the hypergeometric function can be expanded 
in a series as $z^{\prime} \leq 1$.\footnote{Once more, a rigorous approach would involve introducing a cutoff $\lambda$ to prevent reaching the value $\Delta y = 0$, ensuring $z^{\prime} < 1$, and then taking the limit $\lambda \rightarrow 0$ at the end.}
The key point here is that the first term in the square brackets leads to a divergence when integrated on $\Delta y$. 
However, this term cancels against the first term of the hypergeometric series. Doing so and shifting the variable of the series such that it starts at 0 yields
\begin{align}
  J^{\text{coll}} &= 2^{1-2 \, \varepsilon} \, \frac{\Gamma^2 \left( \frac{1}{2} - \varepsilon \right)}{\Gamma(1- 2 \, \varepsilon)} \, \frac{\Gamma(1-\varepsilon)}{\Gamma \left( \frac{1}{2} \right)} \, \sum_{m=0}^{+\infty} \, \frac{\Gamma \left( \frac{3}{2} + m \right) \, \Gamma(2+m)}{\Gamma(2-\varepsilon+m)} \, \frac{1}{(m+1)!} \, \notag \\
  &\quad {} \times \int_0^{\infty} d \Delta y \,  (\cosh(\Delta y))^{-2 \, (m+1)}.
  \label{eqdefhatJcoll3}
\end{align}
Using the eq.~(\ref{eqresUn1}) for the integration on $\Delta y$ leads to a series of the type
\begin{align}
  J^{\text{coll}} &= 2^{-2 \, \varepsilon} \, \sqrt{\pi} \, \frac{\Gamma^2 \left( \frac{1}{2} - \varepsilon \right)}{\Gamma(1- 2 \, \varepsilon)} \, \frac{\Gamma(1-\varepsilon)}{\Gamma \left( \frac{1}{2} \right)} \, \sum_{m=0}^{+\infty} \, \frac{\Gamma^2(1+m)}{\Gamma(2-\varepsilon+m)} \, \frac{1}{m!},
  \label{eqdefhatJcoll30}
\end{align}
which can be expressed as a hypergeometric function whose argument is $1$ and can be rewritten in terms of gamma functions 
as long as $\varepsilon < 0$:
\begin{align}
  J^{\text{coll}} &= 2^{-2 \, \varepsilon} \, \frac{\Gamma^2 \left( \frac{1}{2} - \varepsilon \right)}{\Gamma(1- 2 \, \varepsilon)} \, \frac{\Gamma(1-\varepsilon)}{\Gamma \left( 2 - \varepsilon \right)} \, \Hyp(1,1;2-\varepsilon;1) 
  = \frac{2^{-2 \, \varepsilon}}{-\varepsilon} \,  \, \frac{\Gamma^2 \left( \frac{1}{2} - \varepsilon \right)}{\Gamma(1- 2 \, \varepsilon)}.
  \label{eqdefhatJcoll4}
\end{align}

\section{The collinear integral inside the cone}\label{intir}

Let us evaluate the collinear integral that arises when $i_N$ lies within the cone $\Gamma_i$:
\begin{align}
  J^{\text{coll}}_{\text{out}} &= \iint_{\Gamma_i} \, d \phi_N \, d y_{N} \, \, \frac{\phi_N^{-2 \, \varepsilon}}{(y_i-y_N)^2 + \phi_N^2}.
  \label{eqdefJcollout0}
\end{align}
Similar to Appendix~\ref{collint}, the integration range over $y_N$ is split into two parts. We perform the change of variables $\Delta y = y_N - y_i$ and $\Delta y = y_i - y_N$ in the parts where $y_N \geq y_i$ and $y_N \leq y_i$, respectively, resulting in
\begin{align}
  J^{\text{coll}}_{\text{out}} &= 2 \, \iint_{\Gamma_i} \, d \phi_N \, d \Delta y \, \, \frac{\phi_N^{-2 \, \varepsilon}}{(\Delta y)^2 + \phi_N^2}.
  \label{eqdefJcollout1}
\end{align}
The condition that $i_N$ lies within the cone $\Gamma_i$ is expressed as $(\Delta y)^2 + \phi_N^2 \leq \rth^2$. 
However, by construction, $\Delta y$ must be less than $\min(\yNmax-y_i,y_i-\yNmin)$.
In cases where $\pt{N}$ is large, it's possible that $\min(\yNmax-y_i,y_i-\yNmin) < \rth$, which implies that the integration domain is no longer a disk in the $\Delta y$-$\phi_N$ plane. As a result, the analytical calculation of the integral $J^{\text{coll}}_{\text{out}}$ becomes complicated.
To address this challenge, we adopt the following strategy.
Firstly, in the subtraction term, we use $\ytNmax = {\yNmax}_{|{\phi_N=0}}$ and $\ytNmin = {\yNmin}_{|{\phi_N=0}}$ as bounds for the $y_N$ integration. These bounds are independent of $\phi_N$.
Secondly, we dynamically adjust $\rth$ to $\min(\rth,\Deltaymax)$, where $\Deltaymax = \min(\ytNmax-y_N,y_N-\ytNmin)$ represents the largest value taken by $\Delta y$. In other words, the value of $\rth$ is not fixed but may vary depending on the kinematics.
In summary, by implementing these adjustments, we ensure that our calculation accounts for the varying nature of the integration domain, overcoming the complication introduced by large transverse momenta.
In this way, the analytical integration remains easy and the dependence on $\rth$ is logarithmic.\\

\noindent
To perform the integral analytically, we introduce polar coordinates in the plane $\Delta y$-$\phi$, such that
\begin{align*}
\Delta y &= r \, \cos(\theta), \\
\phi &= r \, \sin(\theta).
\end{align*}
After this change of variable, equation~(\ref{eqdefJcollout0}) becomes
\begin{align}
J^{\text{coll}}_{\text{out}} &= 2 \, \int_0^{\rth} dr \, r^{-1-2 \, \varepsilon} \, \int_0^{\pi/2} d \theta \, (\sin \theta)^{- 2 \, \varepsilon}.
\label{eqdefJcollout11}
\end{align}
Note that, since $\Delta y$ and $\phi$ are positive, the angle $\theta$ runs between $0$ and $\pi/2$. The $r$ integration is trivial and we introduce a new variable $x$ such that $\cos \, \theta = 2 \, x - 1$ leading to
\begin{align}
J^{\text{coll}}_{\text{out}} &= \frac{\rth^{-2 \, \varepsilon}}{-2 \, \varepsilon} \, 2^{1-2 \, \varepsilon} \, \int^1_{1/2} dx \, x^{-1/2-\varepsilon} \, (1-x)^{-1/2-\varepsilon} 
= 2^{-2 \, \varepsilon} \, \frac{\Gamma^2\left( \frac{1}{2} - \varepsilon \right)}{\Gamma(1 - 2 \, \varepsilon)} \; \frac{\rth^{-2 \, \varepsilon}}{-2 \, \varepsilon}.
\label{eqdefJcollout2}
\end{align}

\section{Details of the divergent terms construction} \label{gutsdivterm}
Let us collect the different divergent terms resulting from the analytical integration of the subtraction terms
\begin{align}
  \cTitotdiv &= \cTiindiv{4} + \cTiindiv{3} + \cTiindiv{2}  + \cTiindiv{1} + \cTioutdiv{4} +  \cTioutdiv{3} + \cTioutdiv{2}.
  \label{eqdivtotpart}
\end{align}
Note that the quantities on the right-hand side of eq. (\ref{eqdivtotpart}) are given by eq.~(\ref{eqdefI3480}), eq.~(\ref{eqdefI1id20}), 
eq.~(\ref{eqdefI12d10}), eq.~(\ref{eqdefI3413}) and eq.~(\ref{eqdefT2ijout2}), except $\cTiindiv{3}$ and $\cTioutdiv{3}$ 
which have not been computed explicitly but can be easily obtained from $\cTiindiv{2}$ and $\cTioutdiv{2}$ by changing the signs of all the rapidities and the labels $1 \leftrightarrow 2$. We will distinguish the soft part from the collinear ones. The soft part can be obtained from eq.~(\ref{eqdivtotpart}) by collecting all the terms containing the functions $f_{ij}(0,0,0)$, $f_{1j}(0,0,0)$ and $f_{2j}(0,0,0)$:
\begin{align}
  \cTitotsoftdiv &= 2^{-2 \, \varepsilon} \, \frac{\Gamma^2\left(\frac{1}{2}-\varepsilon\right)}{\Gamma(1-2 \, \varepsilon)}  \notag \\
  &\quad {} \times\left\{ \sum_{i=3}^{N-2} \sum_{j=i+1}^{N-1} f_{ij}(0,0,0) \, \left[ \frac{1}{\varepsilon^2} - \frac{1}{\varepsilon} \, \ln\left( \frac{2 \, p_i \cdot p_j}{Q^2} \right) + \ln^2(\xt{i}) +  \ln^2(\xt{j}) \right. \right. \notag \\
  &\qquad \qquad \qquad \qquad \qquad \qquad {} + \left. 2 \, \ln(\xtm) \, \ln(2 \, \bar d_{ij}) + \frac{1}{2} \, \ln^2(2 \, \bar d_{ij}) - 2 \, \left( \ystij \right)^2  \vphantom{\frac{\Gamma^2(1-\varepsilon)}{\Gamma(1 - 2 \, \varepsilon)}} \right] \notag \\
  &\qquad \quad {} + \sum_{j=3}^{N-1} f_{1j}(0,0,0) \, \left[ \frac{1}{\varepsilon^2} - \frac{1}{\varepsilon} \, \ln\left( \frac{2 \, p_1 \cdot p_j}{Q^2}  \right) + \Upsilon(\bx{1},y_j) + \ln^2(\xt{j}) \right] \notag \\
  &\qquad \quad {} + \sum_{j=3}^{N-1} f_{2j}(0,0,0) \, \left[ \frac{1}{\varepsilon^2} - \frac{1}{\varepsilon} \, \ln\left( \frac{2 \, p_2 \cdot p_j}{Q^2}  \right) + \Upsilon(\bx{2},-y_j) + \ln^2(\xt{j}) \right] \notag \\
  &\qquad \quad {} + \left.  f_{12}\left(0,0,0 \right) \, \left[ \frac{1}{\varepsilon^2} - \frac{1}{\varepsilon} \, \ln\left( \frac{2 \, p_1 \cdot p_2}{Q^2}  \right) + \Upsilon(\bx{1},y_0) + \Upsilon(\bx{2},-y_0) \right] \vphantom{\frac{f_{34} \left( y_3, \frac{1-z_3}{z_3} \, \pt{3}, 0 \right)}{(1-z_3)_{+}}} \right\}.
  \label{eqsofttotpart}
\end{align}
Let us introduce $\sigma_H^{\text{soft}}$ defined as 
\begin{align}
  \sigma_H^{\text{soft}} &\equiv \sum_{\seq{i}{N-1}\in S_p} \, \int d \text{PS}_{N-1 \,\text{h}}^{(n)}(x)  \, \cTitotsoftdiv.
  \label{eqdefhadcrxsoft0}
\end{align}
In order to facilitate the reading, only the case where $\xtm$ fulfils the condition (\ref{eqineq1}) will be presented\footnote{It is easy to have the formulae with the condition (\ref{eqineq2}).}.
Expanding around $\varepsilon=0$ the right hand side of eq.~(\ref{eqsofttotpart}) and keeping only the relevant terms in $\varepsilon$ leads to\footnote{Note that in the soft limit, the variables $x_k$ and $\bx{k}$ are equivalent.}
\begin{align}
  \sigma_H^{\text{soft}} &\equiv \sum_{\seq{i}{N-1} \in S_p} \, K^{(n) \, B}_{i_1 i_2} \, \left(\frac{4 \, \pi \, \mu^2}{Q^2}\right)^{\varepsilon} \, \frac{\alpha_s}{2 \, \pi} \, \frac{1}{\Gamma(1-\varepsilon)} \, \int d \text{PS}_{N-1 \,\text{h}}^{(n)}(\bar{x}) \,  \hphantom{\delta^{n-2}\left( \sum_{l=3}^{N-1} \frac{\vkt{l}}{x_l}\right)} \,\notag \\
  &\quad {} \times  A_{(i)_{N-1}}(\{\bar{x}\}_{N-1}) \,  \delta^{n-2}\left( \sum_{l=3}^{N-1} \frac{\vkt{l}}{\bx{l}}\right) \notag \\
  &\quad {} \times \left\{ \frac{1}{\varepsilon^2} \, \sum_{i=1}^{N-2} \sum_{j=i+1}^{N-1} \Hn{ij}(0) -  \frac{1}{\varepsilon} \, \sum_{i=1}^{N-2} \sum_{j=i+1}^{N-1} \Hn{ij}(0) \, \ln \left( \frac{2 \, p_i \cdot p_j}{Q^2} \right) \right. \notag \\
  &\qquad \quad {} + 2 \, \ln(\xtm) \, \sum_{i=1}^{N-2} \sum_{j=i+1}^{N-1} \Hn{ij}(0) \, \ln \left( \frac{2 \, p_i \cdot p_j}{Q^2} \right)  - 2 \, \ln^2(\xtm) \, \sum_{i=1}^{N-2} \sum_{j=i+1}^{N-1} \Hn{ij}(0) \notag \\
  &\qquad \quad {} + \sum_{i=3}^{N-1} \, \ln^2\left( \frac{\xt{i}}{\xtm} \right) \, \left[ \sum_{j=1}^{i-1} \Hn{ji}(0) + \sum_{j=i+1}^{N-1} \Hn{ij}(0) \right] \notag \\
  &\qquad \quad {} + \ln^2\left( \frac{\bx{1}}{1 - \bx{1}} \, \right) \, \sum_{j=2}^{N-1} \Hn{1j}(0) +  \ln^2\left( \frac{\bx{2}}{1 - \bx{2}} \, \right) \, \left[\Hn{12}(0) + \sum_{j=3}^{N-1} \Hn{2j}(0)\right] \notag \\
  &\qquad \quad {} + \left. \sum_{i=3}^{N-2} \sum_{j=i+1}^{N-1} \Hn{ij}(0) \, \left[ \frac{1}{2} \, \ln^2(2 \, \bd_{ij}) - 2 \, \left( \ystij \right)^2 \right] \right\}.
  \label{eqdefhadcrxsoft1}
\end{align}
The final state collinear singularity related to hard a parton $i_k$ coming from inside the cylinder is given by
\begin{align}
  \calt^{\text{ fin coll}}_{k \, \text{in}}  &= -2^{-2 \, \varepsilon} \, \frac{\Gamma^2\left(\frac{1}{2}-\varepsilon\right)}{\Gamma(1-2 \, \varepsilon)} \, \frac{\xt{k}^{- 2 \, \varepsilon}}{\varepsilon} \, \left\{ \int_{\zm{k}}^{1} \frac{d v_k}{v_k} \, v_k^{2 \, \varepsilon} \, \left[ \frac{1}{(1-v_k)_{+}} - 2 \, \varepsilon \, \left(\frac{\ln(1-v_k)}{1-v_k}\right)_{+} \right] \, \right.  \notag \\
  &\qquad \qquad \qquad \qquad \qquad \qquad {} \times \left[ \calf_{k}\left(y_k,\frac{1-v_k}{v_k} \, \xt{k},0\right) +f_{1k}\left(y_k,\frac{1-v_k}{v_k} \, \xt{k},0\right) \right. \notag \\
  &\qquad \qquad \qquad \qquad \qquad \qquad \qquad {} +  \left. \left. f_{2k}\left(y_k,\frac{1-v_k}{v_k} \, \xt{k},0\right) \right]  \right\},
  \label{eqcollfinpart0}
\end{align}
and the one coming from outside the cylinder is
\begin{align}
  \calt^{\text{ fin coll}}_{k \, \text{out}} &= -2^{-2 \, \varepsilon} \, \frac{\Gamma^2\left(\frac{1}{2}-\varepsilon\right)}{\Gamma(1-2 \, \varepsilon)} \, \frac{\xt{k}^{- 2 \, \varepsilon} \, \rth^{-2 \, \varepsilon}}{\varepsilon} \, \left\{ \int_{\zmin{k}}^{\zm{k}} \frac{d v_k}{v_k} \, v_k^{2 \, \varepsilon} \, \left[ \frac{1}{(1-v_k)_{+}} - 2 \, \varepsilon \, \left(\frac{\ln(1-v_k)}{1-v_k}\right)_{+} \right] \, \right.  \notag \\
  &\qquad \qquad \qquad \qquad \qquad \qquad {} \times \left[ \calf_{k}\left(y_k,\frac{1-v_k}{v_k} \, \xt{k},0\right) +f_{1k}\left(y_k,\frac{1-v_k}{v_k} \, \xt{k},0\right) \right. \notag \\
  &\qquad \qquad \qquad \qquad \qquad \qquad \qquad {} +  \left. \left. f_{2k}\left(y_k,\frac{1-v_k}{v_k} \, \xt{k},0\right) \right]  \right\}.
  \label{eqcollfinpart1}
\end{align}
In eqs.~(\ref{eqcollfinpart0}) and (\ref{eqcollfinpart1}), we changed the name of the integration variable $z_k$ into $v_k$ for reasons which will become clear later on.
The corresponding cross section $\sigma^{\text{fin coll}}_{H \, k}$ is
\begin{align}
  \sigma^{\text{fin coll}}_{H \, k} &\equiv \sum_{\seq{i}{N-1} \in S_p} \, \int d \text{PS}_{N-1 \,\text{h}}^{(n)}(x)  \, \left[ \calt^{\text{ fin coll}}_{k \, \text{in}} + \calt^{\text{ fin coll}}_{k \, \text{out}} \right].
  \label{eqdefhadcrxfincoll0}
\end{align}
Remembering the definition of the functions $f_{ij}$, it is easy to realise that the Dirac distributions coming from the conservation of the transverse momentum will depend on $v_k$ through the product $v_k \, x_k$ suggesting that by a suitable change of variables 
these constraints could be independent of the variable of the "+" distributions.
Let us sketch the structure of eq.~(\ref{eqdefhadcrxfincoll0}) in terms of the variables $x_k$ and $v_k$. It is of the type
\begin{align}
  \int \frac{d x_k}{x_k^{2-2 \, \varepsilon}} \, \int_{\zmin{k}}^1 \frac{d v_k}{v_k} \, v_k^{2 \, \varepsilon} \, G(x_k,v_k),
\end{align}
and the change of variables $v_k = z_k$ and $x_k = \bx{k}/z_k$ whose Jacobian is $1/z_k$ leads to the following structure
\begin{align}
  \int \frac{d \bx{k}}{\bx{k}^{2-2 \, \varepsilon}} \, \int_{\bx{k}}^1 \, d z_k \; G\left(\frac{\bx{k}}{z_k},z_k\right).
\end{align}
Up to vanishing terms when $\varepsilon \rightarrow 0$, the cross section $\sigma^{\text{fin coll}}_{H \, k}$ can be written as
\begin{align}
  \hspace{2em}&\hspace{-2em} \sigma^{\text{fin coll}}_{H \, k} = \sum_{\seq{i}{N-1} \in S_p}  K^{(n) \, B}_{i_1 i_2} \, \left(\frac{4 \, \pi \, \mu^2}{Q^2}\right)^{\varepsilon} \, \frac{\alpha_s}{2 \, \pi} \, \frac{1}{\Gamma(1-\varepsilon)} \, \int d \text{PS}_{N-1 \,\text{h}}^{(n)}(\bar{x}) \, \left(\frac{\Xt{k}}{\bx{k}}\right)^{-2 \, \varepsilon} \notag \\
  &\quad {} \times  \delta^{n-2}\left( \sum_{l=3}^{N-1} \frac{\vkt{l}}{\bx{l}}\right) \left\{ -\frac{1}{\varepsilon} \, \int_{\bx{k}}^{1} \frac{dz_k}{z_k^{1+2 \, \varepsilon}} \, A_{(i)_{N-1}}\left(\seqt{\bar{x}}{\bx{k}}{\frac{\bx{k}}{z_k}}{N-1}\right) \, \right.  \notag \\
  &\quad {} \times \left[  \frac{1}{(1-z_k)_{+}} - 2 \, \varepsilon \, \left(\frac{\ln(1-z_k)}{1-z_k}\right)_{+} \right] \, z_k \,\Xi_k\left((1-z_k) \, \frac{\kt{k}}{\bx{k}}\right)+ \ln \left( \rth^2 \right) \notag \\
  &\quad {} \times \int_{\bx{k}}^{\czm{k}} \frac{dz_k}{z_k^{1+2 \, \varepsilon}} \, A_{(i)_{N-1}}\left(\seqt{\bar{x}}{\bx{k}}{\frac{\bx{k}}{z_k}}{N-1}\right)  \,   \, \frac{1}{(1-z_k)_{+}}  \left. z_k \, \Xi_k\left((1-z_k) \, \frac{\kt{k}}{\bx{k}}\right) \right\},
  \label{eqdefhadcrxfincoll1}
\end{align}
with
\begin{align}
  \Xi_k\left((1-z_k) \, \frac{\kt{k}}{\bx{k}}\right) &= \sum_{j=k+1}^{N-1} \Hn{kj}\left((1-z_k) \, \frac{\kt{k}}{\bx{k}}\right) + \sum_{j=3}^{k-1} \Hn{jk}\left((1-z_k) \, \frac{\kt{k}}{\bx{k}}\right) \notag \\
  &\quad {} +  \Hn{1k}\left((1-z_k) \, \frac{\kt{k}}{\bx{k}}\right) + \Hn{2k}\left((1-z_k) \, \frac{\kt{k}}{\bx{k}}\right).\,
  \label{eqdefXi00}
\end{align}
and $\czm{k}$ given by eq.~(\ref{eqdefczm}).\\

\noindent
The initial state collinear contribution can be split into two parts, a divergent one and a finite one in which the regulator $\varepsilon$ is set to zero:
\begin{align}
  \calt^{\text{ ini coll}}_{\text{div}} &= -2^{-2 \, \varepsilon} \, \frac{\Gamma^2 \left( \frac{1}{2}-\varepsilon \right)}{\Gamma(1 - 2 \, \varepsilon)} \, \frac{1}{\varepsilon} \left\{  \sum_{j=3}^{N-1} \chi(\bx{1},y_j)^{- 2 \, \varepsilon} \, \int_{\bx{1}}^{1} \frac{dz_1}{z_1} \, z_1^{2 \, \varepsilon} \, \frac{f_{1j}^c \left( \frac{\bx{1} \, e^{-y_j}}{\omega} \, \frac{1-z_1}{z_1} \right)}{(1-z_1)_{+}} \right. \notag \\
  &\; \; \; \left. {} + \chi(\bx{1},y_0)^{- 2 \, \varepsilon} \, \int_{\bx{1}}^{1} \frac{dz_1}{z_1} \, z_1^{2 \, \varepsilon} \, \frac{f_{12}^{(1) \, c} \left( \frac{\bx{1} \, e^{-y_0}}{\omega} \, \frac{1-z_1}{z_1} \right)}{(1-z_1)_{+}} + 1 \leftrightarrow 2  \right\},
  \label{eqdefTdinicolld0}
\end{align}
\begin{align}
  \calt^{\text{ ini coll}}_{\text{fin}} &=  2 \, \pi  \,  \left\{  \int_{\bx{1}}^{1} \frac{d z_1}{z_1} \, \left(\frac{\ln(1-z_1)}{1-z_1}\right)_{+} \, f_{12}^{(1) \, c} \left( \frac{\bx{1} \, e^{-y_0}}{\omega} \, \frac{1-z_1}{z_1} \right) \right. \notag \\
  &\qquad  {}\left. + \sum_{j=3}^{N-1}  \int_{\bx{1}}^{1} \frac{d z_1}{z_1} \, \left(\frac{\ln(1-z_1)}{1-z_1}\right)_{+} \, f_{1j}^c \left( \frac{\bx{1} \, e^{-y_j}}{\omega} \, \frac{1-z_1}{z_1} \right) + 1 \leftrightarrow 2  \right\}.
  \label{eqdefTdinicollf0}
\end{align}
The associated cross section is then
\begin{align}
  \hspace{2em}&\hspace{-2em} \sigma^{\text{ini coll}}_{H} = \sum_{\seq{i}{N-1} \in S_p}  K^{(n) \, B}_{i_1 i_2} \, \left(\frac{4 \, \pi \, \mu^2}{Q^2}\right)^{\varepsilon} \, \frac{\alpha_s}{2 \, \pi} \, \frac{1}{\Gamma(1-\varepsilon)} \,\frac{1}{-\varepsilon} \,  \int d \text{PS}_{N-1 \,\text{h}}^{(n)}(\bar{x}) \, \notag \\
  &\quad {} \times   \,  \delta^{n-2}\left( \sum_{l=3}^{N-1} \frac{\vkt{l}}{\bx{l}}\right) \, \left\{  \int_{\bx{1}}^{1} \frac{dz_1}{z_1^2 \, (1-z_1)_{+}} \, A_{(i)_{N-1}}\left(\seqt{\bar{x}}{\bx{1}}{\frac{\bx{1}}{z_1}}{N-1}\right)  \right. \notag \\
  &\qquad {} \times z_1^{1+2 \, \varepsilon} \, \left( \frac{\bx{1} \, \xtm}{1-\bx{1}} \right)^{-2 \, \varepsilon} \, \left[ \Hn{12}((1-z_1) \, p_1) + \sum_{l=3}^{N-1} \Hn{1l}((1-z_1) \, p_1) \right] \notag \\
  &\quad {} -2\varepsilon \, \int_{\bx{1}}^{1} \frac{dz_1}{z_1^2} \, \left( \frac{\ln(1-z_1)}{1-z_1} \right)_{+} \, \, A_{(i)_{N-1}}\left(\seqt{\bar{x}}{\bx{1}}{\frac{\bx{1}}{z_1}}{N-1}\right) \notag \\
  &\qquad {} \times z_1 \, \left[ \Hn{12}((1-z_1) \, p_1) + \sum_{l=3}^{N-1}  \Hn{1l}((1-z_1) \, p_1) \right] \notag \\
  &\qquad {} + \left. 1 \leftrightarrow 2 \vphantom{\int_{\bx{1}}^{1} \frac{dz_1}{z_1^2 \, (1-z_1)_{+}}} \right\}.
  \label{eqdefhadcrxinicoll1}
\end{align}
\\

\noindent
The relations given by eqs.~(\ref{eqrelini1}), (\ref{eqrelini2}) and (\ref{eqrelfink}) must be fulfilled in order to absorb the collinear divergences into a redefinition of the partonic density functions as well as the fragmentation functions. 
They are obtained by comparing the structure of the collinear divergences in the initial and final states, as derived 
in sec.~\ref{secLO} (see eq.~(\ref{eqdefhadcrxg4})), with the results 
delineated form the subtraction terms after the integration over the phase space of the soft/collinear 
parton, cf.~eqs.~(\ref{eqdefhadcrxfincoll1}) and (\ref{eqdefhadcrxinicoll1}).
When the variables $z_l$ ($l \in {1,2,\ldots,N-1}$) approach one, according to the definition of the functions $a^{(n)}_{ij}(z)$, only the diagonal cases survive, namely when $i=j$ and the relations (\ref{eqrelini1}), (\ref{eqrelini2}) and (\ref{eqrelfink}) become
\begin{align}
   \Hn{12}(0) + \sum_{l=3}^{N-1} \Hn{1l}(0) &= a^{(n)}_{i_1 i_1}(1) \, |M^n_{[i]_{N-1}}|^2, \label{eqrelini11} \\
   \Hn{12}(0) + \sum_{l=3}^{N-1} \Hn{2l}(0) &= a^{(n)}_{i_2 i_2}(1) \, |M^n_{[i]_{N-1}}|^2, \label{eqrelini21} \\
  \Xi_k\left(0\right) &= a^{(n)}_{i_k i_k}(1) \, |M^n_{[i]_{N-1}}|^2 \qquad \text{for each $k$ in $S_f$}. \label{eqrelfink1}
\end{align}
With these relations in hand, eqs.~(\ref{eqdefhadcrxinicoll1}) and (\ref{eqdefhadcrxfincoll1}) can be expressed in terms of DGLAP kernels. 
To achieve this, we enforce the appearance of the quantity $\calh$ for initial state and final state radiations, utilizing 
eqs.~(\ref{eqdefHij}) and (\ref{eqdefapndim0}), and expanding the rest around $\varepsilon=0$. 
This leads to the eq.~(\ref{eqdefhadcrxinicoll3}) presented in sec.~\ref{divterm}.
The terms containing the final state collinear divergences can be treated in the same manner, yielding eq.~(\ref{eqdefhadcrxfincoll3}).\\

\noindent
Note that in order to enforce the appearance of a DGLAP kernel, we must include additional soft terms that are not explicitly 
expressed in eqs.~(\ref{eqdefhadcrxinicoll3}) and (\ref{eqdefhadcrxfincoll3}). These terms will be incorporated into the soft part
which now reads
\begin{align}
  \sigma_H^{\text{soft}} &\equiv \sum_{\seq{i}{N-1} \in S_p} \, K^{(n) \, B}_{i_1 i_2} \, \left(\frac{4 \, \pi \, \mu^2}{Q^2}\right)^{\varepsilon} \, \frac{\alpha_s}{2 \, \pi} \, \frac{1}{\Gamma(1-\varepsilon)} \, \int d \text{PS}_{N-1 \,\text{h}}^{(n)}(\bar{x}) \,\notag \\
  &\quad {} \times  A_{(i)_{N-1}}(\{\bar{x}\}_{N-1}) \,  \delta^{n-2}\left( \sum_{l=3}^{N-1} \frac{\vkt{l}}{\bx{l}}\right) \notag \\
  &\quad {} \times \left\{ \frac{1}{\varepsilon^2} \, \sum_{i=1}^{N-2} \sum_{j=i+1}^{N-1} \Hn{ij}(0) -  \frac{1}{\varepsilon} \, \sum_{i=1}^{N-2} \sum_{j=i+1}^{N-1} \Hn{ij}(0) \, \ln \left( \frac{2 \, p_i \cdot p_j}{Q^2} \right) \right. \notag \\
  &\qquad \quad {} + \frac{1}{\varepsilon} \, \left[ \, \sum_{k=1}^{2} b_{i_k i_k} \, \left( \frac{M^2}{Q^2} \right)^{-\varepsilon}  + \, \sum_{k=3}^{N-1} b_{i_k i_k} \,  \left( \frac{M_f^2}{Q^2} \right)^{-\varepsilon} \; \right] \, |M^{(n)}_{[i]_{N-1}}|^{2} \notag \\
  &\qquad \quad {} + \left. \text{finite terms of eq.~(\ref{eqdefhadcrxsoft1})} \vphantom{\frac{1}{\varepsilon^2} \, \sum_{i=1}^{N-2} \sum_{j=i+1}^{N-1} H_{ij}(0)} \right\}.
  \label{eqdefhadcrxsoft2}
\end{align}
The soft divergences cancel out with the real emission due to the relations (\ref{eqrels3}), which are valid in $n$ dimensions. Following this cancellation, some finite terms remain, as shown in equation (\ref{eqdefhadcrxsoft3}).
Note that by using eqs.~(\ref{eqrelini11}), (\ref{eqrelini21}) and (\ref{eqrelfink1}), we get that
\begin{align}
  \left[ a^{(n)}_{i_1 i_1}(1) + a^{(n)}_{i_2 i_2}(1) \right] \, |M^{(n)}_{[i]_{N-1}}|^{2} &= \left[ 2 \, \Hn{12}(0) + \sum_{k=3}^{N-1} \left(  \Hn{1k}(0) + \Hn{2k}(0)  \right) \right], \label{eqrelso21} \\
  \left[ \sum_{k=3}^{N-1} a^{(n)}_{i_k i_k}(1) \right] \, |M^{(n)}_{[i]_{N-1}}|^{2} &= \sum_{i=3}^{N-2} \left[ \sum_{j=1}^{i-1} \Hn{ji}(0) + \sum_{j=i+1}^{N-1} \Hn{ij}(0) \right], \label{eqrelso31} \\
  \cala^{(n)} &= \frac{1}{2} \, \sum_{k=1}^{N-1} a^{(n)}_{i_k i_k}(1). \label{eqrelso11}
\end{align}

\section{A non trivial example to illustrate the Born colour connected amplitudes}\label{qqbgg}

In this appendix, we compute explicitly the soft approximation of the reaction $q(p_1) + \bar{q}(p_2) \rightarrow g(p_3) + g(p_4) (+ g(p_5))$, that is to say the coefficient of the eikonal factors taken at $p_5 =0$. We compare them to the result of the one loop correction of $q(p_1) + \bar{q}(p_2) \rightarrow g(p_3) + g(p_4)$. This gives an example of the relations among the coefficients of the eikonal factors in order to have a cancellation of the IR divergences.   This reaction is taken because it contains both external fermions and gluons.\\

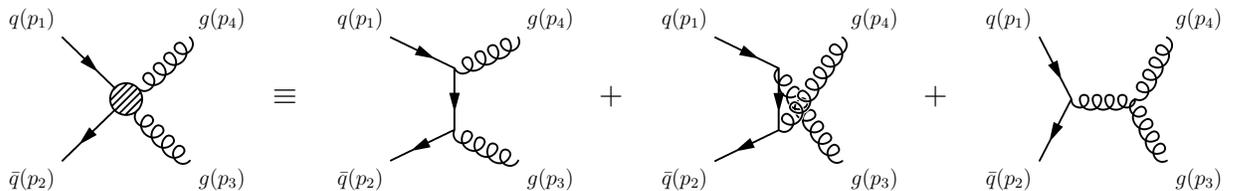
\begin{figure}[h!]
\[
\resizebox{2.15cm}{!}{
\parbox{30mm}{\begin{fmfgraph*}(31,24)
  \fmfleftn{i}{2} \fmfrightn{o}{2}
  \fmflabel{$q(p_1)$}{i2}
  \fmflabel{$\bar{q}(p_2)$}{i1}
  \fmflabel{$g(p_4)$}{o2}
  \fmflabel{$g(p_3)$}{o1}
  \fmf{fermion}{v1,i1}
  \fmf{fermion}{i2,v1}
  \fmf{gluon}{v1,o1}
  \fmf{gluon}{v1,o2}
  \fmfblob{0.20w}{v1}
\end{fmfgraph*} }}
\qquad \equiv \qquad
\resizebox{2.15cm}{!}{
\parbox{30mm}{\begin{fmfgraph*}(31,24)
  \fmfleftn{i}{2} \fmfrightn{o}{2}
  \fmflabel{$q(p_1)$}{i2}
  \fmflabel{$\bar{q}(p_2)$}{i1}
  \fmflabel{$g(p_4)$}{o2}
  \fmflabel{$g(p_3)$}{o1}
  \fmf{fermion}{i2,v2}
  \fmf{fermion}{v2,v1}
  \fmf{fermion}{v1,i1}
  \fmf{gluon}{v1,o1}
  \fmf{gluon}{v2,o2}
\end{fmfgraph*} }}
\qquad + \qquad
\resizebox{2.15cm}{!}{
\parbox{30mm}{\begin{fmfgraph*}(31,24)
  \fmfleftn{i}{2} \fmfrightn{o}{2}
  \fmflabel{$q(p_1)$}{i2}
  \fmflabel{$\bar{q}(p_2)$}{i1}
  \fmflabel{$g(p_4)$}{o2}
  \fmflabel{$g(p_3)$}{o1}
  \fmf{fermion}{i2,v2}
  \fmf{fermion}{v2,v1}
  \fmf{fermion}{v1,i1}
  \fmf{phantom}{v1,o1}
  \fmf{phantom}{v2,o2}
  \fmffreeze
  \fmf{gluon}{v2,o1}
  \fmf{gluon,rubout}{v1,o2}
\end{fmfgraph*} }} 
\qquad {} +\qquad
\resizebox{2.15cm}{!}{
\parbox{30mm}{\begin{fmfgraph*}(31,24)
  \fmfleftn{i}{2} \fmfrightn{o}{2}
  \fmflabel{$q(p_1)$}{i2}
  \fmflabel{$\bar{q}(p_2)$}{i1}
  \fmflabel{$g(p_4)$}{o2}
  \fmflabel{$g(p_3)$}{o1}
  \fmf{fermion}{i2,v2}
  \fmf{gluon}{v2,v1}
  \fmf{fermion}{v2,i1}
  \fmf{gluon}{v1,o1}
  \fmf{gluon}{v1,o2}
\end{fmfgraph*} }}
\]
\caption{\small Born Feynman diagrams of the reaction $q + \bar q\rightarrow g + g$.}
\label{graphborn}
\end{figure}

\noindent
The Born amplitude is depicted in fig.~\ref{graphborn}.
At the Born level, the sum of the three diagrams is represented by the product of a string of $\gamma$ matrices carrying two Lorentz indices as well as string of colour matrices carrying colour indices $\Gamma^{\mu_3 \mu_4}_{i_2 i_1 ; a_3 a_4}$ 
multiplied by the Dirac spinors and the polarisation vectors describing the external particles.

\noindent
The Born amplitude can be written as
\begin{align}
  \mathcal{M}^B_{q\bar{q}\rightarrow 2g} &= \bar{v}_{i_2}(p_2) \, \Gamma^{\mu_3 \mu_4}_{i_2 i_1 ; a_3 a_4} \, u_{i_1}(p_1) \, \varepsilon_{\mu_3}^{a_3}(p_3) \, \varepsilon_{\mu_4}^{a_4}(p_4). 
  \label{eqdefmbc}
\end{align}

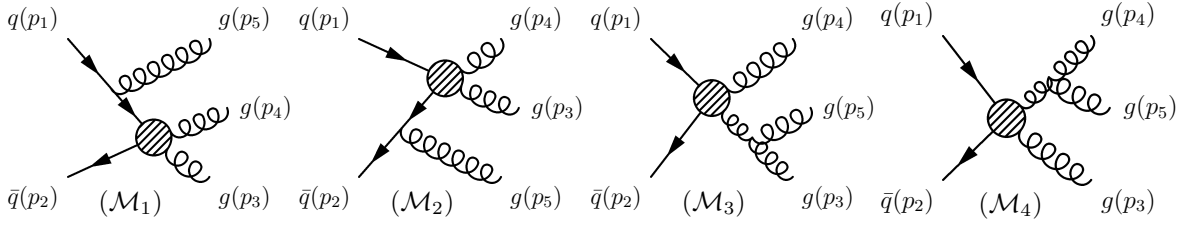
\begin{figure}[h!]
\[
\subfloat[${(\mathcal{M}_1)}$]{
\resizebox{2.5cm}{!}
{
\parbox{30mm}{\begin{fmfgraph*}(31,24)
  \fmfleftn{i}{2} \fmfrightn{o}{3}
  \fmflabel{$q(p_1)$}{i2}
  \fmflabel{$\bar{q}(p_2)$}{i1}
  \fmflabel{$g(p_5)$}{o3}
  \fmflabel{$g(p_4)$}{o2}
  \fmflabel{$g(p_3)$}{o1}
  \fmf{fermion}{v1,i1}
  \fmf{fermion}{i2,v3,v1}
  \fmf{gluon}{v1,o1}
  \fmf{gluon}{v1,o2}
  \fmffreeze
  \fmf{gluon}{v3,o3}
  \fmfblob{0.20w}{v1}
\end{fmfgraph*} }
}}
\qquad\quad
\subfloat[${(\mathcal{M}_2)}$]{
\resizebox{2.5cm}{!}
{
\parbox{30mm}{\begin{fmfgraph*}(31,24)
  \fmfleftn{i}{2} \fmfrightn{o}{3}
  \fmflabel{$q(p_1)$}{i2}
  \fmflabel{$\bar{q}(p_2)$}{i1}
  \fmflabel{$g(p_4)$}{o3}
  \fmflabel{$g(p_3)$}{o2}
  \fmflabel{$g(p_5)$}{o1}
  \fmf{fermion}{v1,v3,i1}
  \fmf{fermion}{i2,v1}
  \fmf{gluon}{v1,o2}
  \fmf{gluon}{v1,o3}
  \fmffreeze
  \fmf{gluon}{v3,o1}
  \fmfblob{0.20w}{v1}
\end{fmfgraph*} }
}}
\qquad\quad
\subfloat[${(\mathcal{M}_3)}$]{
\resizebox{2.5cm}{!}
{
\parbox{30mm}{\begin{fmfgraph*}(31,24)
  \fmfleftn{i}{2} \fmfrightn{o}{3}
  \fmflabel{$q(p_1)$}{i2}
  \fmflabel{$\bar{q}(p_2)$}{i1}
  \fmflabel{$g(p_4)$}{o3}
  \fmflabel{$g(p_5)$}{o2}
  \fmflabel{$g(p_3)$}{o1}
  \fmf{fermion}{v1,i1}
  \fmf{fermion}{i2,v1}
  \fmf{gluon}{v1,v3,o1}
  \fmf{gluon}{v1,o3}
  \fmffreeze
  \fmf{gluon}{v3,o2}
  \fmfblob{0.20w}{v1}
\end{fmfgraph*} }
}}
\qquad\quad
\subfloat[${(\mathcal{M}_4)}$]{
\resizebox{2.5cm}{!}
{
\parbox{30mm}{\begin{fmfgraph*}(31,24)
  \fmfleftn{i}{2} \fmfrightn{o}{3}
  \fmflabel{$q(p_1)$}{i2}
  \fmflabel{$\bar{q}(p_2)$}{i1}
  \fmflabel{$g(p_4)$}{o3}
  \fmflabel{$g(p_5)$}{o2}
  \fmflabel{$g(p_3)$}{o1}
  \fmf{fermion}{v1,i1}
  \fmf{fermion}{i2,v1}
  \fmf{gluon}{v1,o1}
  \fmf{gluon}{v1,v3,o3}
  \fmffreeze
  \fmf{gluon}{v3,o2}
  \fmfblob{0.20w}{v1}
\end{fmfgraph*} }}
}
\]
\caption{\small {Feynman diagrams depicting the soft gluon emission associated to $q+\bar{q}\rightarrow g+g$.}}
\label{reals}
\end{figure}

\noindent
In the soft approximation, it is necessary to consider only the emission of an extra gluon on the external legs as depicted in fig. \ref{reals}.
\captionsetup[subfigure]{labelformat=empty}

\noindent
The different amplitudes are
\begin{align}
  \mathcal{M}_1 &= -g_s \, \frac{p_1^{\mu_5}}{p_1 \cdot p_5} \, \bar{v}_{i_2}(p_2) \, \Gamma^{\mu_3 \mu_4}_{i_2 j_1 ; a_3 a_4} \, \left( T^{a_5} \right)_{j_1 i_1} \, u_{i_1}(p_1) \, \varepsilon_{\mu_3}^{a_3}(p_3) \, \varepsilon_{\mu_4}^{a_4}(p_4) \, \varepsilon_{\mu_5}^{a_5}(p_5),
  \label{eqdefM1}\\
  \mathcal{M}_2 &= g_s \, \frac{p_2^{\mu_5}}{p_2 \cdot p_5} \, \bar{v}_{i_2}(p_2) \, \Gamma^{\mu_3 \mu_4}_{j_2 i_1 ; a_3 a_4} \, \left( T^{a_5} \right)_{i_2 j_2} \, u_{i_1}(p_1) \, \varepsilon_{\mu_3}^{a_3}(p_3) \, \varepsilon_{\mu_4}^{a_4}(p_4) \, \varepsilon_{\mu_5}^{a_5}(p_5),
  \label{eqdefM2}\\
  \mathcal{M}_3 &= -g_s \, \frac{p_3^{\mu_5}}{p_3 \cdot p_5} \, \bar{v}_{i_2}(p_2) \, \Gamma^{\mu_3 \mu_4}_{i_2 i_1 ; b_3 a_4} \, i \, 
  f^{b_3 a_5 a_3} \, u_{i_1}(p_1) \, \varepsilon_{\mu_3}^{a_3}(p_3) \, \varepsilon_{\mu_4}^{a_4}(p_4) \, \varepsilon_{\mu_5}^{a_5}(p_5),
  \label{eqdefM3}\\
  \mathcal{M}_4 &= -g_s\, \frac{p_4^{\mu_5}}{p_4 \cdot p_5} \, \bar{v}_{i_2}(p_2) \, \Gamma^{\mu_3 \mu_4}_{i_2 i_1 ; a_3 b_4} \, i \, 
  f^{b_4 a_5 a_4} \, u_{i_1}(p_1) \, \varepsilon_{\mu_3}^{a_3}(p_3) \, \varepsilon_{\mu_4}^{a_4}(p_4) \, \varepsilon_{\mu_5}^{a_5}(p_5).
  \label{eqdefM4}
\end{align}

\noindent
In the Born amplitude, using colour decomposition, the string $\Gamma^{\mu_3 \mu_4}_{i_2 i_1 ; a_3 a_4}$ can be written as
\begin{align}
  \Gamma^{\mu_3 \mu_4}_{i_2 i_1 ; a_3 a_4} &= \left( T^{a_4} \, T^{a_3} \right)_{i_2 i_1} \, A^{\mu_3 \mu_4}(p_3,p_4) + \left( T^{a_3} \, T^{a_4} \right)_{i_2 i_1} \, A^{\mu_4 \mu_3}(p_4,p_3),
  \label{eqdefGamma340}
\end{align}
with
\begin{align}
  A^{\mu_3 \mu_4}(p_3,p_4) &= \left( -i \, g_s^2 \right) \, \left[ \gamma^{\mu_4} \, \frac{\psla_1 - \psla_3}{(p_1-p_3)^2} \, \gamma^{\mu_3} - \gamma^{\rho} \, \frac{V^{\mu_3 \mu_4 \rho}(-p_3,-p_4,p_3+p_4)}{(p_1+p_2)^2} \right]. \label{eqdefA}
\end{align}
The tensor $V^{\mu \nu \rho}(p,q,r)$ is the usual tensor appearing in the QCD Feynman rules for the triple gluon vertex
\begin{equation}
  V^{\mu \nu \rho}(p,q,r) = g^{\mu \, \nu} \, (p-q)^{\rho} + g^{\nu \, \rho} \, (q-r)^{\mu} + g^{\rho \, \mu} \, (r-p)^{\nu}. 
  \label{eqdefV}
\end{equation}

\noindent
Let us define the soft amplitude $\mathcal{M}$ as $\mathcal{M} = \mathcal{M}_1 + \mathcal{M}_2 + \mathcal{M}_3 + \mathcal{M}_4$. Because of the gauge invariance,
replacing $\varepsilon^{\mu_5}(p_5)$ by $p_5^{\mu_5}$ in the soft amplitude leads to the following relation
\begin{align}
  \Gamma^{\mu_3 \mu_4}_{i_2 j_1 ; a_3 a_4} \, \Tmat{a_5}{j_1}{i_1} - \Gamma^{\mu_3 \mu_4}_{j_2 i_1 ; a_3 a_4} \, \Tmat{a_5}{i_2}{j_2}
  + \Gamma^{\mu_3 \mu_4}_{i_2 i_1 ; b_3 a_4} \, i \, f^{b_3 a_5 a_3} + \Gamma^{\mu_3 \mu_4}_{i_2 i_1 ; a_3 b_4} \, i \, f^{b_4 a_5 a_4} &= 0.
  \label{eqrelat0}
\end{align}
This relation can be checked explicitly by injecting eq.~(\ref{eqdefGamma340}) into eq.~(\ref{eqrelat0}) and by using $i \, f^{a b c} \, T^a = [T^b,T^c]$. The decomposition given by eq.~(\ref{eqdefGamma340}) yields for the soft amplitude
\begin{align}
  \mathcal{M} &= -g_s \,  \bar{v}_{i_2}(p_2) \, \left\{ \left( T^{a_4} \, T^{a_3} \, T^{a_5} \right)_{i_2 i_1} \, \left[ \frac{p_1^{\mu_5}}{p_1 \cdot p_5} - \frac{p_3^{\mu_5}}{p_3 \cdot p_5} \right] \, A^{\mu_3 \mu_4}(p_3,p_4) \right. \notag \\
  &\qquad \qquad \qquad \quad {} + \left( T^{a_5} \, T^{a_4} \, T^{a_3} \right)_{i_2 i_1} \, \left[ \frac{p_4^{\mu_5}}{p_4 \cdot p_5} - \frac{p_2^{\mu_5}}{p_2 \cdot p_5} \right] \, A^{\mu_3 \mu_4}(p_3,p_4) \notag \\
  &\qquad \qquad \qquad \quad {} + \left( T^{a_4} \, T^{a_5} \, T^{a_3} \right)_{i_2 i_1} \, \left[ \frac{p_3^{\mu_5}}{p_3 \cdot p_5} - \frac{p_4^{\mu_5}}{p_4 \cdot p_5} \right] \, A^{\mu_3 \mu_4}(p_3,p_4) \notag \\
  &\qquad \qquad \qquad \quad {} + \left. 3 \leftrightarrow 4 \vphantom{ \left( T^{a_4} \, T^{a_3} \, T^{a_5} \right)_{i_2 i_1} \, \left[ \frac{p_1^{\mu_5}}{p_1 \cdot p_5} - \frac{p_3^{\mu_5}}{p_3 \cdot p_5} \right]} \right\} \,  \, u_{i_1}(p_1) \, \varepsilon_{\mu_3}^{a_3}(p_3) \, \varepsilon_{\mu_4}^{a_4}(p_4) \, \varepsilon_{\mu_5}^{a_5}(p_5).
  \label{eqdefsoftamp0}
\end{align}
Under this form, the invariance $\varepsilon^{\mu_5}(p_5) \rightarrow \varepsilon^{\mu_5}(p_5) + \lambda \, p_5^{\mu_5}$ is explicit.\\

\noindent
Squaring the soft amplitude generates terms which are the product of a trace on the colour matrices times a trace on the $\gamma$ matrices times an eikonal factor $E_{ij} \equiv p_i \cdot p_j/(p_i \cdot p_5 \, p_j \cdot p_5)$. There are four independent types of colour matrices which can be easily evaluated. We want to enforce the appearance of the Casimirs of the fundamental and the adjoint representations of the $SU(N_c)$ Lie group (resp. $C_F$ and $C_A$) in the colour factors for reasons that will be clear later on. To do so, we make use of 
\begin{align}
(T^a\, T^a)_{ij}&=C_F\, \delta_{ij},&
f_{abc}\, f_{dbc}&= C_A\, \delta_{ad},
\label{forms}
\end{align}
where $C_F=(N_c^2-1)/(2N_c)$ and $C_A=N_c$. We get
\begin{align}
  \text{Tr} \left[ T^{a_4} \, T^{a_3} \, T^{a_5} \, T^{a_5} \, T^{a_3} \, T^{a_4} \right] &= C_F^3\, N_c, \label{eqtracecol1} \\
  \text{Tr} \left[ T^{a_4} \, T^{a_3} \, T^{a_5} \, T^{a_5} \, T^{a_4} \, T^{a_3} \right] &= C_F^2\, N_c\, \left(C_F-\frac{C_A}{2}\right), \label{eqtracecol2} \\
  \text{Tr} \left[ T^{a_4} \, T^{a_3} \, T^{a_5} \, T^{a_4} \, T^{a_3} \, T^{a_5} \right] &= \frac{1}{2}\, C_F\, N_c\, (C_A-C_F)\, (C_A-2\, C_F),\label{eqtracecol3} \\
  \text{Tr} \left[ T^{a_4} \, T^{a_3} \, T^{a_5} \, T^{a_3} \, T^{a_4} \, T^{a_5} \right] &= \frac{1}{4}\, C_F\, N_c\, (C_A-2\, C_F)^2.
\label{eqtracecol4}
\end{align}
\noindent
Concerning the traces of the $\gamma$ matrices, there is no new trace to compute. It is rather clear from eqs.~(\ref{eqdefmbc}), (\ref{eqdefM1}), (\ref{eqdefM2}), (\ref{eqdefM3}), and (\ref{eqdefM4}) that they will be the same as those appearing in the computation of the Born amplitude squared, namely
\begin{align}
  \text{Tr} \left[ \psla_2 \, A^{\mu_3 \mu_4}(p_3,p_4) \, \psla_1 \, A^{\dagger \,\nu_3 \nu_4}(p_3,p_4) \right] \, \Pi_{\mu_3 \nu_3}(p_3,q_3) \, \Pi_{\mu_4 \nu_4}(p_4,q_4) &= 8 \, g_s^4 \, \left( \frac{u}{t} - 2 \, \frac{u^2}{s^2} \right), \label{eqtracegamma1} \\
  \text{Tr} \left[ \psla_2 \, A^{\mu_4 \mu_3}(p_4,p_3) \, \psla_1 \, A^{\dagger \,\nu_4 \nu_3}(p_4,p_3) \right] \, \Pi_{\mu_3 \nu_3}(p_3,q_3) \, \Pi_{\mu_4 \nu_4}(p_4,q_4) &= 8 \, g_s^4 \, \left( \frac{t}{u} - 2 \, \frac{t^2}{s^2} \right), \label{eqtracegamma2} \\
  \text{Tr} \left[ \psla_2 \, A^{\mu_3 \mu_4}(p_3,p_4) \, \psla_1 \, A^{\dagger \,\nu_4 \nu_3}(p_4,p_3) \right] \, \Pi_{\mu_3 \nu_3}(p_3,q_3) \, \Pi_{\mu_4 \nu_4}(p_4,q_4) &= 8 \, g_s^4 \, \frac{t^2+u^2}{s^2}, \label{eqtracegamma3}
\end{align}
where the symbols $s$, $t$ and $u$ are the usual Mandelstam variables: $s=(p_1+p_2)^2$, $t=(p_1-p_3)^2$ and $u=(p_2-p_3)^2$ and 
\begin{align}
 \Pi_{\nu \tau}(p_i,q_i) &= \left( -g_{\nu \tau} + \frac{p_{i \, \nu} \, q_{i \, \tau} + p_{i \, \tau} \, q_{i \, \nu}}{p_i \cdot q_i} \right) \quad \text{for $i=3,4$}. 
 \label{eqdefPimunu}
\end{align}
The arbitrary four-momentum $q_i$ in eq.~(\ref{eqdefPimunu}) is light-like and not collinear to $p_i$. Also, the square matrix element does not depend on the choice of $q_3$ and $q_4$ but the different traces do. To get the right hand side of eqs.~(\ref{eqtracegamma1}), (\ref{eqtracegamma2}) and (\ref{eqtracegamma3}), we have used that $q_3 = p_4$ and $q_4=p_3$.
Putting everything together and after some algebra, we end up with the following result for the squared amplitude not averaged over the initial polarisations and colours
\begin{align}
|\mathcal{M}_{q\bar{q}\rightarrow 3g}|^2 &= 4\, g_s^6\, C_F\, N_c \, \left\{ E_{12} \, (u^2+t^2)\, (C_A-2\, C_F)\left[\frac{C_A-2\, C_F}{t\, u}+\frac{2\, C_A}{s^2}\right]\right.\nonumber\\
&\qquad \qquad \qquad \qquad + \left( E_{13} + E_{24} \right)\,  C_A\, \left[2\, C_F\left(\frac{t}{u}+\frac{u}{t}\right)-C_A\left(\frac{t}{u}+2 \, \frac{u^2}{s^2}\right)\right]\nonumber\\
&\qquad \qquad \qquad \qquad + \left( E_{14} + E_{23} \right)\, C_A\,\left[2\, C_F\left(\frac{t}{u}+\frac{u}{t}\right)-C_A\left(\frac{u}{t}+2 \, \frac{t^2}{s^2}\right)\right]\nonumber\\
&\qquad \qquad \qquad \qquad + \left. E_{34}\, (t^2+u^2)\, C_A^2\, \left[\frac{1}{t\, u}-\frac{2}{s^2}\right]\right\}. 
  \label{eqdefsoftamp1}
\end{align}
\noindent
This result is in agreement with the one given in ref. \cite{Ellis:1986bv}.
The coefficients $H_{ij}(0)$ can be read directly from eq.~(\ref{eqdefsoftamp1}) \footnote{The factor $g_s^6$ is not included in the coefficients $H_{ij}(0)$, cf. eq.~(\ref{eqdefhadcrxg5}).}. 
Note that the QED case ($q + \bar{q} \rightarrow \gamma + \gamma + \gamma$) is obtained by setting $C_A \equiv 0$, $C_F\equiv 1$ and replacing $g_s$ by $e\, Q_f$, where $Q_f$ is the electric charge of the initial fermions\footnote{Note that there remains a global factor $N_c$, which corresponds to the number of colours of the initial state $q \, \bar{q}$.}. We get:
\begin{align*}
|\mathcal{M}_{{\tiny q\bar{q}\rightarrow 3\gamma}}|^2 &=16\, e^6\, Q_f^6\, N_c\, E_{12}\, \left(\frac{t}{u}+\frac{u}{t}\right).
\end{align*} 
After that, the only surviving factor is $H_{12}(0)$ which is proportional to the Born squared amplitude. 
In the QCD case, as discussed previously, due to the fact that the soft gluon carries colour charge, these coefficients $H_{ij}(0)$ are not always proportional to the Born squared amplitude as seen in eq.~(\ref{eqdefsoftamp1}).
Having the different $H_{ij}(0)$ at our disposal, we can verify the cancellation of soft divergences between the real emission and the virtual ones, as described in equation (\ref{eqrels3}). We can extract the values for the factors $\cala^{(n)}$ and $\calc^{(n)}_{ij}$ in front of the divergences from reference \cite{Ellis:1985er}, considering the structure of the virtual corrections outlined in equation (\ref{eqvirtstruct}).
We indeed verify that
\begin{align}
\Hq{12}(0) + \Hq{34}(0) 
&=8\, C_F\, N_c\, \left[\frac{(C_A-C_F)^2+C_F^2}{t\, u}-\frac{2\, C_A\, C_F}{s^2}\right]\, (t^2+u^2),\\
\Hq{13}(0) + \Hq{24}(0) &=8\, C_F\, C_A\, N_c\, \left[2\, C_F\, \left(\frac{t}{u}+\frac{u}{t}\right)-C_A\, \left(\frac{t}{u}+2\frac{u^2}{s^2}\right)\right],\\
\Hq{14}(0) + \Hq{23}(0) &=8\, C_F\, C_A\, N_c\, \left[2\, C_F\, \left(\frac{t}{u}+\frac{u}{t}\right)-C_A\, \left(\frac{u}{t}+2\frac{t^2}{s^2}\right)\right],\\
\sum_{i=1}^{3} \sum_{j=i+1}^{4} \Hq{ij}(0)
&=2\, \, (C_A+C_F)\, |M_{q \bar{q} \rightarrow gg}^B|^2,
\end{align}
where $|M_{q \bar{q} \rightarrow gg}^B|^2$ is the squared Born amplitude not averaged over spins and colours and  stripped from the coupling constant of the reaction $q \bar{q} \rightarrow gg$. It is given by
\begin{align}
|M_{q \bar{q} \rightarrow gg}^B|^2&=8\, C_F\, N_c\, \left[\frac{C_F}{t\, u}-\frac{C_A}{s^2}\right]\, (t^2+u^2),
\end{align}
which is in agreement with ref.~\cite{Ellis:1985er}.

\section{Notations \label{notation}}

In this article, we introduced new notations to maintain the formulas as compact as possible. This appendix serves as a glossary where most of the notations used are summarised.
We have denoted by $\{x\}_N$ a sequence of labelled symbols $x_i$ where the index $i$ runs from 1 to $N$
\begin{align}
    \seq{x}{N} \equiv x_1, \, x_2, \cdots, \, x_{N}.
\end{align}
We also denoted by $(x)_N$ a collection (a sequence without comma) of symbols indexed by an integer 
\begin{align}
    \indic{x}{N} \equiv x_1 \, x_2 \cdots \, x_{N}.
\end{align}
And a symbol to describe a generic reaction where two bodies produce $N-2$ bodies, with all the bodies labelled by $i$ 
indexed by an integer running from 1 to $N$, is introduced as follows:
\begin{align}
    \reac{i}{N} \equiv i_1 + i_2 \rightarrow i_3 + \cdots  + i_{N}.
\end{align}
Furthermore, we also introduced a notation to describe a sequence, collection, or reaction where one symbol is replaced by another, for example:
\begin{align}
    \seqt{x}{x_i}{z_i}{N} &\equiv x_1, \, x_2, \cdots, \, x_{i-1}, z_i, \, x_{i+1}, \cdots, \, x_{N}, \\
    \indict{i}{i_k}{j_k}{N} &\equiv i_1 \, i_2 \cdots \, i_{k-1} \, j_k \, i_{k+1} \, \cdots \, i_N, \\
    \react{i}{i_k}{j_k}{N} &\equiv  i_1 + i_2 \rightarrow i_3 + \cdots + i_{k-1} + j_k + i_{k+1} + \cdots + i_{N}.
\end{align}
We also needed a special notation for the convolution. 
Specifically, when considering the convolution of two multivariate functions $f(a_1,\ldots,a_N)$ and $g(b_1,\ldots,b_K)$ where the convolution involves only the variables $a_k$ and $b_l$, we denoted it as:
\begin{align}
    \hspace{2em}&\hspace{-2em}\left[ f\left(a_1, \cdots, a_{k-1}, *, a_{k+1}, \cdots a_N\right) \otimes g(b_1, \cdots, b_{l-1}, * , b_{l+1}, \cdots , b_K)  \right]_{\eta} (x) \notag \\
    &\equiv \int_x^1 \frac{dz}{z^{\eta}} \, f\left(a_1, \cdots, a_{k-1}, z, a_{k+1}, \cdots a_N\right) \, g\left(b_1, \cdots, b_{l-1}, \frac{x}{z} , b_{l+1}, \cdots , b_K\right).
    \label{eqdefconv1}
\end{align}
Note that we also adhere to the convention that if a function $h$ involved in the convolution has only one argument, we write it in our special notation as simply $h$ instead of $h(*)$.
A word of caution regarding eq.~(\ref{eqdefconv1}): when $\eta \neq 1$, the convolution product defined in this equation is non-commutative. Therefore, in such cases, the function on the right-hand side of the product corresponds to the PDFs, while the one on the left-hand side corresponds to the DGLAP kernels. \\

\noindent
To describe the measures for the different phase spaces, we introduced the notations
\begin{align}
  d \text{PS}_{N-1 \,\text{h}}^{(n)}(\bar{x}) &\equiv \prod_{l=3}^{N-1} \frac{d \bx{l}}{\bx{l}^{n-2}} \, d y_l \, d^{n-2} \kt{l},
\end{align}
which is the measure of the phase space of the hadrons and
\begin{align}
   d \text{PS}_N^{(n)} &\equiv d y_N \, d \xt{N} \, \xt{N}^{n-5} \, d \phi_N \, (\sin \phi_N)^{n-4},
\end{align}
which is the measure of the phase space of the parton which can be soft and/or collinear to the other ones divided by $\xt{N}^2$.\\

\noindent
In addition, we provide a summary of the various variables used in this article.
Let us recall that $K_i$ represents the four-momentum of the hadron labelled by $i$, which is parameterised by the rapidity $y_i$, the 
magnitude of the transverse momentum $\kt{i}$, and the azimuthal angle between the transverse momentum $\vkt{i}$ and a reference 
direction in the transverse plane.
We also introduced an arbitrary energy scale $Q$ in order to build dimensionless variables. Furthermore, the parton $i$ which fragments into the hadron $i$ carrying a fraction of four-momentum $x_i$ has a four-momentum $p_i = K_{i}/x_i$.
Additional variables introduced include:
\begin{align*}
\omega &= Q/\sqrt{s}, &\Xt{i}&=\frac{\kt{i}}{Q},& \xt{i} =  \frac{\pt{i}}{Q} =\frac{\kt{i}}{x_i \, Q}.
\end{align*}
Due to the conservation of the energy and longitudinal momentum, the fractions of four-momentum $x_1$ and $x_2$ carried away by the initial partons from their parent hadrons are fixed
\begin{align*}
  x_1 & = \hat{x}_1 + \omega \, \xt{N} \, e^{y_N}, &
  x_2 & = \hat{x}_2 + \omega \, \xt{N} \, e^{-y_N},
\end{align*}
with
\begin{align*}
    \hat{x}_1&=\sum_{l=3}^{N-1} \frac{\kt{l}}{\sqrt{s} \, x_l} \, e^{y_l}, & \hat{x}_2&=\sum_{l=3}^{N-1} \frac{\kt{l}}{\sqrt{s} \, x_l} \, e^{-y_l}.
\end{align*}
Note that the constraint coming from the conservation of the transverse momentum has not been taken into account. This constraint 
can be expressed as:
\begin{align*}
    \sum_{l=3}^{N-1} \frac{\vkt{l}}{x_l} + \vpt{N} &= 0.
\end{align*}
At leading order, these fractions of four-momentum are denoted $\bx{1}$ and $\bx{2}$, and they are given by:
\begin{align*}
        \bx{1} &= \sum_{l=3}^{N-1} \frac{\kt{l}}{\sqrt{s} \, \bx{l}} \, e^{y_l}, & \bx{2} &= \sum_{l=3}^{N-1} \frac{\kt{l}}{\sqrt{s} \, \bx{l}} \, e^{-y_l}.
\end{align*}
It's worth noting that while the definition of $\bx{1}$ (and $\bx{2}$) may resemble that of $\hat{x}_1$ (and $\hat{x}_2$), the constraint on the transverse momenta is different at this order. Specifically, it reads:
\begin{align*}
    \sum_{l=3}^{N-1} \frac{\vkt{l}}{\bx{l}} &= 0,
\end{align*}
where we have named $\bx{i}$ the fraction of momentum carried by the hadron from its parent parton by similarity.\\

\noindent
The fact that $x_1$ and $x_2$ must be less than one determines the bounds on the rapidity of the parton $i_N$ which can be soft and/or collinear.
These bounds are given by
\begin{align*}
      \yNmax &= \ln \left( \frac{1-\hat{x}_1}{\omega \, \xt{N}} \right), & \yNmin &= \ln \left( \frac{\omega \, \xt{N}}{1-\hat{x}_2} \right).
\end{align*}
For convenience while integrating over $y_N$, we have defined
\begin{align*}
     \dyM &= \yNmax-y_j, &\dym &= y_j-\yNmin.
\end{align*}
When the parton $i_N$ is emitted collinearly to another parton $i_k$, a new variable $z_k$ is introduced which is bounded from above by $\zm{k}$ outside the cylinder of size $\xtm$:
\begin{align*}
    z_k &= \frac{\xt{k}}{\xt{k}+\xt{N}}, &\zm{k} &= \frac{\xt{k}}{\xt{k} + \xtm}.
\end{align*}
To enforce that the coefficient of the collinear singularity is a convolution of a partonic cross section times a DGLAP kernel, we trade $x_k$ for $\bx{k}=z_k \, x_k$. This bound becomes, for a hadron or a jet,
\begin{align*}
    \czm{k} &= \frac{\Xt{k}-\bx{k} \, \xtm}{\Xt{k}}, & \zmj &= \frac{\Xtj - \xtm}{\Xtj}.
\end{align*}
To conclude this appendix, we also needed to define quantities that enable us to distinguish between two regimes in the case of initial state singularities. Specifically,
\begin{align*}
    \beta_{1j} &= \omega \, \exp(y_j)/(1-\bx{1}), & \beta_{2j} &= \omega \, \exp(-y_j)/(1 - \bx{2}).
\end{align*}

\bibliographystyle{unsrt}
\bibliography{biblio1,biblio_jph,publi}
\end{fmffile}

\end{document}